\documentclass[prd,aps,tightenlines,superscriptaddress,showpacs,nofootinbib,
eqsecnum,amsfonts,amsmath, twocolumn]{revtex4}
\usepackage{amssymb,amsfonts,latexsym, soul}
\usepackage{fancyhdr}
\usepackage{acronym, multirow}
\usepackage[style=altlist]{glossary}
\usepackage{dcolumn}
\usepackage{subfigure, psfrag}

\def\RCS$#1: #2 ${\expandafter\def\csname RCS#1\endcsname{#2}}
\RCS$Date: 2007-06-29 14:47:45 $
\RCS$Revision: 1.10 $
\usepackage{graphicx}

\usepackage{hyperref}

\newacronym{GW}{gravitational wave}{}
\newacronym{BBH}{binary black hole}{}
\newacronym{BNS}{binary neutron star}{}
\newacronym{PBH}{primordial black hole}{}
\newacronym{SNR}{signal-to-noise ratio}{}
\newacronym{LIGO}{Laser Interferometer Gravitational wave Observatory}{}
\newacronym{LSC}{LIGO scientific collaboration}{}
\newacronym{GRB}{gamma-ray bursts}{}
\newacronym{ICB}{inspiralling compact binaries}{}
\newacronym{PSD}{power spectral sensitivity}{}
\newacronym{BHNS}{black hole - neutron star}{}
\newacronym{SPA}{stationary phase approximation}{}

\begin{document}
\bibliographystyle{unsrt}
\pagestyle{plain}
\fancypagestyle{plain}
\rfoot{}
\cfoot{\arabic{page} of \pageref{theend}}
\lfoot{}
\pagestyle{plain}
\lhead{Template placement -- v.\RCSRevision, \RCSDate}
\rhead{Template placement for inspiral searches}

\title{Gravitational waves  from inspiralling compact binaries:
hexagonal template placement and its efficiency in detecting physical signals.}

\author{ T. Cokelaer}
\affiliation{School of Physics and Astronomy, Cardiff University, Cardiff CF24 3YB, UK}

\begin{abstract}
Matched filtering is used to search for gravitational waves emitted by 
inspiralling compact binaries in data from the ground-based
interferometers.  One of the key aspects of the
detection process is the design of a \emph{template bank} that covers the
astrophysically pertinent parameter space. In an earlier
paper, we described a  template bank that is based on a \emph{square lattice}.
Although robust, we showed that the square placement is over-efficient, with
the implication that it is computationally more demanding than required.
In this paper, we present a template bank based on an \emph{hexagonal 
lattice}, which size is reduced by 40\% with respect to the proposed square
placement. We describe the practical aspects of the hexagonal template
bank implementation, its size, and computational cost. We have also performed
exhaustive simulations to characterize its \emph{efficiency} and \emph{safeness}.  We
show that the bank is adequate to search for a wide variety of binary systems (primordial
black holes, neutron stars and stellar mass black holes) and in data from both
current detectors (initial LIGO, Virgo and GEO600) as well as future 
detectors (advanced LIGO and EGO). Remarkably, although our
template bank placement uses a metric arising from a particular template
family, namely stationary phase approximation, we show that it can be used
successfully with other template families (e.g.,  Pad\'e resummation
and effective one-body approximation). This quality of being effective for different
template families makes the proposed bank suitable for a search that would use several of
them in parallel (e.g., in a binary black hole search). The hexagonal template bank
described in this paper is  currently used to search for non-spinning inspiralling
compact binaries in data from the Laser Interferometer Gravitational-Wave Observatory
(LIGO).
\end{abstract}

\maketitle

\section{Introduction}
Ground-based laser interferometer detectors such as {\LIGO}~\cite{LIGO} or
Virgo~\cite{VIRGO} are expected to detect {\GW}-signals in 
 data that have been, or will soon be, collected. The most promising and
well-understood astrophysical sources of gravitational waves are {\ICB} in close
orbits~\cite{Thorne}, which consist of two compact objects such as primordial
black holes,  neutron stars and/or stellar-mass black holes. 

Potentiality of a detection  verges towards one event per year. However, the
detection rate strongly depends on the {\ICB} coalescence rate
~\cite{kalogera:2004nt,kalogera:2004tn,OShaughnessy:2006b}  and the
volume of universe that detectors can probe. While we cannot
influence  the coalescence rates, we can increase the volume or distance at
which a signal can be detected, which highly depends on (i)~the design of the
detectors and their sensitivities, and (ii)~on the detection technique that is
used. Detector sensitivity can be increased most certainly; but once data have
been recorded, only the deployment of an optimal method of detection can ensure
the highest
detection probability, and that is a passport, not only to probe the largest
volume of Universe possible, but also to detect a GW-signal directly
for the first time. Fortunately enough, altough the two body problem cannot be
solved exactly in general relativity, post-Newtonian (hereafter PN)
approximation have been used to obtain  accurate
\emph{models} of the late-time dynamics of {\ICB}~\cite{LB}. Therefore, we can deploy a matched filtering technique,
which is an optimal method of detection when the signal buried in Gaussian and
stationary noise is known exactly. The models that we used for detection are
also called \emph{template families}.

The shape of the incoming {\GW}-signals depends on
various parameters, which are not known \textit{a priori} (e.g., the masses
of the two component stars in the case of a search for non-spinning binaries). Thus, we
have no choice but to
filter the data through a set of templates, which is also
called a \emph{template bank} and must cover the parameter
space that is astrophysically relevant. Since we cannot filter the data through an
infinitely large number of templates the bank is essentially discrete.
Consequently, the mismatch between any signal and the nearest
template in the discrete template bank will cause reduction in the {\SNR}.
Spacing between templates must be chosen so as to render acceptable this SNR
reduction as well as the computational demand required by the cross
correlation of the data with the entire discrete template bank. As we shall see,
the spacing between templates is set by specifying a \emph{minimal match}
between any signal and the template bank. In practice, template families are
approximation of the true gravitational wave signal, and no true signal will
perfectly match any of the template families. However, in this paper we
shall consider that template and simulated signal belong to the same template
family.

The template bank placement is one of the key aspects of the detection process.
Nonetheless, its design is not unique. There are essentially two types of
template bank placements. The first one does not assume any knowledge on the
signal manifold; the second does. The first type of placement computes matches
between surrounding templates until two templates have a match close to the
requested minimal match, and computes matches repeatedly over the entire
parameter space until it is fully populated. Using geometrical considerations,
an efficient instance of this technique has been developed~\cite{damir}.  A
second approach, described in various
papers~\cite{Owen96,OwenSathyaprakash98,squarebank},
utilizes a metric that is defined on the signal manifold. It
uses local flatness theorem to place templates at proper
distances~\cite{Owen96} over the parameter space. We developed a template bank
placement in~\cite{squarebank} that was implemented and fully
tested within the LIGO algorithm library~\cite{LAL}. This template bank was used
in the analysis of data from different LIGO science runs
~\cite{LIGOS1iul, LIGOS2iul,LIGOS2macho, LIGOS2bbh, LIGOS3S4}. We also shown
that although robust with respect to the requirement (matches should be
above the minimal match), it is over-efficient. This result was expected
because we used a square lattice to place templates over the parameter space.

In this paper, we fully describe and validate a hexagonal template bank
placement that is currently used by the {\LIGO} scientific collaboration so as to analyze
the most recent science runs.  In Section~\ref{sec:formalism}, we
recapitulate some fundamental techniques and notions that are needed to describe
the bank placement, and previous results on the square template
bank placement. We also  provide a framework to validate a template bank. In
Section~\ref{sec:algorithm}, we describe the
algorithm that places templates on a hexagonal lattice. 
In Section~\ref{sec:simulation}, we summarize the outcome of the simulations
performed to test the hexagonal bank. We envisage various parameter spaces
that allows to search for {\BNS}, {\PBH}, {\BHNS} and, or {\BBH}
signals. We also considered design sensitivity curves for the
current and advanced generation of ground-based detectors. In Section~\ref{sec:simspa},
we show that the proposed hexagonal template bank has the required specifications.

Finally, in addition to the case of a template family based on the stationary phase
approximation, we also investigate in Section~\ref{sec:simother} the possibility to use
the same hexagonal bank placement with other template families including Pad\'e
resummation and effective one-body approximation. We show that there is no need to
construct specific template bank for each template family: the proposed bank can be used
for the different families that we looked at in this paper.

\section{Formalism and template bank validation}\label{sec:formalism}
Matched filtering and template bank placement use formalisms that are 
 summarized in this Section. We also review the main results of
the square placement, and recapitulate the framework introduced
in~\cite{squarebank} that allows us to validate a template bank. 

\subsection{Signal and Metric}
The matched filtering technique is an optimal method to detect a known signal,
$s(t)$, that is buried in a stationary and Gaussian noise,
$n(t)$~\cite{Helmstrom}. The method
performs a correlation of the data $x(t)=n(t)+s(t)$ with a template $h(t)$.
In this paper, we shall assume that $s(t)$ and $h(t)$ are generated with the
same model so that a template can be an exact copy of the signal. Matched
filtering of the data $x(t)$ with a template $h(t)$ can be expressed via the
inner product weighted by the noise {\PSD}, $S_h(f)$,  and is given by
\begin{equation}\label{eq:innerproduct}
(x,h) = 2 \int_0^\infty \frac{\tilde{x}^*(f)
\tilde{h}(f)+\tilde{x}(f)\tilde{h}^*(f)}{S_h(f)}df \,	.
\end{equation}
Note that for simplicity, we will ignore the time $t$ within the inner product
expressions. A template and a signal can be normalized according to 
\begin{equation}
\hat{h} = \frac{h}{\sqrt{(h,h)}},\, \hat{s} = \frac{s}{\sqrt{(s,s)}} \,.
\end{equation}
The {\SNR} after filtering by $h(t)$ is 
\begin{equation}
\rho = \frac{(x,h)}{\sqrt{(h,h)}} = (x, \hat{h}).
\end{equation}
The simulations that we will perform assume that template and
signal are normalized, that is $(\hat{h},\hat{h})=1$, and
$(\hat{s},\hat{s})=1$. In this paper, we are interested in the fraction of the optimal SNR
obtained by filtering the signal $x(t)$ with a set of template $h(t)$, therefore, we can
ignore the noise $n(t)$, and $(x, \hat{h})$ becomes $(\hat{s}, \hat{h})$. Strictly
speaking, $(\hat{s},\hat{h})$ does not refer to a SNR anymore, but to the ambiguity
function, which is by definition always less than or equal to unity if the two waveforms
are normalized. In the following, we shall use the notion of \emph{match} introduced
in \cite{Owen96}; the match between two templates is the inner product between two
templates that is maximized over the time (using the inverse Fourier transform) and the 
initial orbital phase (using a quadrature matched filtering).

The incoming signal has unknown parameters and one needs to filter the
data through a set of templates, i.e., a template bank. The templates
are characterized by a set of $p$ parameters $\vartheta^{\mu},\;\mu=0,1,..,
p-1$. The templates in the bank are copies of the signal corresponding
 to a set of values $\vartheta^{\mu}_i,\;i=0,1,..,N_b-1$, where $N_b$ is
the total number of templates. A template bank is optimally designed if $N_b$ is
minimal and if for any signal there always exists at least one template in the
bank such
that \begin{eqnarray}\label{eq:mm}
\min_{\vartheta^{'\mu}} \max_{i}
(\hat{s}(\vartheta^{'\mu}), \hat{h}(\vartheta^{\mu}_i)) \ge MM,
\end{eqnarray}
where $MM$ is the minimal match mentioned earlier. Usually, in searches for
{\ICB}, the value of the minimal match is
set by the user to 95\% or 97\%, which corresponds to a decrease in
detection rate of 15\% and 9\%, respectively. Nevertheless, the minimal match
may have a much smaller value for the first stage of a
hierarchical search (e.g., 80\%), or for a one-stage search of periodic signals (e.g.,
70\% or lower).

The distance between two infinitesimally separated normalized templates on the
signal manifold is given by~\cite{Owen96, OwenSathyaprakash98}
\begin{eqnarray}
||h(\vartheta^{\mu} + d\vartheta^{\mu}) - h(\vartheta^{\mu}) ||^2
& = &  \left|\left| h_{\mu} d\vartheta^{\mu} \right|\right|^2 
       \nonumber \\
& = & \left( h_{\mu}, h_{\nu}\right)\,d\vartheta^{\mu}
      d\vartheta^{\nu} \nonumber \\
& \equiv & g_{\mu\nu}d\vartheta^{\mu}
      d\vartheta^{\nu},\label{norm}
\end{eqnarray}
where $h_\mu$ is the partial derivative of the signal $h$ with respect to the
parameter $\vartheta^{\mu}$. So, the quadratic form
\begin{equation}\label{eq:metric}
g_{\mu\nu} = (h_\mu, h_\nu)
\end{equation}
defines the $g_{\mu\nu}$ metric induced on the signal manifold. The
metric is used to place
templates at equal distance in the parameter space. The distance $dx_i$ between
templates in each dimension is given by
\begin{equation}\label{eq:step}
dx_i = 2 \sqrt{\frac{1-MM}{g_{ii}} }, i=0,1,..,p-1\,.
\end{equation}
In practice, using such $dx_i$ leaves a fraction of the parameter space
uncovered, and overlap between templates is required (e.g., in the square
placement, spacing is actually set to $dx_i/\sqrt{2}$).

Since we restrict ourself to the case of non-spinning waveforms, $h(t)$ depends 
on 4 parameters only: the two component masses, $m_1$ and $m_2$
which may vary from sub-solar mass to tens of solar mass systems, the initial
orbital phase $\varphi_C$, and the time of coalescence $t_C$. We can maximized
over $t_C$  and $\varphi_C$  analytically, therefore the parameter space that we need
to cover with our template bank is a 2-dimensional space only. For
conciseness, we can represent the
GW-waveform with a simplified expression given by 
\begin{equation}
h(t) = \frac {4A\eta M}{D} \left[\pi M f(t)\right]^{2/3}
\cos [\varphi(t) + \varphi_C],
\label{eq:waveform1}
\end{equation}
where $f(t)$ is the (invariant) instantaneous frequency of the signal measured
by a remote observer, the phase of the signal $\varphi(t)=2\pi\int^{t_C} f(t)
dt$ is defined so that it is zero when the binary coalesces at time
$t=t_C$, and $A$ is a numerical constant representing the amplitude~\cite{DIS2}.
The asymmetric mass ratio
is $\eta=m_1m_2/M^2$, where $M=m_1+m_2$ is the total mass of the system.
There exist amplitude corrections up to 2.5PN~\cite{AmplitudeCorrection},
the importance of which for detection and estimation is shown in~\cite{VDB}.
However, in
this work, we use
restricted post-Newtonian models only and limit PN-expansion of the phase to
2PN order. Moreover, in the template bank placement, namely for the metric
computation, we consider the {\SPA}~\cite{SD91},
for which the metric can be derived analytically~\cite{squarebank}.
Nevertheless, other template families can be used both for injection and
filtering (see Section~\ref{sec:models}).

\subsection{Example : the Square Template Bank}

The placement that we proposed in~\cite{squarebank} uses the metric based
 on the {\SPA} model, and the spacing
$dx_i$, as defined in Eq.~(\ref{eq:step}). Since the model $h$ explicitly
depends on the two mass parameters $M$ and $\eta$, then the spacing $dx_i$ are
function of these two
quantities as well. However, the metric expressed in these two coordinates is
not a constant; it is not a constant either if we were to use the component
masses, $m_1$ and $m_2$. The
preference of chirptimes, denoted $\tau_0$ and $\tau_3$ (see appendix
\ref{annex:tools}, Eqs.~\ref{eq:t0t3}) as coordinates on the
signal manifold is indeed more practical because these variables are almost
Cartesian~\cite{DS92,SD91}. Although not perfectly constant for PN-order larger
than 1PN, we shall assume that the metric is essentially constant in the local
vicinity of every point on the manifold. We could use any
 combinations of chirptimes, but using the pair $\tau_0-\tau_3$, there exists
analytical inversion with the pair $M-\eta$ (see appendix \ref{eq:Meta}).

\begin{figure}[t]
\centering
{\includegraphics[width=0.4\textwidth]{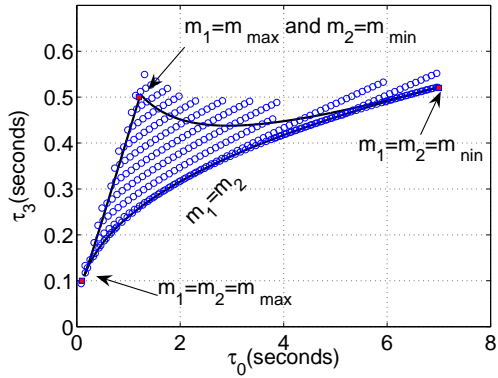}}
\caption{Example of parameter space and template bank placement.
The parameter space is defined by
the individual mass components (from $3~M_\odot$ to $30~M_\odot$) and the lower
cut-off frequency ($f_L=40$~Hz). The bottom line corresponds to $m_1=m_2$, or
equivalently to $\eta=0.25$. The two other boundaries meet where
$m_1=m_{\rm min}$ and $m_2=m_{\rm max}$. The bottom left point of the
parameter
space corresponds to $m_1=m_2=m_{\rm max}$ whereas the top right point
corresponds to $m_1=m_2=m_{\rm min}$. The gray/blue circles give the position
of each template that is needed to cover the entire parameter space (black
curves). Even though some templates lie outside the parameter space boundaries,
these are required to fully cover the parameter space. }
\label{fig:space}
\end{figure}

The parameter space to be covered is defined by the minimum and maximum
component masses of the systems considered ($m_{\rm min}$ and $m_{\rm
max}$), and possibly the minimum and maximum total mass ($M_{\rm min}$ and
$M_{\rm max}$) as shown in Fig.~\ref{fig:space}. The lower cut-off frequency
$f_L$, at
which the template starts in
frequency, sets the length of the templates and therefore directly influences
the metric components, the
parameter space, and the number of templates $N_b$. In~\cite{squarebank}, we
showed how the size of the template bank changes with $f_L$. We also investigated
the loss of match due to the choice of $f_L$. We generally set $f_L$ so
that the loss of match is of the order of a percent. 

\begin{figure}[tbh]
\centering
{\includegraphics[width=0.45\textwidth]{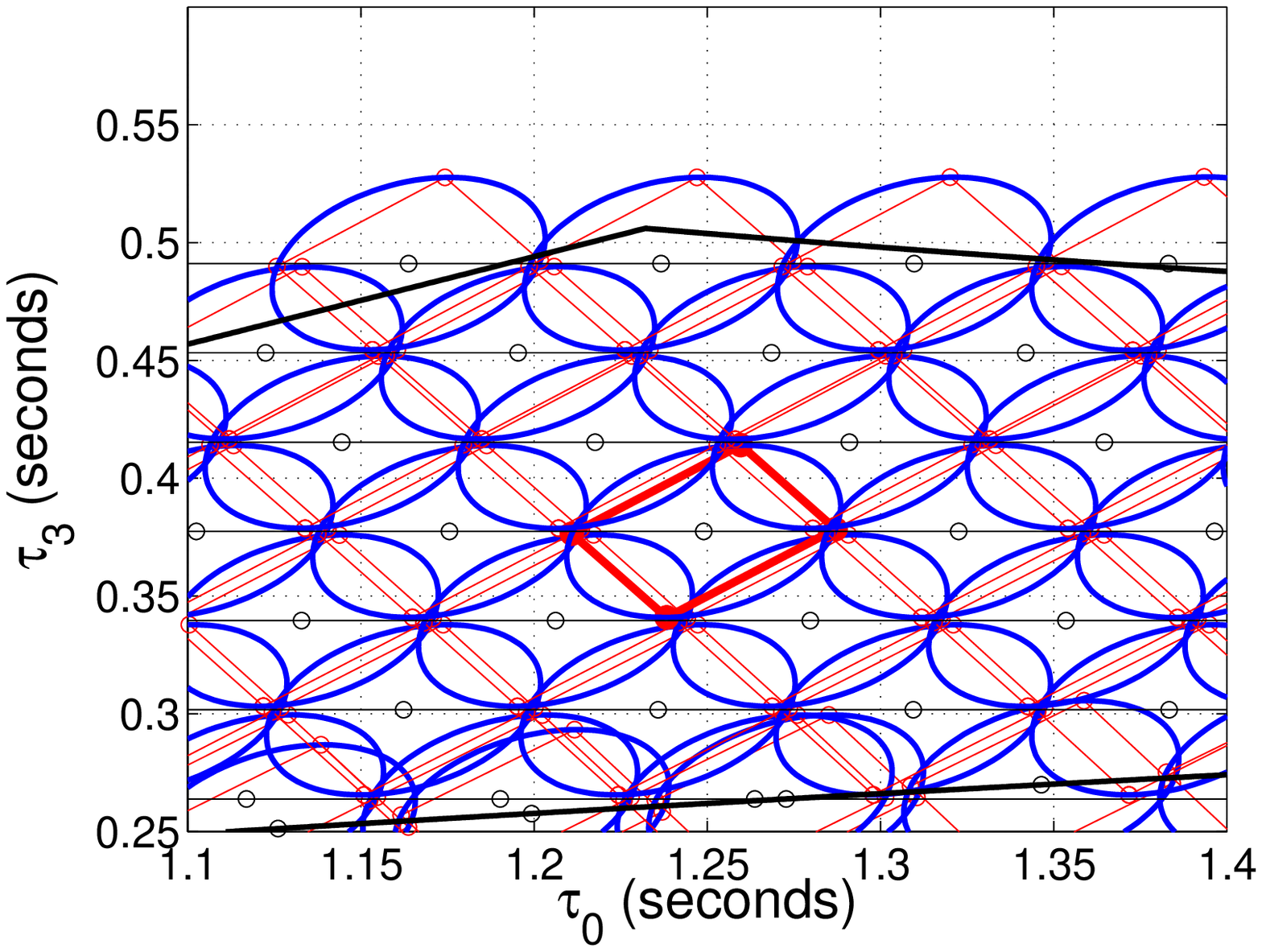}}\\
{\includegraphics[width=0.45\textwidth]{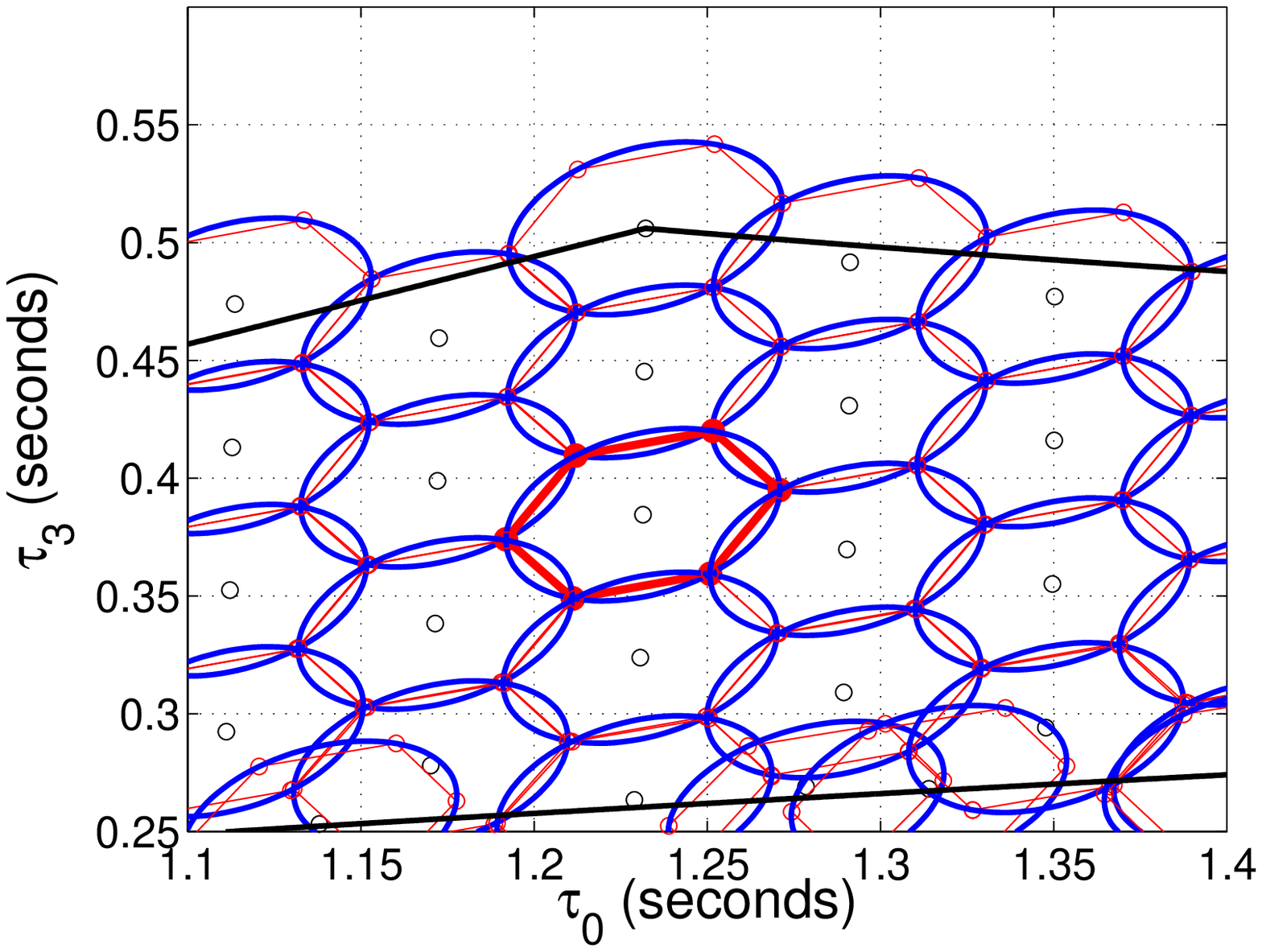}}
\caption{\label{fig:squarevshexa} Two instances of template bank placements. In
the two plots, we focus on a small area of the parameter space presented in
figure~\ref{fig:space}. We used a square (top panel) and hexagonal (bottom
panel) placement. For convenience, we re-scale the metric components so that,
$g_{00}\sim g_{11}$. Each template position is represented by a small circle. 
Around each template position, we plot an ellipse that represents iso-match
contour of $MM=0.95\%$. Each ellipse contains an inscribed square or
hexagon which emphasizes how ellipses overlap each other. We can see
that squares (top) slightly overlap each other. This is because templates are layed along $\tau_3$ equal constant line and not along the eigen-vector directions, which change over the parameter space. In the hexagonal
placement, we take care of this problem short-coming, and therefore hexagons are
perfectly adjacent to each other: the placement is optimal.}
\end{figure}

We briefly remind how the proposed square template bank works. First,
templates are placed along the $m_1=m_2$ or $\eta=0.25$ line starting from the
minimum to the
maximum mass. Then, additional templates are placed so as to cover the remaining
part of the parameter space, in rows, starting at $\eta=0.25$ along lines of
constant $\tau_3$ until a template lies outside the parameter space. The
spacing between lines is set adequately.  Distances
between templates are based on a square lattice. An example of such a
placement
is shown in Fig.~\ref{fig:squarevshexa}. One of the
limitations of the placement is that templates are not placed along the
eigenvectors of the metric but along the standard basis vectors that
describe the $\tau_0,\tau_3$ space. This approximation make the ellipses
slightly more overlapping than requested and may also create holes when the
orientation of the ellipses varies significantly (i.e., at high mass regime).
The square placement is also over-efficient as compared to a hexagonal
placement (see Fig.~\ref{fig:square}).

\subsection{Bank Efficiencies}\label{sec:be}
Independently of the template bank placement, the
template bank must be validated to check whether it fulfills the requirements
(e.g., from Eq.~\ref{eq:mm}). First, we perform Monte-Carlo simulations so as to
compute the \emph{efficiency} vector, $\mathcal{E}$,  given by
\begin{equation}
\mathcal{E}\left(
\mathcal{\chi}_{s},
 \mathcal{\chi}_{h}\right)
= \left\{
\max_{j}\left(\hat{s}(\vartheta^s_i), \hat{h}(\vartheta^h_j)  
\right)
\right\}_{i=1..N_s, j=1..N_b}
\end{equation}
where $N_b$ is the number of templates in the bank, $N_s$ the number of
injections.

The vectors $\vartheta^{s}$ and $\vartheta^{h}$ correspond
to
the parameters of the simulated signals and the templates, $\chi_s$
and $\chi_h$ are the models used in the generation
of the signal and template, respectively.  In all the simulations, we set 
$\vartheta^{s} = \{m_1, m_2, \varphi_C, t_C\}$. Furthermore, we  can
analytically maximize over the unknown orbital phase $\varphi_C$ and, therefore,
$\vartheta^{h} = \{m_1, m_2\}$. 

\begin{figure}[tbh]
\centering
{\includegraphics[width=0.4\textwidth,height=0.35\textwidth]
{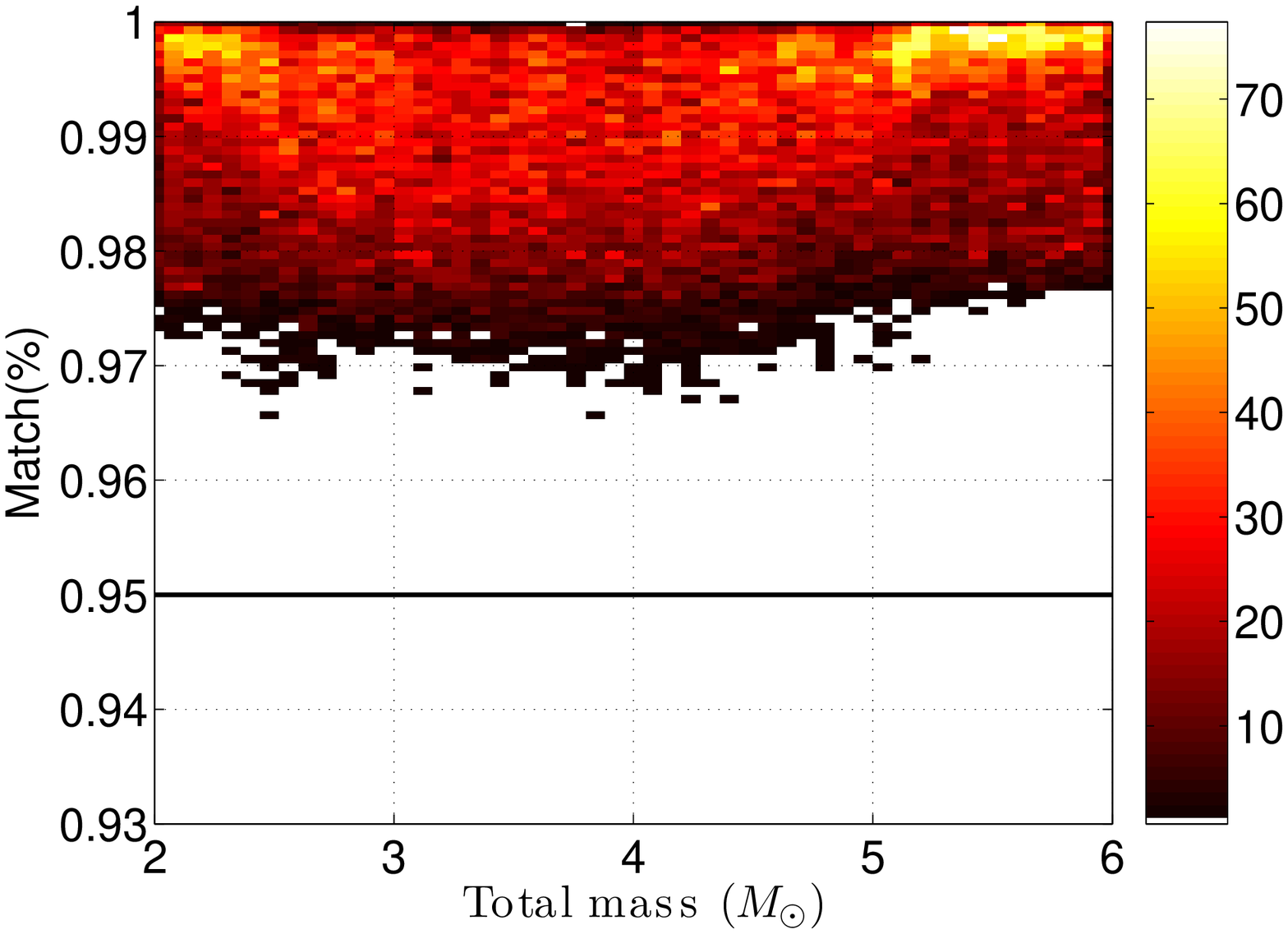}}\\
{\includegraphics[width=0.4\textwidth,height=0.35\textwidth]
{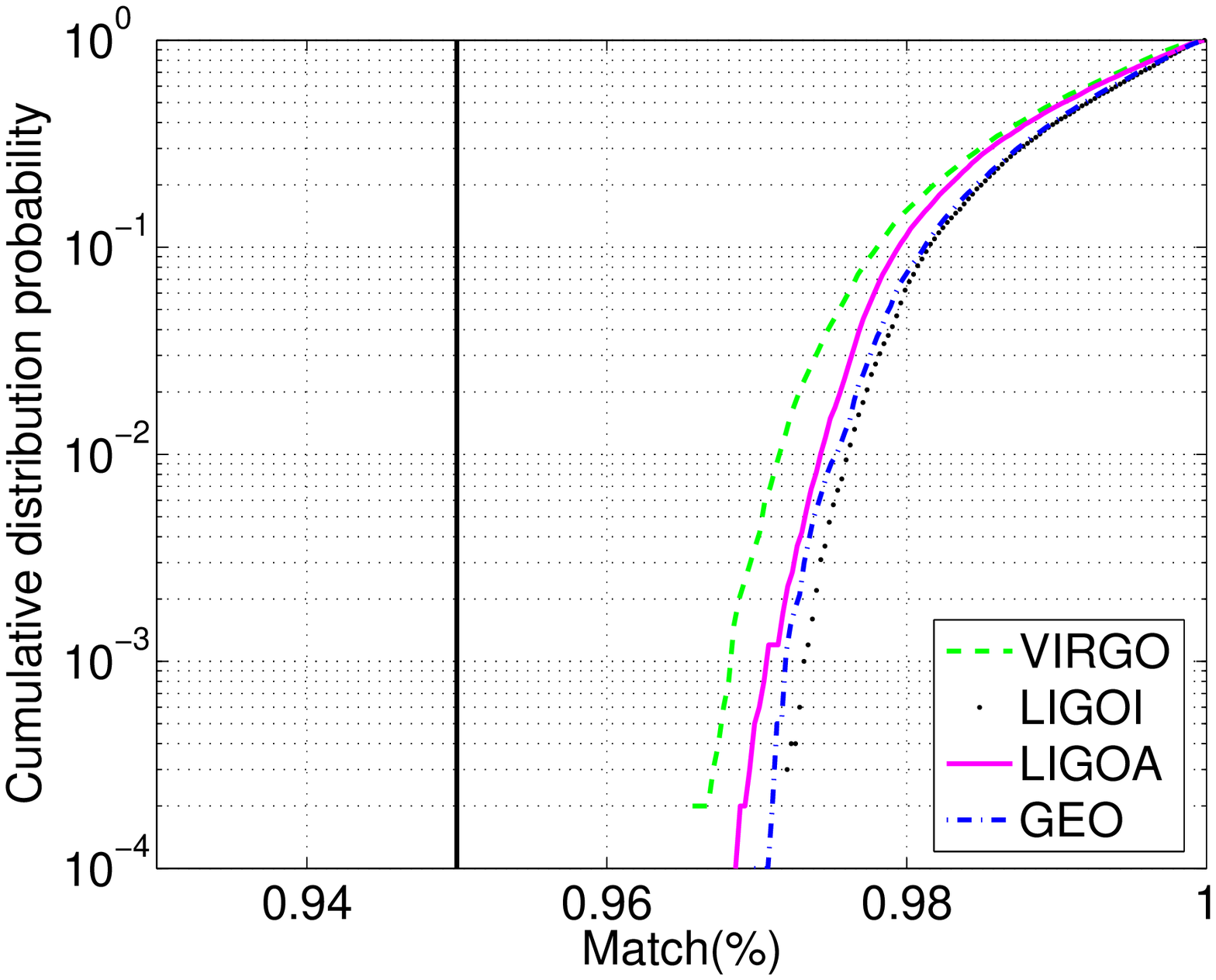}}
\caption{\label{fig:square} Efficiencies of the square template bank. For
convenience we remind the reader of some results of the square template bank
provided in~
\cite{squarebank}. 
In the simulations, we used stationary phase approximant models for both injections and
templates.
Injections consist of  binary neutron stars. We used 4  design sensitivity
curves (LIGOI, advanced LIGO, VIRGO and GEO), and for each of them we
performed
10,000 injections. In the top panel, we show all the results together: 
all injections are recovered with a match higher than 95\%, as requested. In the
bottom panel, we decomposed the 4 simulations and show that all of them behave
similarly. Actually, we can see that most of the injections are recovered with
even higher matches (above 97\%) showing the over-efficiency of the placement.}
\end{figure}

The efficiency vector $\mathcal{E}$ and the signal
parameter vector    $\vartheta^s$ are useful to derive several figures of
merit. The cumulative distribution of $\mathcal{E}$ (Fig.~\ref{fig:square},
bottom panel) indicates how quickly matches drop as the minimal match is
reached. Nevertheless, the cumulative distribution function of $\mathcal{E}$
hides the dependency of the matches upon masses. Therefore, we also need
to look at the distribution of  $\mathcal{E}$
versus total mass $M$ (e.g., Fig.~\ref{fig:square}, top panel), or
versus $\eta$, or chirp mass, $\mathcal{M}$ (see appendix for an exact definition).
Usually, we look at
$\mathcal{E}_M$ only. Indeed, in most
cases, the dynamical range of $\eta$ is small (from 0.1875 to 0.25 in the {\BNS}
case). Finally, we can quantify the efficiency of a template bank with a unique
 value, that is the {\it safeness}, $\mathcal{S}$, given by
    \begin{equation}
    \mathcal{S}_{\mathcal{R}}\left(
    \mathcal{\chi}_{s},
    \mathcal{\chi}_{h}\right) = \min  \mathcal{E}(\chi_s, \chi_h).
    \end{equation}
Ideally, we should have a template bank such that $\mathcal{S}_\mathcal{R} \approx 
MM$. $\mathcal{S}_\mathcal{R}$ is a generalization of the left hand side of
Eq.~\ref{eq:mm} on $N_s$ injections. The higher $N_s$ is, the more
confident we are with the value of the safeness. Ideally, the number $N_s$
should be several times the size of the template bank that is $N_s\gg N_b$, so
that statistically we have at least one injection per template. The sub-index
$\mathcal{R}$ of the safeness is the ratio between $N_s$ and $N_b$ and indicates
the relevance of the simulations. The safeness provides also a way of characterizing the
template bank: if $\mathcal{S}_\mathcal{R}$ is less than the expected minimal match
$MM$, then the bank is \emph{under-efficient}. Conversely, a template bank can be
over-efficient like in Fig.~\ref{fig:square}.

\section{Hexagonal placement based on the metric}\label{sec:algorithm}
 In the $\tau_0,\tau_3$ basis vectors, both amplitude and orientation of the
eigenvectors change, which may imply a laborious placement. In this Section, we
describe the hexagonal placement that is conceptually different from the square
placement and takes into account the eigenvectors change throughout the
parameter space. 

\begin{figure*}[tbh]
\centering
\subfigure[Initialization]
{\includegraphics[width=0.29\textwidth,height=0.2\textheight,angle=0]
{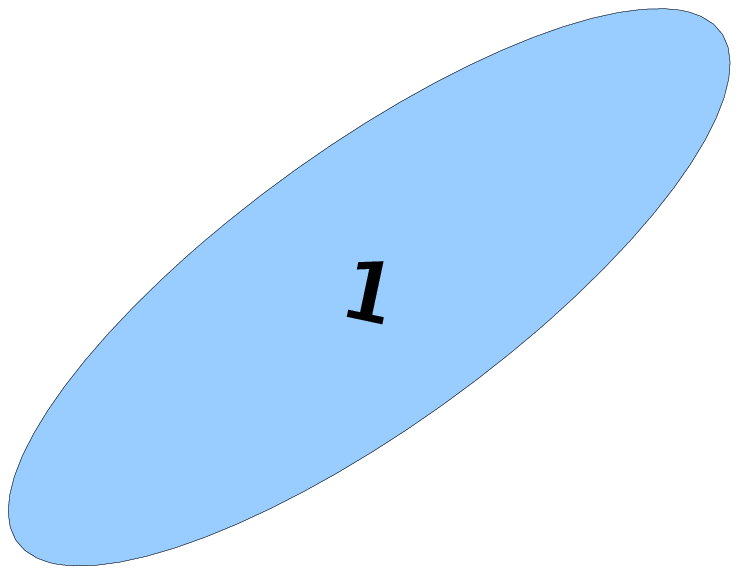}}
\subfigure[Reproduction cell 1]
{\includegraphics[width=0.29\textwidth,height=0.2\textheight,angle=0]
{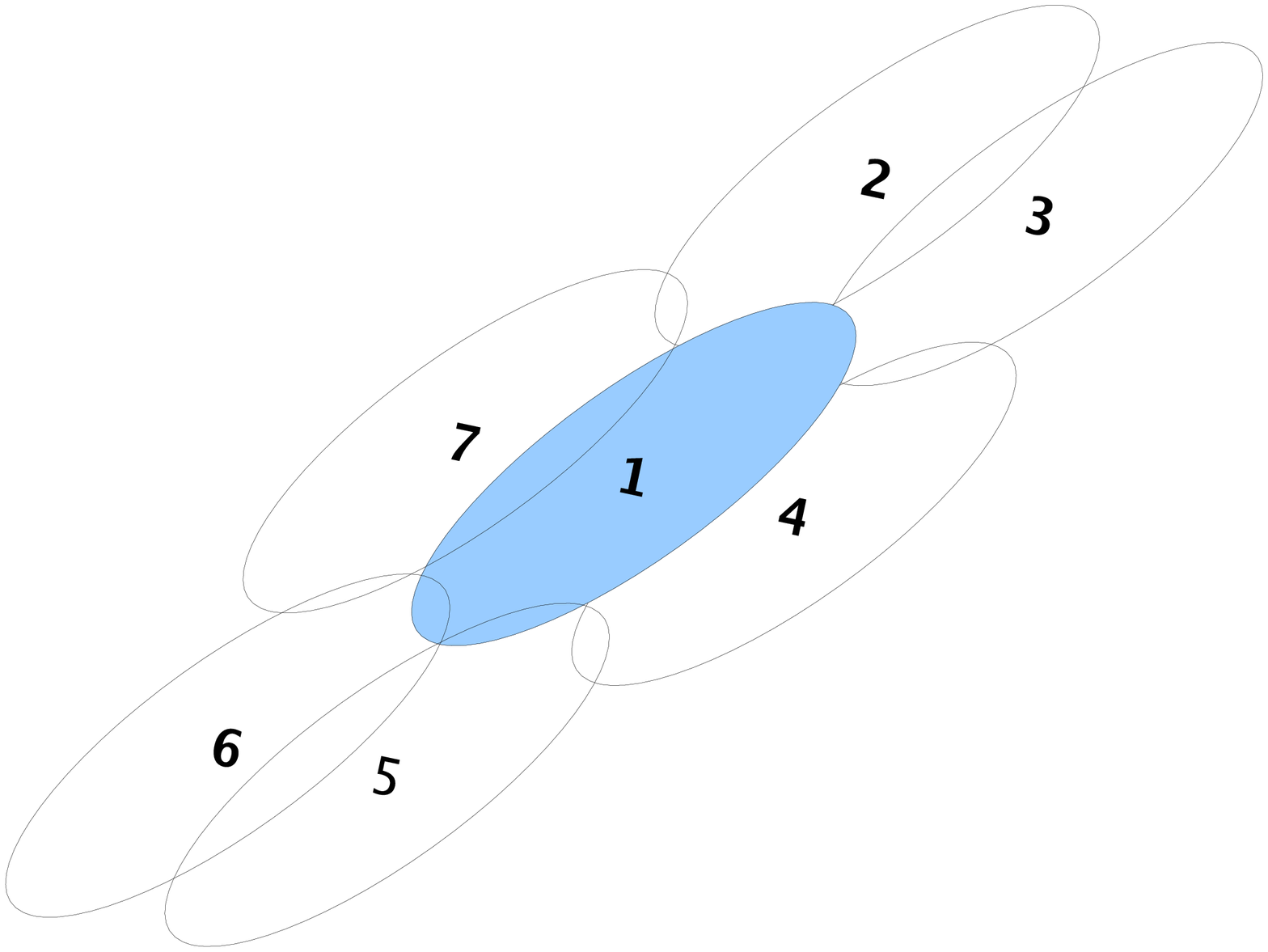}}
\subfigure[Connections]
{\includegraphics[width=0.29\textwidth,height=0.2\textheight,angle=0]
{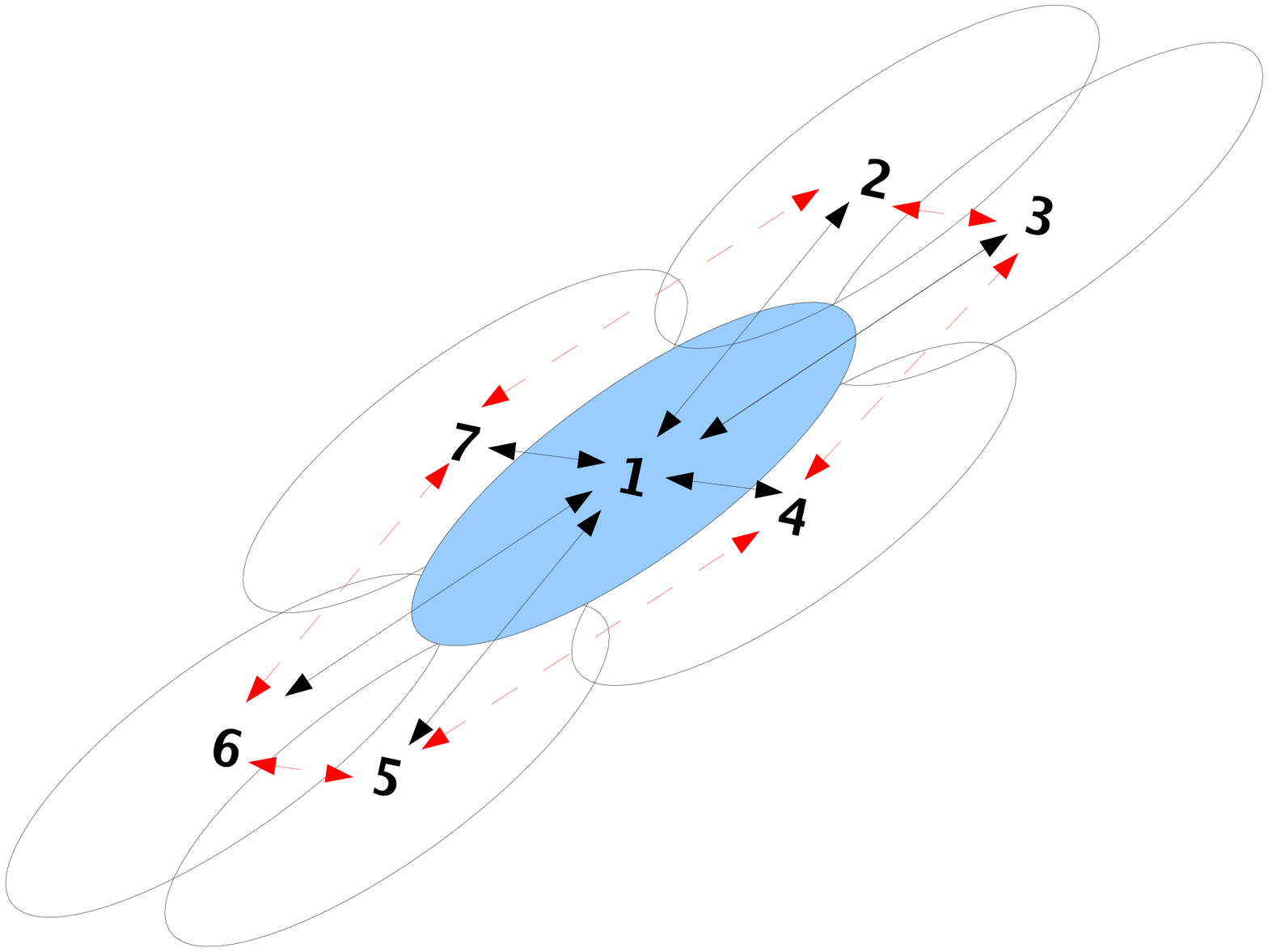}} \\
\subfigure[Initialization]
{\includegraphics[width=0.29\textwidth,height=0.2\textheight,angle=0]
{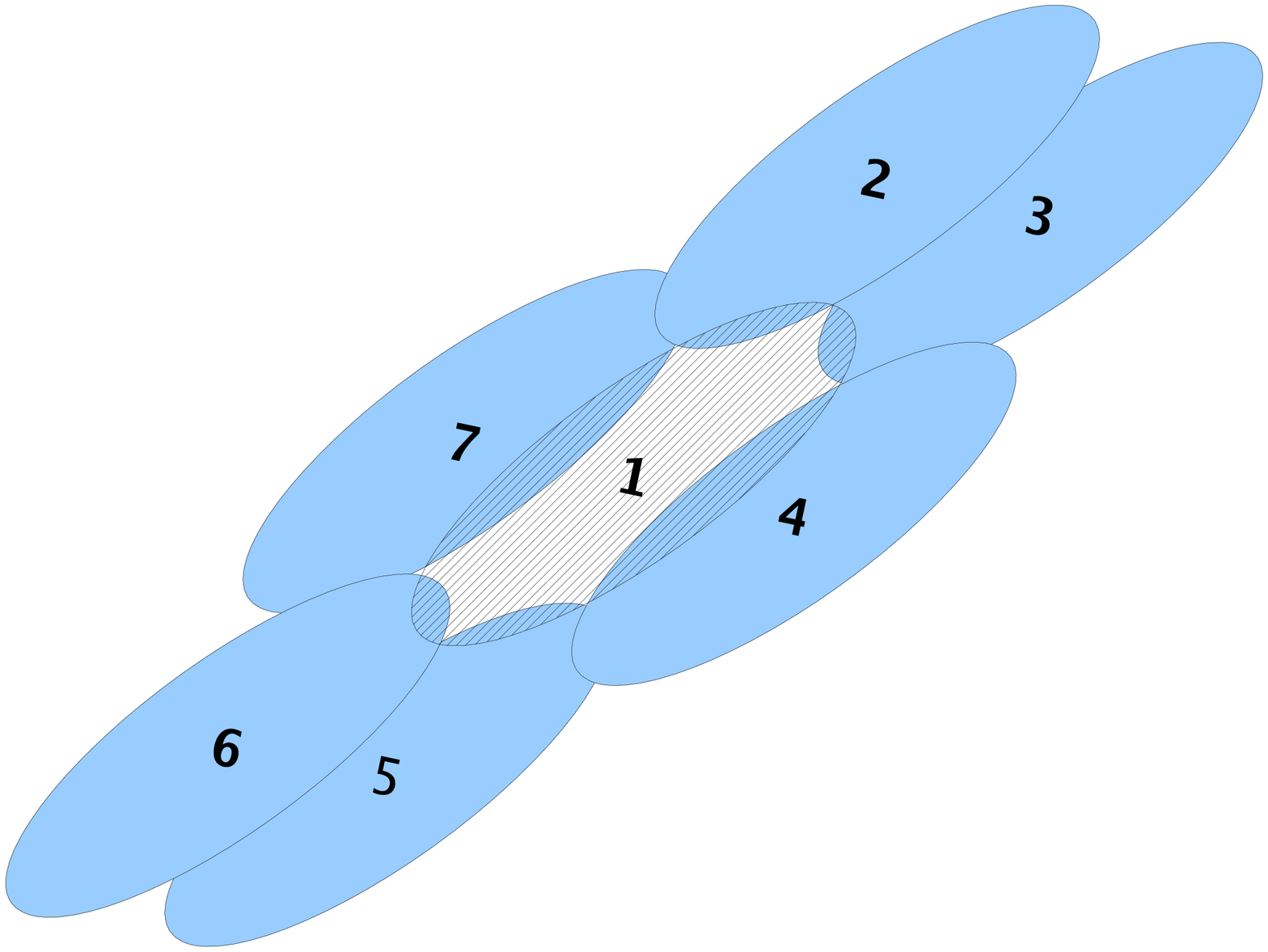}}
\subfigure[Reproduction cell 2]
{\includegraphics[width=0.29\textwidth,height=0.2\textheight,angle=0]
{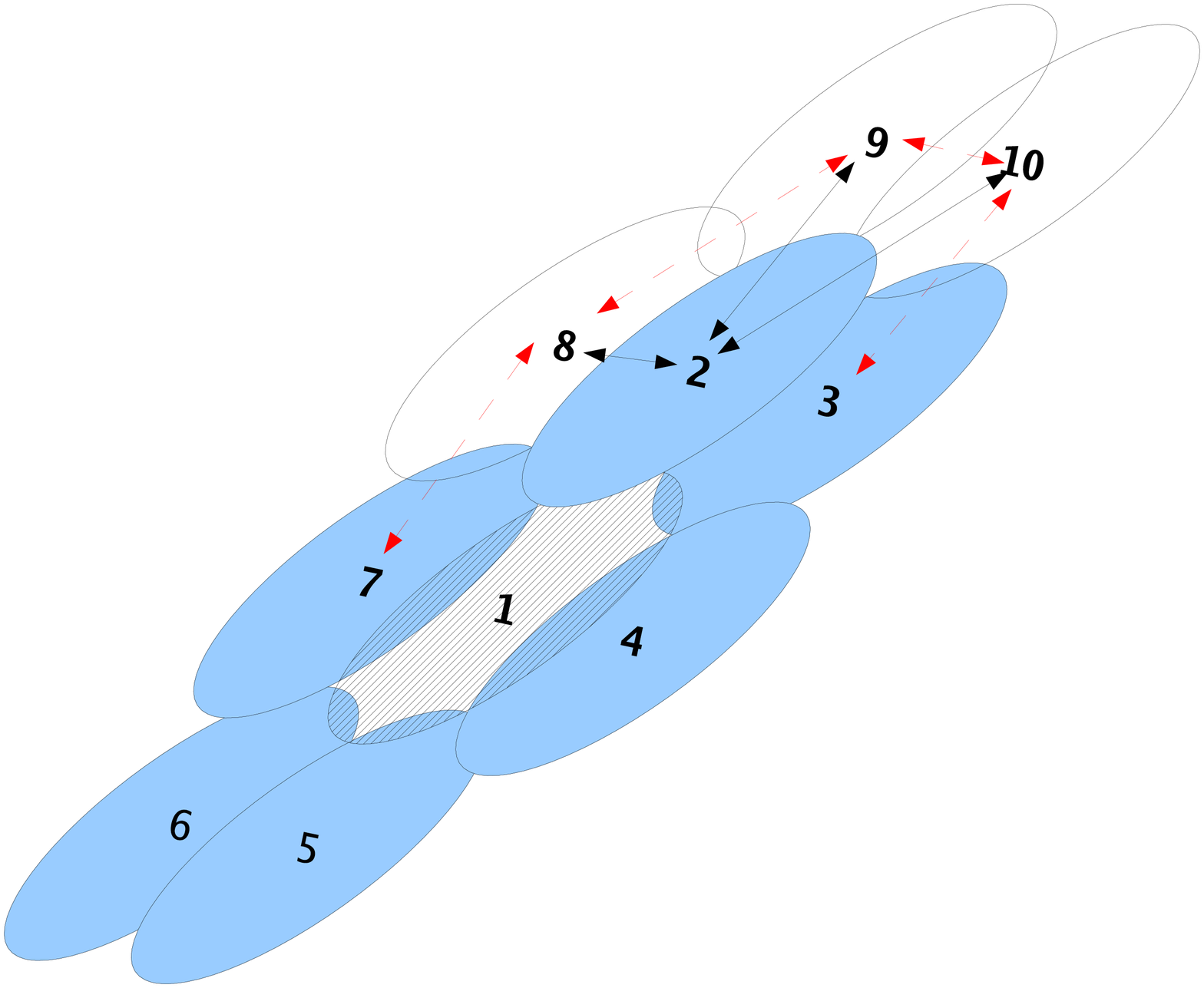}}
\subfigure[Reproduction cell 3]
{\includegraphics[width=0.29\textwidth,height=0.2\textheight,angle=0]
{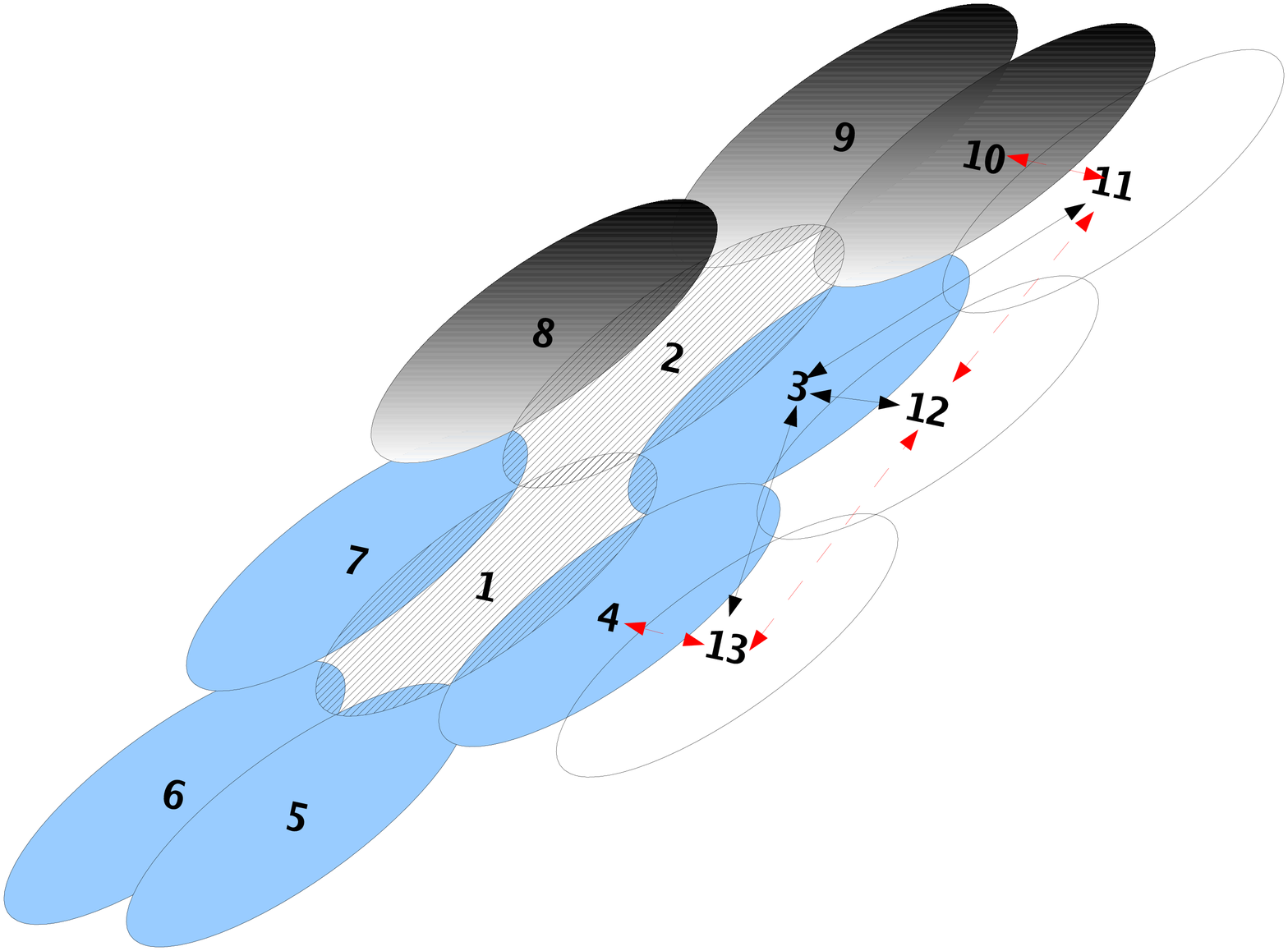}}
\caption{Hexagonal template bank placement. Using the terminology that is
introduced in the text, we can describe the template bank placement algorithm as follows.
First, in sub-figure~(a), a cell/template is arbitrary placed in the parameter space. Its
coordinates correspond to $m_1=m_{\rm min}$ and $m_2=m_{\rm max}$, and its Id is 1. In
sub-figure~(b), the cell with ID=1, which is also a mother cell, reproduces 6 new cells
according to an optimal hexagonal lattice that takes into account its metric components.
In sub-figure~(c) the connection between the offsprings and the mother cell are created
(black
arrows). Cells that belong to the same generation (white ellipses)
 are also connected if there are adjacent to each other (red arrows). In sub-figure~(d),
the new cells can start to reproduce. However, the
reproduction is exclusive: reproduction takes place cell after cell, and the cell
with lowest ID is chosen to reproduce first.  In sub-figure~(e), therefore, the
cell with ID equals 2 starts to reproduce. Because connections already exist with other
cells, this cell will reproduce only 3 directions (i.e., 8, 9 and 10). In sub-figure~(f),
the cell with ID equals 3 starts to populate. The 3 new cells (gray ellipses) created by
the cell with ID=2  have to wait for the current generation (blue ellipse) to fully
reproduce. The cells spread until
the boundary of the parameter space is reached. \label{fig:hexalgo}}
\end{figure*}    

\subsection{Algorithm}
Although the hexagonal placement algorithm is independent of any genetic
 or evolutionary algorithms, it can be  compared to biological process,
and we will use this analogy to explain the placement. First, let us introduce
a \emph{cell} that contain a template position (e.g., $\tau_0-\tau_3$), the
metric components defined at this position, and a unique identification
 number that we refer to as an ID. A cell covers an area defined by an
ellipse with semi-axis equal
to $dx_i/2$. The goal of a cell is to populate the parameter space with an
offspring of at most 6 cells (hexagonal placement). 
 A cell can be
characterized by the following principles: 

\begin{description}
\item[1-Initialization] A cell is created at a given position in 
$\tau_0-\tau_3$ plane, not necessarily at a physical place (i.e, $\eta$ can 
be less than 1/4). The initialization requires that
\begin{itemize}
 \item metric components at $(\tau_0-\tau_3)$ are calculated,
 \item a unique ID number is assigned,
 \item 6 connectors are created and set to zero.
\end{itemize}
Finally, if the cell area intersects with the parameter space, then it has the
ability to survive in its environment: it is \emph{fertile}.
Conversely, a cell whose coverage is entirely outside the parameter space
is
\emph{sterile}.

\item[2-Reproduction] A fertile cell can reproduce into 6 positions that are
the corner of a hexagon inscribed in the ellipse whose semi-axes
are derived from the metric components $dx_i$'s. A
cell that has reproduced is a \emph{mother cell} and its offspring is composed
of 6 \emph{daughter cells}. Once 
a daughter cell is initialized, it cannot reproduce in place of its mother. This
is taken into account via the connection principle.

\item[3-Connection] Following the reproduction process, a mother cell
sets the connections with its daughter cells by sharing their IDs.
Therefore, a mother cell knows the IDs of its daughter cells and vice-versa.
Moreover, when a mother cell reproduces,  it also sets up the connections
between two adjacent daughters so that they both know their IDs. These
connections prevent cells to reproduce in a direction that is already
populated.

\item[4-Sterility] A cell becomes sterile (cannot reproduce anymore) 
when both reproduction and connection principles have been applied. A cell
that is outside the parameter space is also sterile (checked during the
initialization). 

\item[5-Exclusivity] The reproduction process is \emph{exclusive}: only one cell
at a time can reproduce. It is exclusive because a
cell cannot start to reproduce while another cell is
still reproducing.

\end{description}

The cell population evolves by the reproduction of their individuals over as
many \emph{generations} as needed to cover the entire parameter space. The first
generation is composed of one cell only. The position of
this first cell corresponds to  $m_1=m_{\rm min}, m_2=m_{\rm max}$. We could
start at any place in the
parameter space. However, local flatness is an approximation and the author
thinks it is better NOT to start at $m_1=m_{\rm max}, m_2=m_{\rm max}$ where the
metric evolves quicker (highest mass). The first cell is initialized
(first principle). Then, the cell reproduces into 6 directions (second
principle). 
Once the reproduction is over, the connectors between the mother cell and its
daughters are set (third principle), and finally, the cell becomes sterile
(fourth principle). This loop over the first cell has created a new generation
of 6 cells, and each cell will now follow the four principles again.  However,
the
new generation of cells will not be able to reproduce in 6 directions.
Indeed, connectors between the first mother cell and its daughters have been
 set, and therefore the new cell generation cannot propagate towards the
mother direction. Furthermore, the 6 new cells have already 2
other adjacent cells. Therefore, each cell of the second generation can
reproduce 3 times only.
Moreover, some of the cells might be outside the parameter space and are
sterile by definition.  Once a new generation has been created, the previous
generation must contain sterile cells only. The algorithm
loops over the new generation while there exists fertile cells. The first
generation is a particular case since it contains only one cell. However, the following 
generations are not necessarily made of a unique cell, and the reproduction
warrants a careful procedure: the reproduction takes place cell after cell starting
from the smallest ID.
Moreover, in agreement with the
fifth principle, the cells of the newest generation wait until all the cells of
the previous generation have reproduced. The reproduction over generations stops
once no more fertile cells are present within the population. Since the parameter space
 is finite, the reproduction will automatically stop. Figure
\ref{fig:hexalgo} illustrates how the first 3 generations populate the parameter
space.

Once the reproduction is over, some cells might be outside the physical
parameter space, or outside the mass range requested. An optional final step
consists in``pushing back" the corresponding cells inside the parameter space. 
First, we can  push back the non-physical cells only, that is the cells that
are below the $\eta=1/4$ line towards the relevant eigen-vector directions onto
the $\eta=1/4$ line. Second, there are other cells for which mass parameters
correspond to physical masses but that are outside the parameter space of interest.
Nothing
prevents us from pushing these cells back into the parameter space as well.
This procedure is especially important in regions where the masses of the component
objects are large. Indeed, keeping templates of mass larger than a certain value causes
problems owing to the fact that the search pipeline uses a fixed lower cut-off
frequency and the waveforms of mass greater than this value cannot be generated.
In the simulations presented in this paper, we move the cells that are
below the $\eta=1/4$ boundary, and keep the cells that are outside the parameter space
but with $\eta>1/4$. Useful equations that characterize the boundaries of the
parameter space are provided in appendix~\ref{annex:tools}. A flow chart of the
algorithm  is also presented in appendix~\ref{annex:algorithm}.

An example of the proposed hexagonal placement is
shown in Fig.~\ref{fig:space}. In this example, the minimum and maximum
individual mass component are $3~M_\odot$ and $30~M_\odot$, and the lower
cut-off frequency is of $40$~Hz. We can see that none of the templates are
placed below the equal mass line whereas some are placed outside the parameter
space. Figure~\ref{fig:squarevshexa} gives another placement example.

\subsection{Size, Gain and Computational Cost}
The ratio of a circle's surface to the area of a square inscribed within this circle is
$\pi R^2 / (2 R^2)\approx 1.57$, where $R$ is the circle's radius. The ratio of the same
circle's surface to an inscribed
hexagon equals $\pi R^2 / (3\sqrt{3}R^2/2)\approx 1.21$. The ratio of the square surface
to the hexagon surface is therefore about 29\%, which means that about 29\% less
templates are needed to cover a given surface when a hexagonal lattice is used instead of
a square lattice; computational cost could be reduced by the same amount. Tables
\ref{tab:squaresize} and \ref{tab:hexasize} summarize the sizes of the proposed square
and hexagonal template bank placements. The hexagonal template bank reduces the number of
templates by about 40\% (see Table.~\ref{tab:gain}). This gain is larger than
the expected 29\%, and is related to the fact that we take into account the
evolution of the metric (orientation of cells/ellipses) on the parameter space.

Computational time required to generate a hexagonal bank appears to be smaller  than the square
bank. In Table~\ref{tab:compcost}, we record the approximate time needed to generate
each template bank, which is of the order of a few seconds even for template
banks as large as 100,000 templates. It is also
interesting to note that most of the computational time is spent in the
computation of the moments (used by the metric space) rather than in the
placement algorithm. 

The template bank size depends on various parameters such as the minimal
match and lower cut-off frequency that strongly influence the template bank
size. Other parameters such as the final frequency at which moments are
computed, or the sampling frequency may also influence the bank size.  There are
also refinements that can be made on the placement itself. Two main issues arise
from our study. First, the hexagonal placement populates the entire parameter
space. Yet, parameter space
is not a square but rather a triangular shape. In the corner of the parameter
space, a hexagonal placement is not needed anymore: a single template
overlaps with two boundary lines. In this case, hexagonal placement can be
switched to a bisection placement that places templates at equal distances from
the two boundary lines. A secondary issue is that the hexagonal placement is
aligned along an eigenvector direction. Nothing prevents us to place templates
along the other eigenvector direction. It seems that this choice affects
neither the efficiencies nor the template bank size significantly.

\begin{table}[th]
\caption{\label{tab:square} Typical square template bank size. We summarize the
number of templates of typical square template banks. We consider several design
sensitivity curves such as LIGO, VIRGO, \dots (see appendix~\ref{annex:PSDs} for
analytical expressions and lower cut-off frequencies), and 4 typical parameter spaces
(see Section~\ref{sec:simulation} for the mass range.}
\label{tab:squaresize}
\begin{ruledtabular}
\begin{tabular}{cccccc}
Bank size & EGO & GEO\,600 & LIGO-I & LIGO-A&
Virgo\\\hline
BBH	&   5582 &  1229 &   744 &   2238 &   4413 \\
BHNS	&  94651 & 16409 &  9964 &  35869 &  74276 \\
BNS	&  22413 &  5317 &  3452 &   9743 &  17764 \\
PBH	& 303168 & 62608 & 39118 & 122995 & 242609
\end{tabular}
\end{ruledtabular}
\end{table}

\begin{table}[th]
\caption{\label{tab:hexagonal}Typical hexagonal template bank size. We summarize the
number of templates of typical hexagonal template banks. We consider several design
sensitivity curves such as LIGO, VIRGO, \dots (see appendix~\ref{annex:PSDs} for
analytical expressions and lower cut-off frequencies), and 4 typical parameter spaces
(see Section~\ref{sec:simulation} for the mass range.}
\label{tab:hexasize}
\begin{ruledtabular}
\begin{tabular}{cccccc}
Bank size & EGO & GEO\,600 & LIGO-I & LIGO-A&
Virgo \\\hline
BBH	&   4109 &   838 &   532 &  1712 &  3283 \\
BHNS	&  71478 & 12382 &  7838 & 27511 & 57557 \\
BNS	&  16036 &  3576 &  2319 &  6969 & 12958 \\
PBH	& 205439 & 41354 & 26732 & 84154 & 167725
\end{tabular}
\end{ruledtabular}
\end{table}

\begin{table}[th]
\caption{\label{tab:gain}Size reduction between the square and hexagonal
template banks. We summarize the template bank size ratio (in percentage)
between the hexagonal and square placements. The ratios are calculated with the
numbers provided in Tables.~\ref{tab:squaresize} and~\ref{tab:hexasize}. For
various PSDs and parameter spaces, we can see that on average the gain
is about 40\%. }
\label{tab:banksize2}
\begin{ruledtabular}
\begin{tabular}{l|ccccc|c}
 & EGO & GEO\,600 & LIGO-I & LIGO-A &Virgo&average  \\\hline
BBH     & 36 & 47    &40    &31    &34 &37.6\\
BHNS    & 32 & 33    &27    &30    &29 &30.2\\
BNS     & 40 & 49    &49    &40    &37 &43.0\\
PBH     & 48 & 51    &46    &46    &45 &47.2\\
average & 39 & 45    &40.5  &36.75 &36.25 &39.5\\\hline
\end{tabular}
\end{ruledtabular}
\end{table}

\begin{table}[th]
\caption{\label{tab:compcost}Computational cost for different template banks. We
assume $MM=95\%$, a Vigo-like PSD with $f_L=30$~Hz, a segment duration of
$256$s, and a sampling of 4~kHz. Most of the computation time is spent in the
computation of the moments, that depend on the duration of the segment. Using
short
duration vector of a couple of seconds, the computational time decreases by
about 6 seconds showing that the time spent in the placement itself is
negligible even for large template banks.}
\begin{ruledtabular}
\begin{tabular}{cccccc}
$m_{\textrm{min}}$ & $m_{\textrm{max}}$ & $N_{\rm square}$ & Time(s)& $N_{\rm
hexa}$& Time(s)\\\hline
0.5 & 30 &  182136 & 25.0 & 124652 & 9.5\\
1   & 3  &  10187  & 7.5 & 7251 &6.3\\
1   & 30 &  34095  & 9 & 24501 & 7\\
3   & 30 &   2422  & 6.3 & 1764 & 6.1\\
\end{tabular}
\end{ruledtabular}
\end{table}

\section{Simulations}\label{sec:simulation}
The proposed square and hexagonal template bank placements are used to search
for various {\ICB} in the LIGO and GEO~600
GW-data. They are used to search for primordial black holes, binary
neutron stars, binary black holes and a mix of neutron stars and black holes.
In the past, the parameter space was split into sub-spaces that encompass
different astrophysical binary systems such as {\PBH}, {\BNS}, {\BBH}, and/or
{\BHNS}~\cite{LIGOS1iul, LIGOS2iul,LIGOS2macho, LIGOS2bbh, LIGOS3S4}.
We can filter the data through a unique template bank that covers the
different types of binaries, however, we split the parameter space
into the same 4 sub-spaces that have been used to validate the square template
bank placement so that we can compare results together. We use the same mass
range as in our companion paper, that is
PBH binaries with component masses in the range $[0.3-1]M_\odot$,
{\BNS}
$[1-3]M_\odot$, {\BBH} $[3-30]M_\odot$, and {\BHNS} with one neutron star with
component mass in the range
$[1-3]M_\odot$ and a black hole with component mass in the range $[3-60]M_\odot$, in
which case the template
bank must cover $[4-63]M_\odot$. We also use the same PSD by incorporating
the design sensitivities of current detectors (GEO, VIRGO and LIGO-I) and
advanced detectors (advanced LIGO (or LIGO-A), and EGO). Each of the PSDs has a
design sensitivity curve, provided in Appendix~\ref{annex:PSDs}. The lower
cut-off frequencies  are the same as in~\cite{squarebank} and are summarized in
the appendix as well. In the case of the EGO PSD, which we have not used
previously, we set the lower cut-off frequency $f_L=20$~Hz.
Actually, this value
can be decreased to about 10~Hz for the {\BBH} case, increasing the template
bank sizes.

In all the simulations, we tend to use common parameters so as to simplify the
interpretation. We use a sampling frequency of 4096~Hz over all simulations because
the last stable orbit $f_{\rm LSO}=1/(6^{3/2}2\pi M)$ is less than 
the Nyquist frequency of 2048~Hz for most of the {\BBH}, {\BHNS}, and {\BNS} signals. The
computational time
is strongly related to the size of the vectors, whose length depends
on the time duration of the template/signal used in our simulations. In order
to optimize the computational cost, in each search, we extract the longest
template duration that we round up to the next power of 2. The vector duration
is then multiply by 2 for safety. We set the minimal match to 95\%. We
considered 5 types of template families that are described later. We can estimate
the number of simulations. For instance, using $N_s$ injections,  with 5
different PSDs, 4 searches ({\BNS}, {\BBH}, \dots), and 5 template families
, we have a total of $N_s \times 5 \times 4 \times 5  = 100
\times N_s$ injections, which need to be filtered through $N_b$ templates. If we
approximate $N_b$ to be 10,000 and $N_s$ to be 10,000 as well, it is clear that
computational cost is huge. In order to speed up the simulations, we chose not to
filter signals with all the available templates, but only a relevant fraction of
them around the injected signal; this selection is
trivial since template and signal are based on the same model.

\begin{figure*}[tbh]
\centering
{\includegraphics[width=0.35\textwidth]
{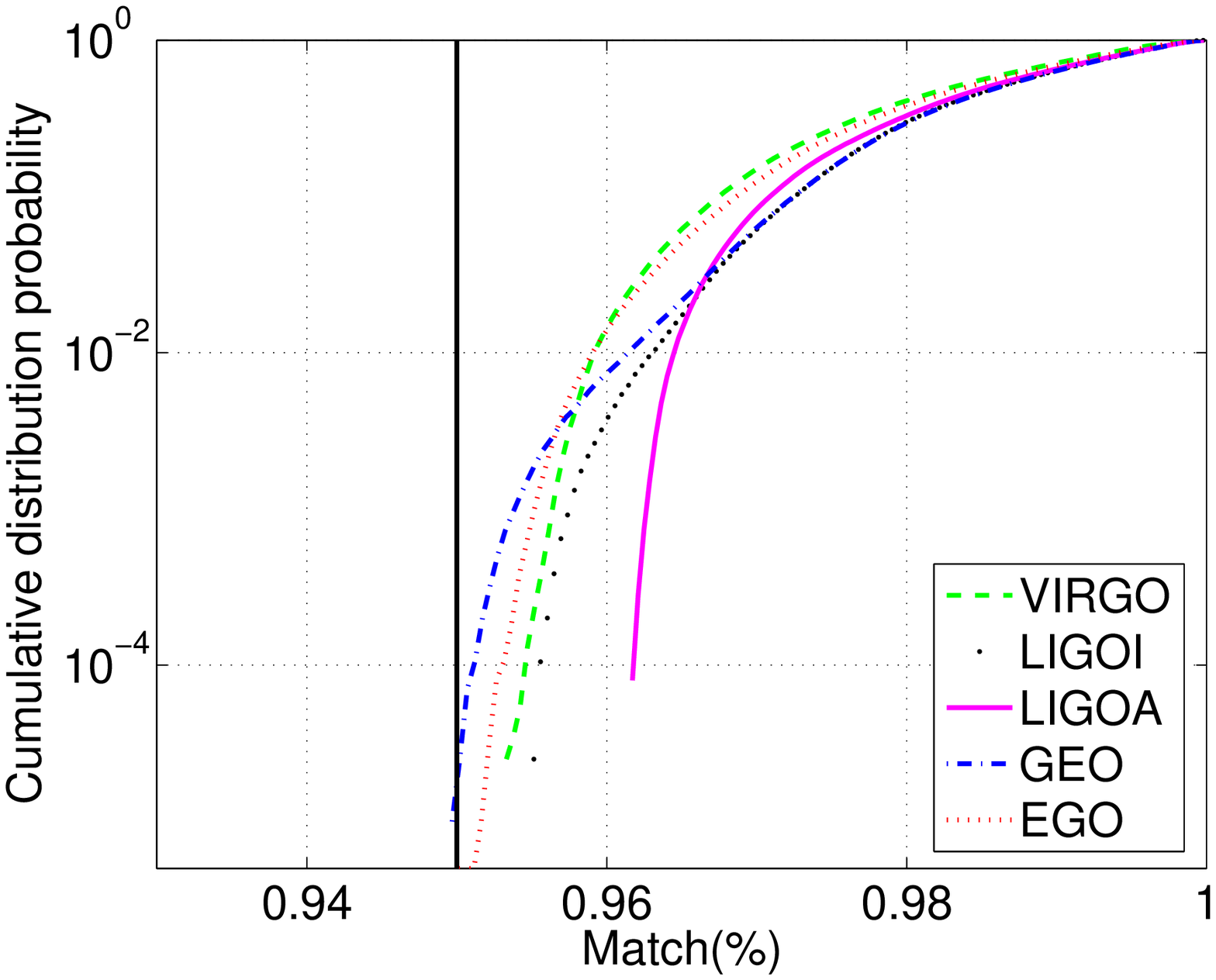}}
{\includegraphics[width=0.35\textwidth]
{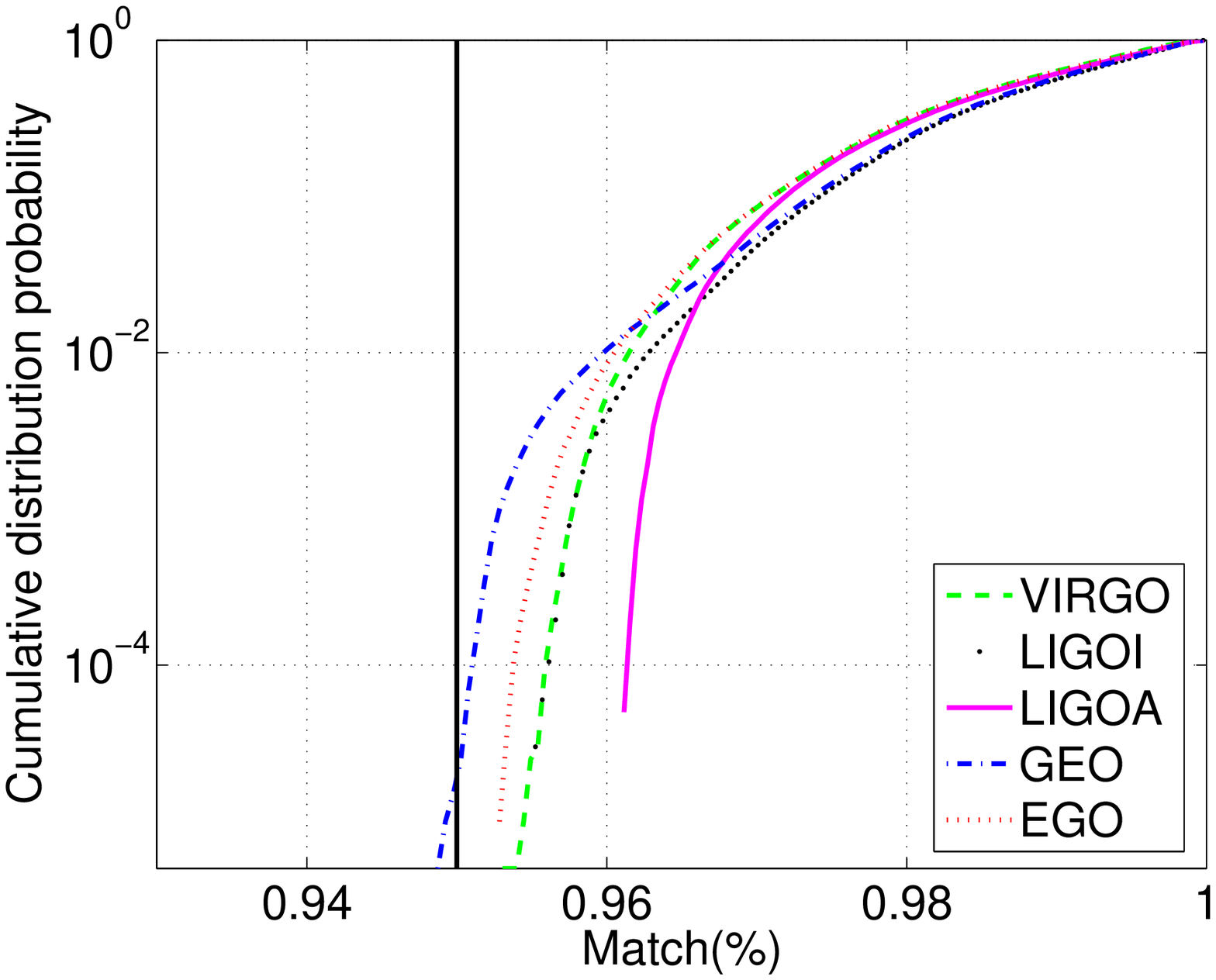}}\\
{\includegraphics[width=0.35\textwidth]
{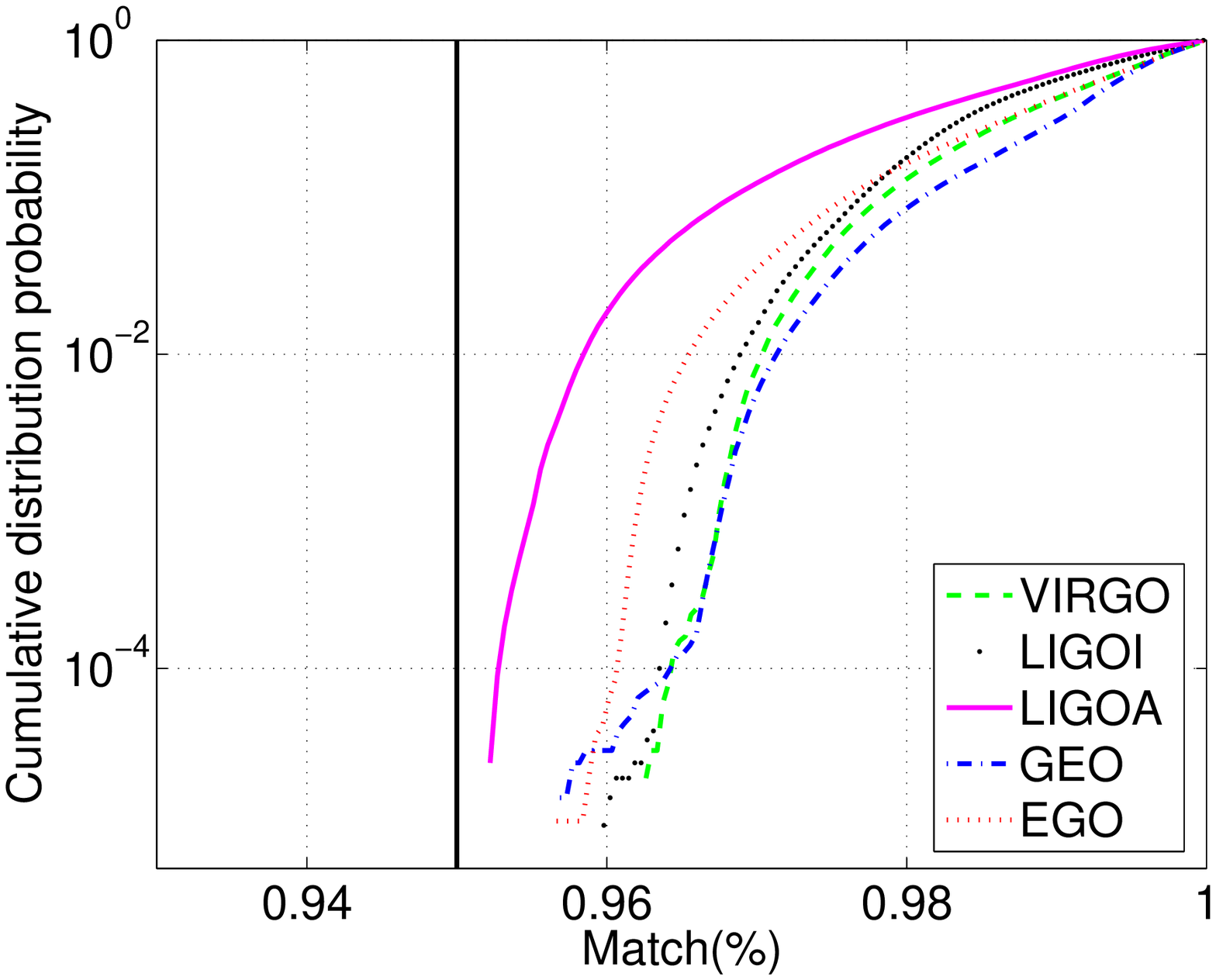}}
{\includegraphics[width=0.35\textwidth]
{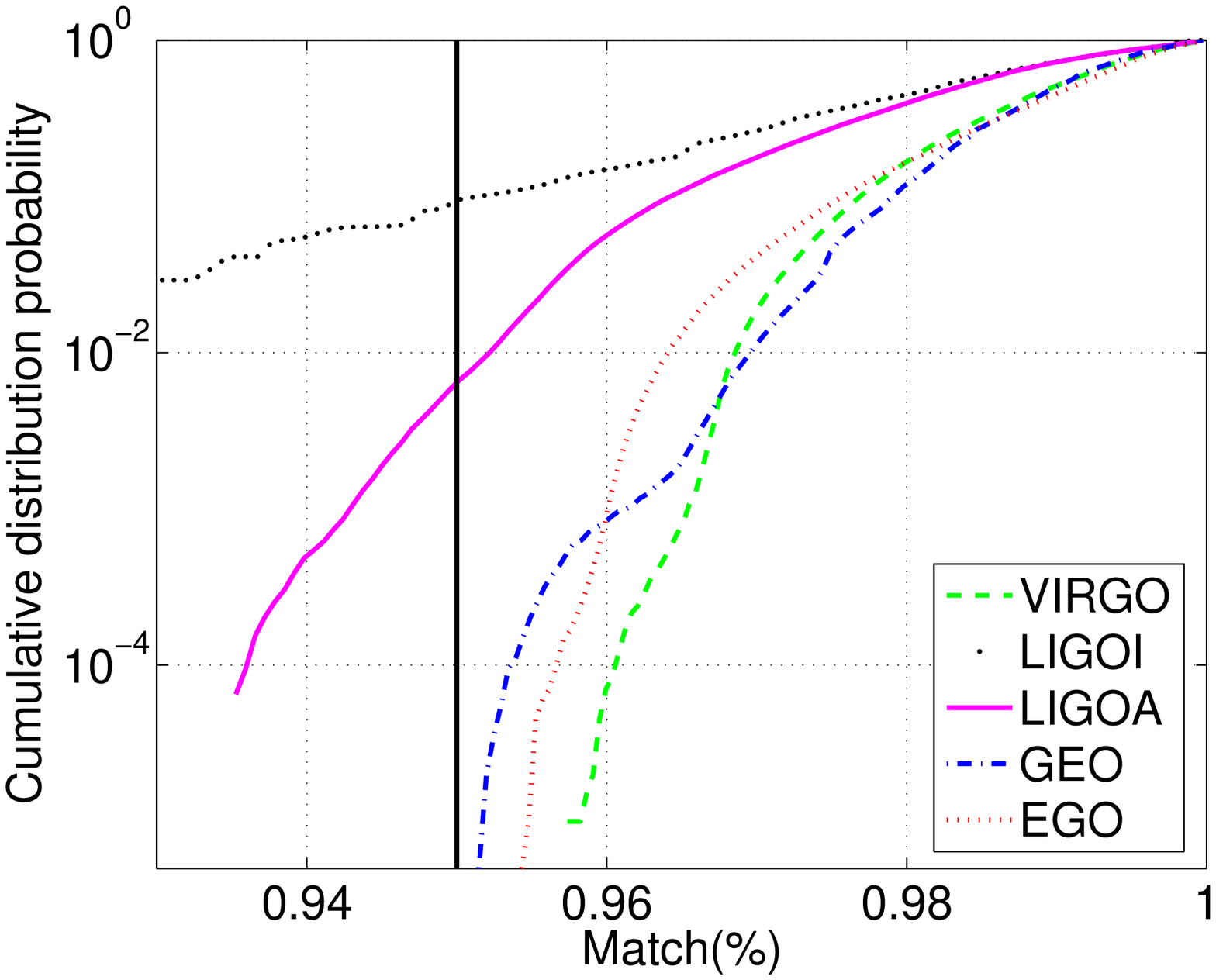}}
\caption{Cumulative efficiencies of the hexagonal template bank. Both template
and signal are based on TaylorF2 model. From top left to bottom right (clockwise),
injection and template bank cover the PBH binary, BNS, BHNS, and BBH
inspiralling compact binaries. 
\label{fig:TaylorF2_1}}
\end{figure*}

\begin{figure*}[tbh]
\centering
{\includegraphics[trim =0 55mm 0 0, clip, width=0.3\textwidth]
{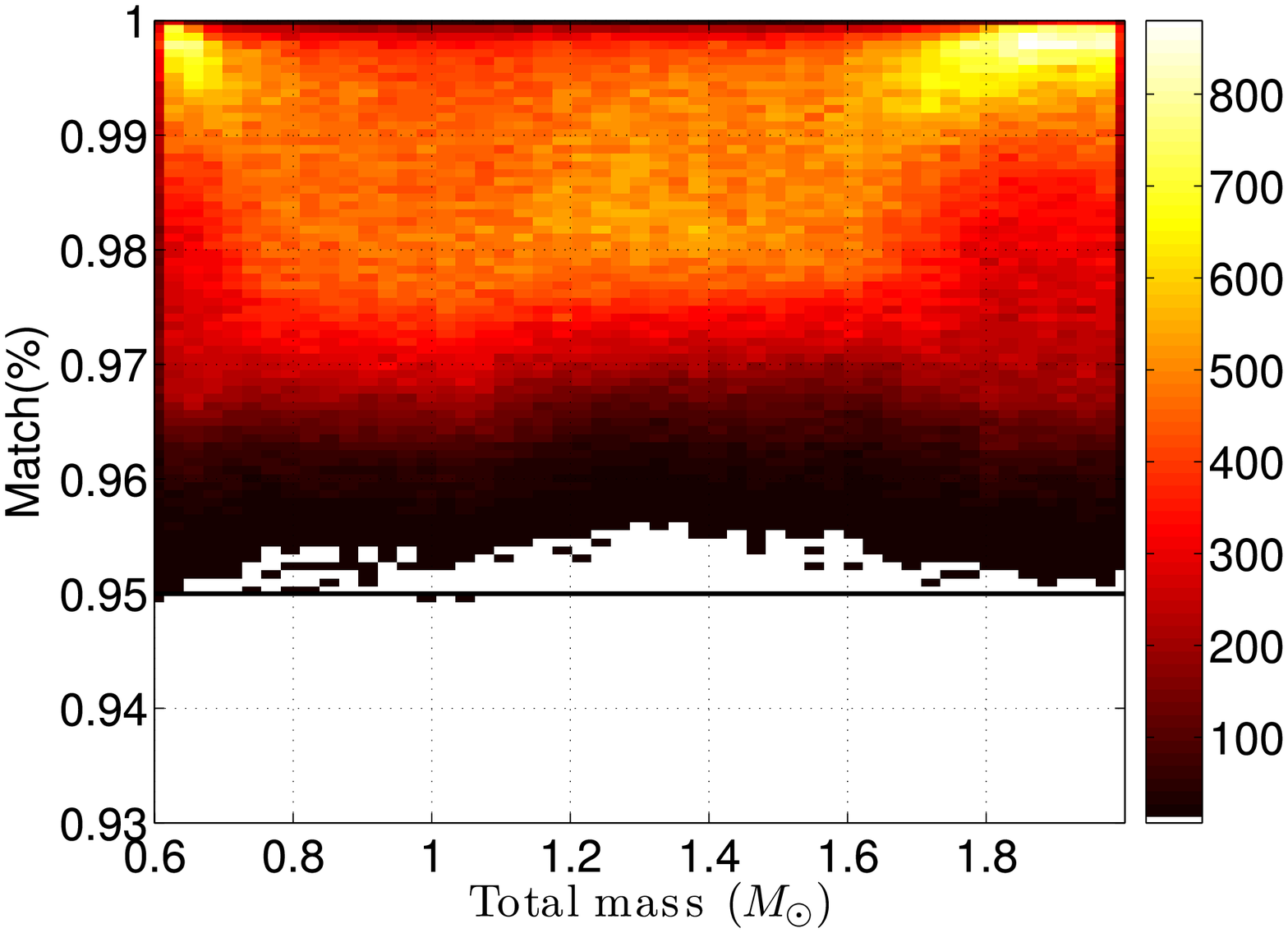}}
{\includegraphics[trim =0 55mm 0 0, clip, width=0.3\textwidth]
{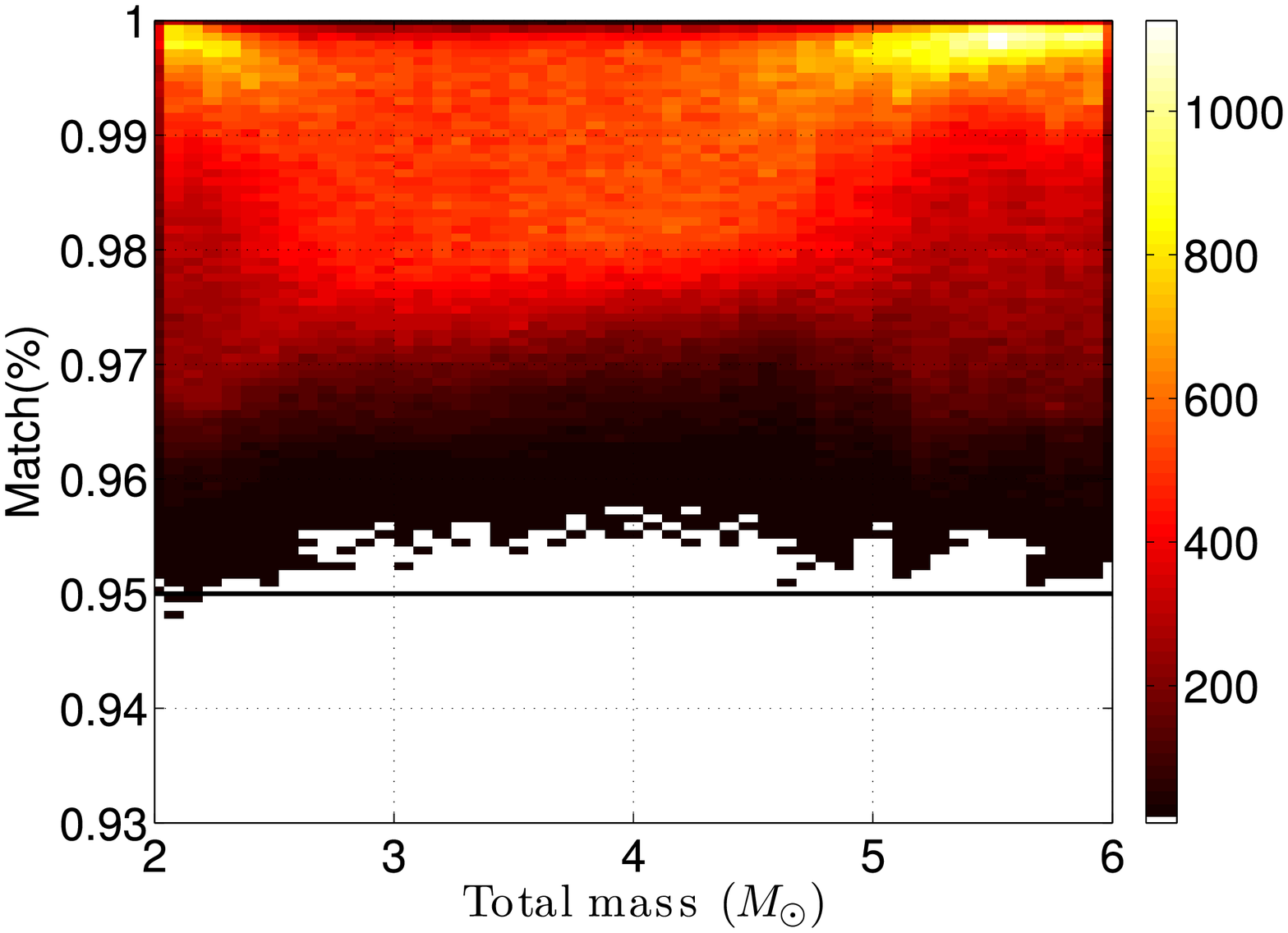}}\\
{\includegraphics[trim =0 55mm 0 0, clip, width=0.3\textwidth]
{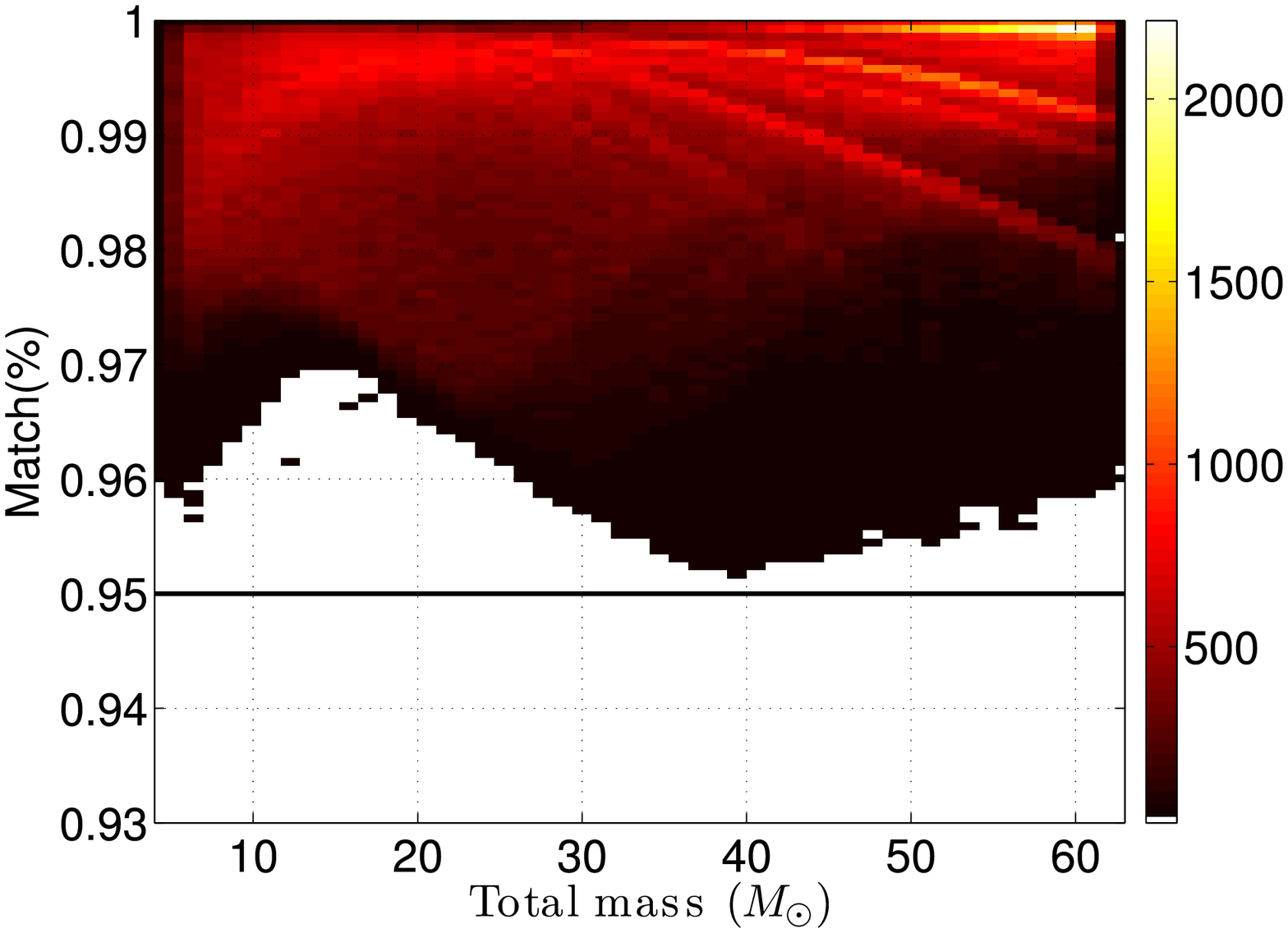}}
{\includegraphics[trim =0 55mm 0 0, clip, width=0.3\textwidth]
{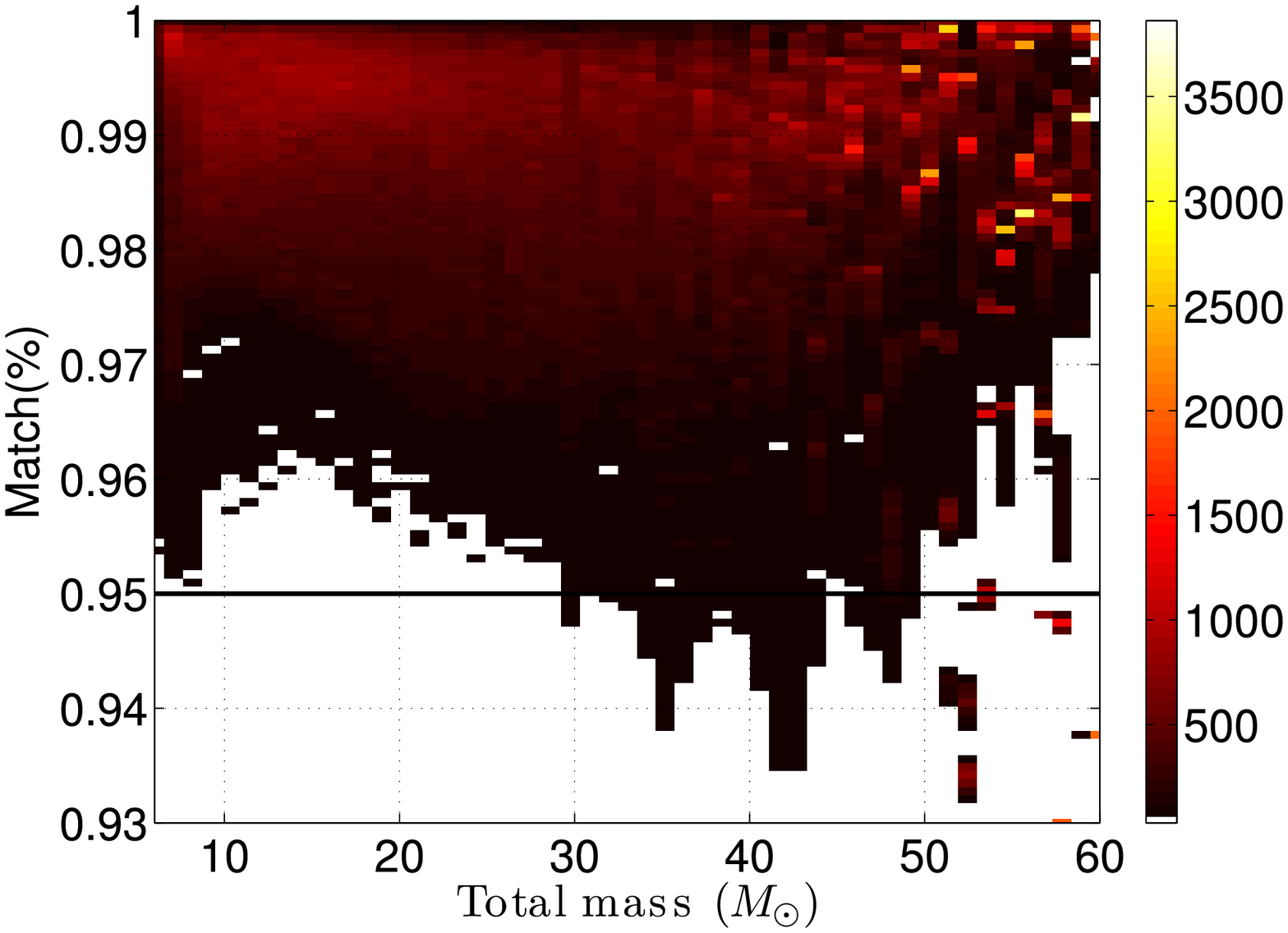}}
\caption{Efficiencies of the hexagonal template bank. Both
template and signal are based on TaylorF2 model. From top left to bottom right
(clockwise), injection and template bank cover the PBH binary, BNS, BHNS, and BBH
inspiralling compact binaries.}
\label{fig:TaylorF2_2}
\end{figure*}

\subsection{Description of the Physical Models}\label{sec:models}

Theoretical calculations using post-Newtonian
approximation of General Relativity give waveforms as expansions in the orbital
velocity $v$, where $v=\left(2\pi M f(t) \right)^{1/3}$. The PN expansions are
known up to order $v^5$ in amplitude and $v^7$ in phase. However, we limit
this study to restricted post-Newtonian, that is all amplitude corrections are
discarded. Moreover, we expand the flux only to 2PN order. The energy function
$E(v)$
and the flux $\mathcal{F}(v)$ are given by
\begin{equation}
E(v) = E_N \sum_k E_k v^{2k}, \mathcal{F}(v) = \mathcal{F}_N
\sum_j\mathcal{F}_jv^{j}.
\end{equation}
We can obtain the phase starting from the kinematic equations
$dt=(dt/dE)(dE/dv)dv$ and $d\phi/dt = 2\pi f(t)$ and the change of
binding energy $\mathcal{F}=-dE/dt$ giving a phasing formula of the
form~\cite{DIS3}. 
\begin{eqnarray}
t(v) &=&t_{\rm ref} + m \int_v^{v_{\rm ref}} \,
\frac{E'(v)}{{\cal F}(v)} \, dv, \nonumber \\
\phi (v) & = & \phi_{\rm ref} + 2 \int_v^{v_{\rm ref}}  v^3 \,
\frac{E'(v)}{{\cal F}(v)} \, dv\,.
\label{InspiralWavePhasingFormula}
\end{eqnarray}
There are different ways in which the above equations can be solved. For
convenience, we introduce
labels so as to refer to different physical template families that are used
within the gravitational wave community and in our simulations. 

\begin{description}
\item[TaylorT1]
If we  integrate the equations
(\ref{InspiralWavePhasingFormula})
numerically, we obtain the so-called TaylorT1 model. If instead, we use the
P-approximant for the energy and flux functions~\cite{DIS1, DIS2}, then one
generates the PadeT1 model.
\item[TaylorT2] We can also expand $E'(v)/\mathcal{F}(v)$ in a Taylor
expansion in which case the integrals can be solved analytically to obtain the
phase in terms of polynomial expressions as a function of $v$, which corresponds
to TaylorT2 model~\cite{DIS1}. This model is not used in this paper but results
are very similar to the TaylorT3 model.
\item[TaylorT3] From TaylorT2, $t(v)$ can be inverted and the polynomial
expression of $v$ used within the expression for $\phi(v)$ to obtain an
explicit time-domain
phasing formula in terms of $t$. This corresponds to the TaylorT3 model.
\item[EOB]  The non-adiabatic models directly integrate the
equations of motion (as opposed to using the energy balance equation) and there
is no implicit conservation of energy  used in the orbital dynamics
approach~\cite{EOB1,EOB2,EOB3,DIS2}.  The EOB maps the real two-body
conservative
dynamics onto an effective one-body problem wherein a test mass moves in an
effective background metric.

\item[TaylorF2] The phasing formula is expressed in the Fourier domain, and is
equivalent to the {\SPA} case already mentioned. 
\end{description}

\subsection{SPA Model Results}\label{sec:simspa}
First, we validate the hexagonal template bank with a model based
on the {\SPA} (also labelled TaylorF2),  used to
compute the metric components. We set
$\mathcal{\chi}_{s} = \mathcal{\chi}_{h} = {\rm
TaylorF2}$, and compute $\mathcal{E}$ and $\mathcal{S}_\mathcal{R}$. We
intensively tested this bank by setting $N_s = 200,000$ for each PSD and each
parameter space considered. Using the template bank size from Table
\ref{tab:hexasize}, the ratio $\mathcal{R}$ between template bank size and
number of simulations varies from 1.7 to 375, which is much larger
than unity in agreement with discussions that arose in Sec.~\ref{sec:be}. The
results are summarized in Fig.~\ref{fig:TaylorF2_1} and \ref{fig:TaylorF2_2}.

 In Fig.~\ref{fig:TaylorF2_1}, we notice that the hexagonal bank is efficient over
the
entire range of {\PBH} binary, {\BNS}, and {\BHNS} searches. Moreover, the
safeness is close to the minimal match ($\mathcal{S}_\mathcal{R} \in
[95\%\textrm{--}96\%]$); by looking at the cumulative efficiencies, the bank
seems to be neither under or over-efficient. However, looking more closely at
$\mathcal{E}_M$ (see Fig.~\ref{fig:TaylorF2_2}), we
can identify a small over-efficient region in the BHNS case, where the
efficiency is always larger than 97\% for signals with total
mass between $[4\textrm{--}20] M_\odot$. 

In the {\BBH} case, the bank is
also efficient for the various PSDs with total mass between $[6-40] M_\odot$,
and similarly to the BHNS case, it is over-efficient (above 97\%) for systems
with total mass between $[8\textrm{--}20] M_\odot$. The bank is also
under-efficient
with matches as low as 93\% but for very high mass systems above $40~M_\odot$.
 The match below the minimal match are related to the LIGO-I PSD
only, for which the lower cut-off frequency is 40~Hz. For high mass and
nearly equal mass systems, the waveforms tend to be very short and contain only
a few
cycles: the metric is not a good approximation anymore. It also explains the
feature seen at high mass, that shows some oscillations in the matches: a single
template matches with many different injected signals. One solution to prevent
matches to drop below the minimal match is to refine the grid for high mass
range by decreasing the distances (i.e, increasing $MM$) between templates in this part
of the parameter space. However, the high mass also correspond
to the shortest
waveforms which lead to a high rate of triggers in real data
analysis. Therefore it is advised not to over-populate the high mass region.
Overall, the hexagonal placement has the same behavior as in~\cite{squarebank}
but the bank is not over-efficient anymore in most cases.

\subsection{Non SPA Model Results}\label{sec:simother}
The square and hexagonal template banks are designed for TaylorF2 model. Yet,
models presented in Section~\ref{sec:models} do not differ from each other
significantly so long as $v\ll c$, which is the case for {\PBH}, {\BNS}
waveforms and most of the {\BHNS} and {\BBH} waveforms. Therefore,
we expect the efficiencies of the template banks to
be equivalent to the SPA-model results. 

The models used in this Section have the same PN-order (i.e., 2PN) as in the TaylorF2
model. The simulation parameters are identical except the number of simulations
that is restricted to $N_s=10,000$ for computational reasons. Finally, we
tested only the {\BNS}, {\BHNS} and {\BBH} searches. The {\PBH} using {\SPA}
model being sufficient for a detection search. 

\subsubsection{TaylorT1, TaylorT3, PadeT1}
\begin{figure*}[tbh]
\centering
{\includegraphics[trim =0 55mm 0 0, clip, width=0.3\textwidth]
{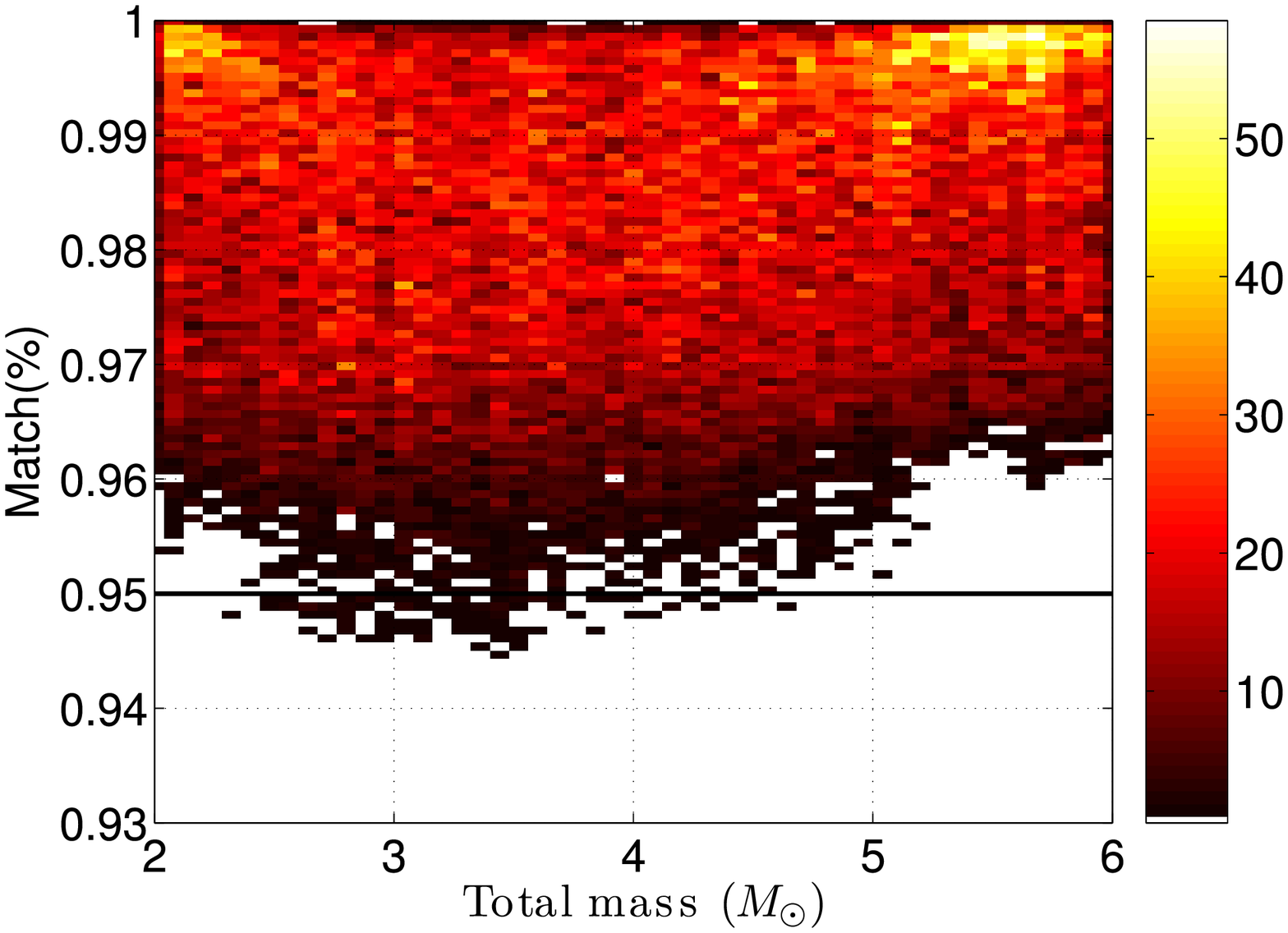}}
{\includegraphics[trim =0 55mm 0 0, clip, width=0.3\textwidth]
{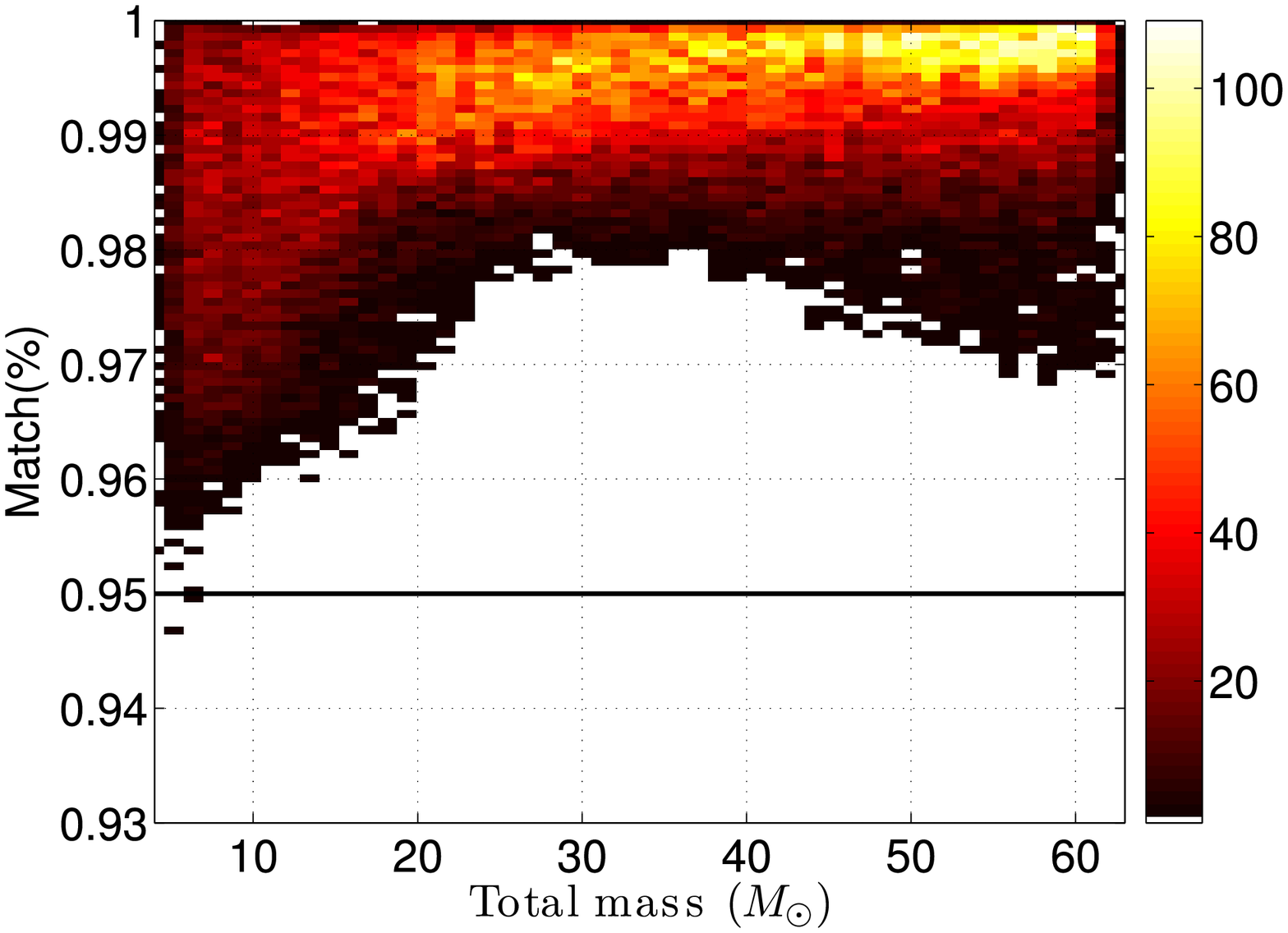}}
{\includegraphics[trim =0 55mm 0 0, clip, width=0.3\textwidth]
{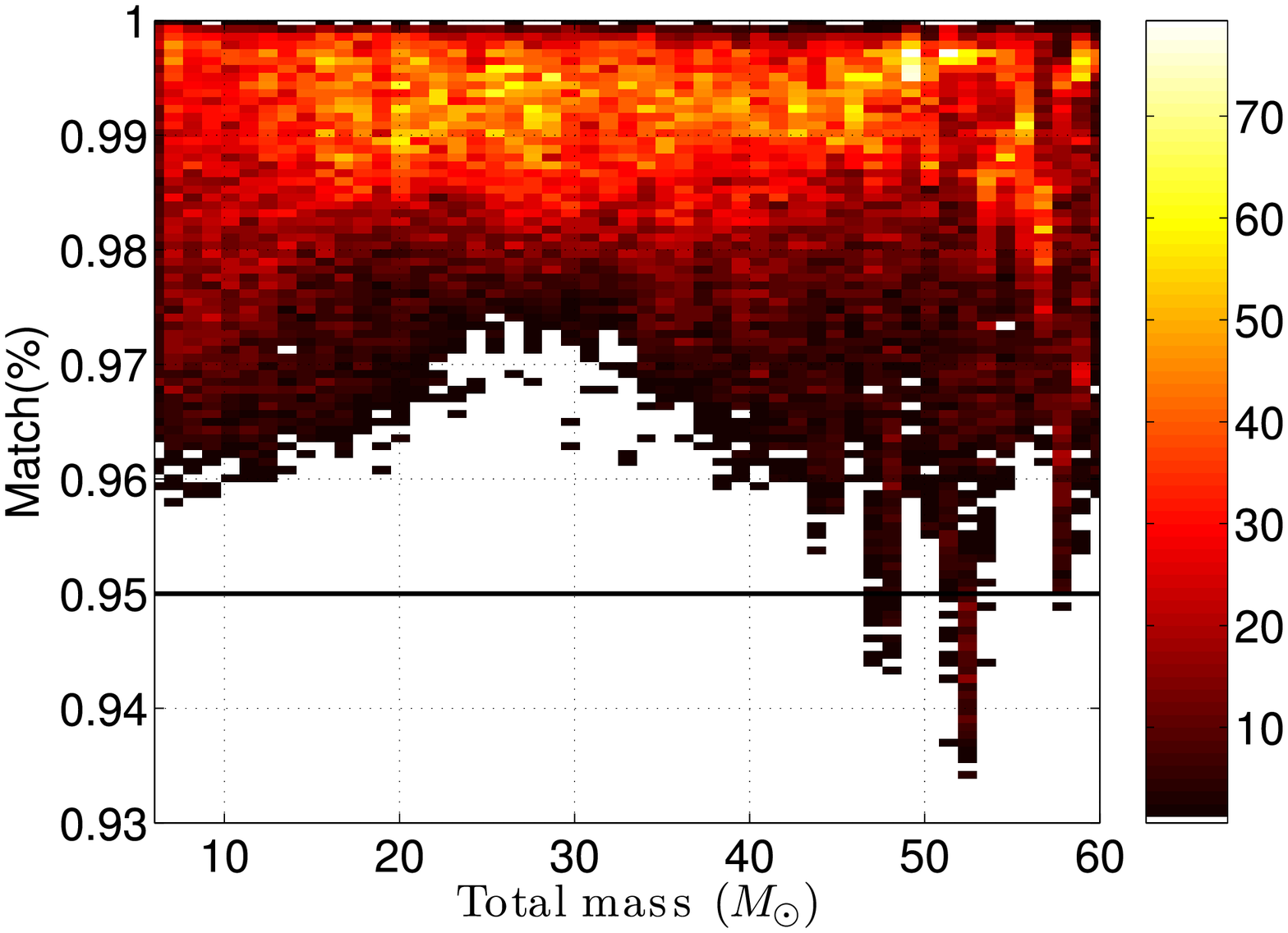}}\\
{\includegraphics[width=0.3\textwidth]
{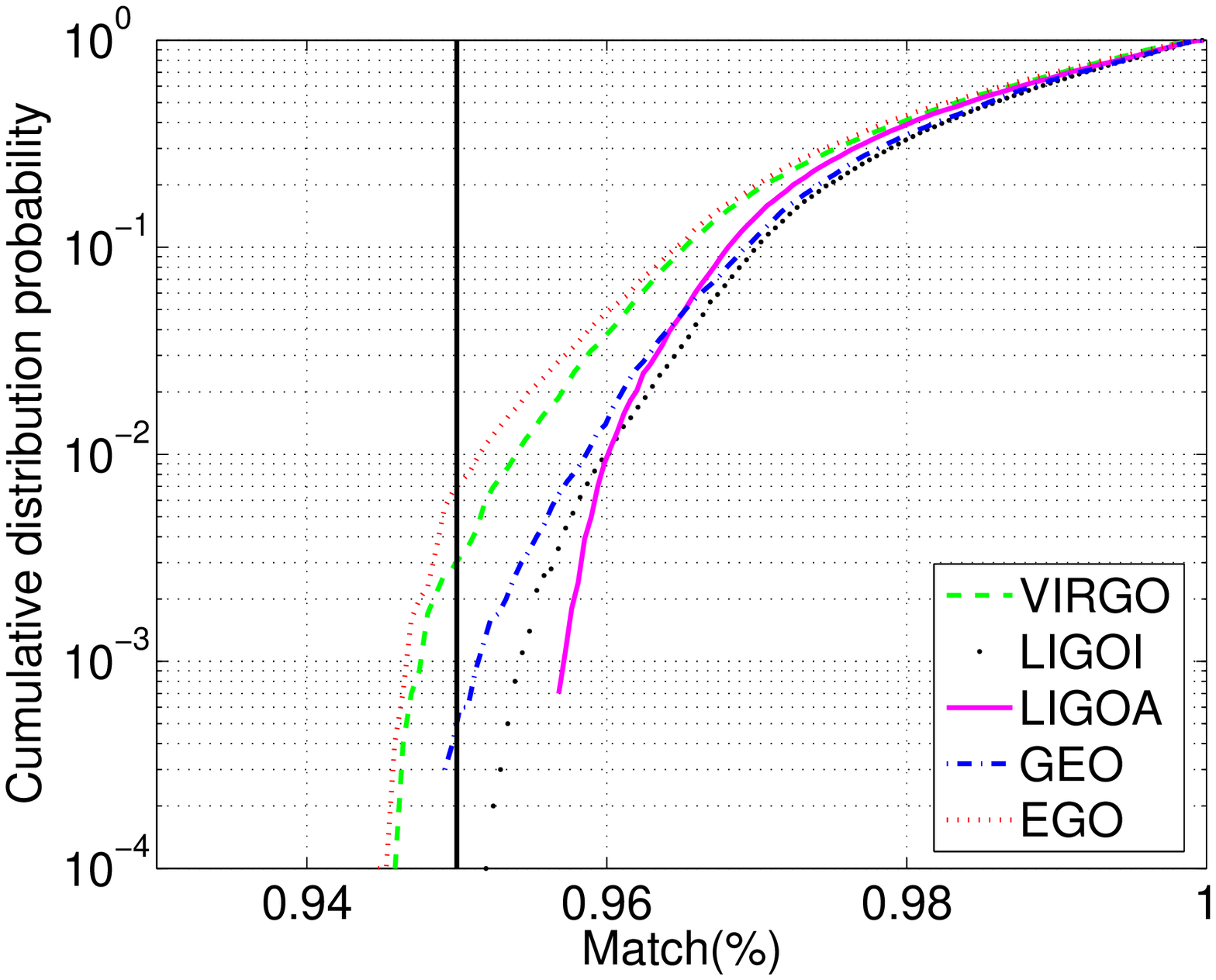}}
{\includegraphics[width=0.3\textwidth]
{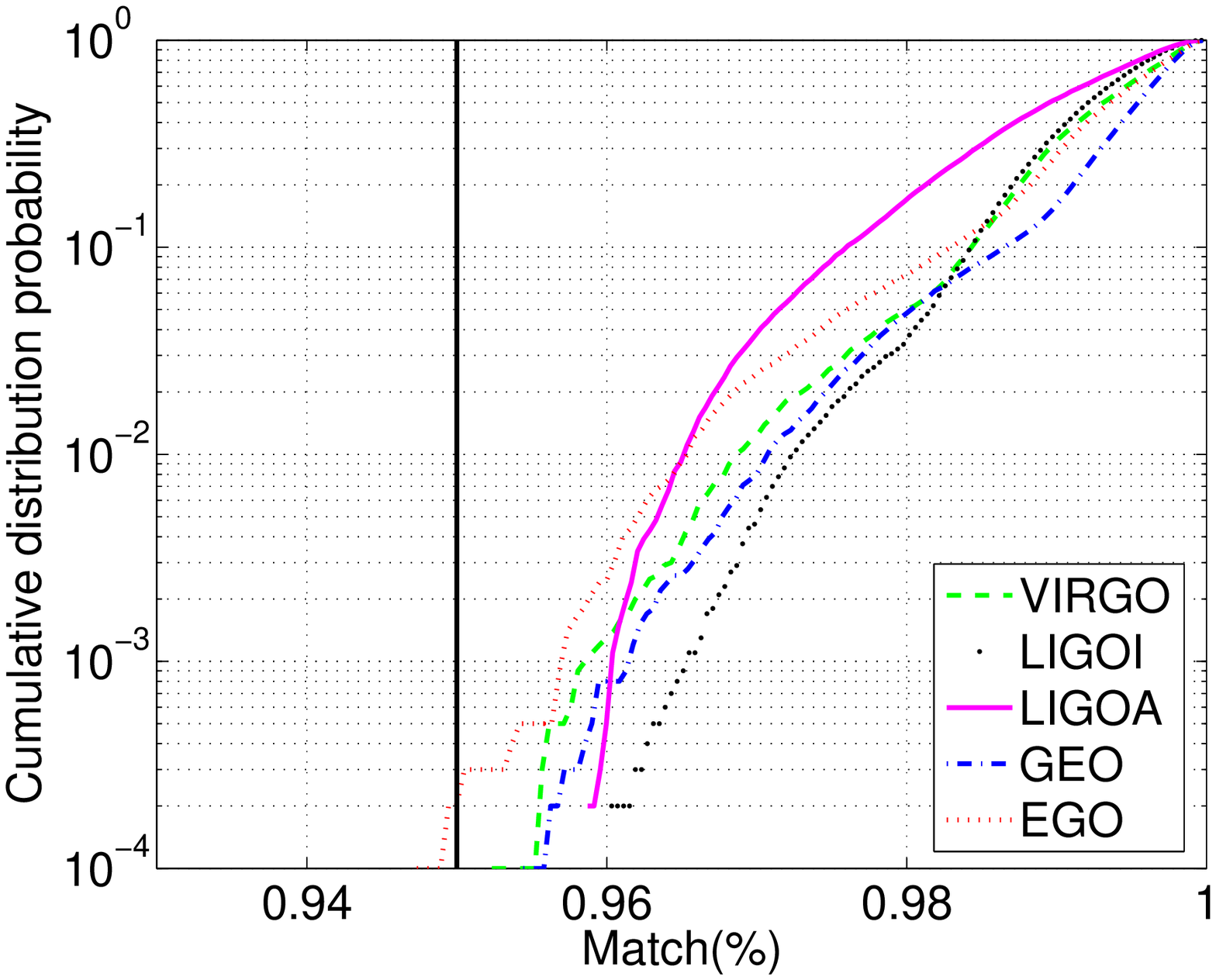}}
{\includegraphics[width=0.3\textwidth]
{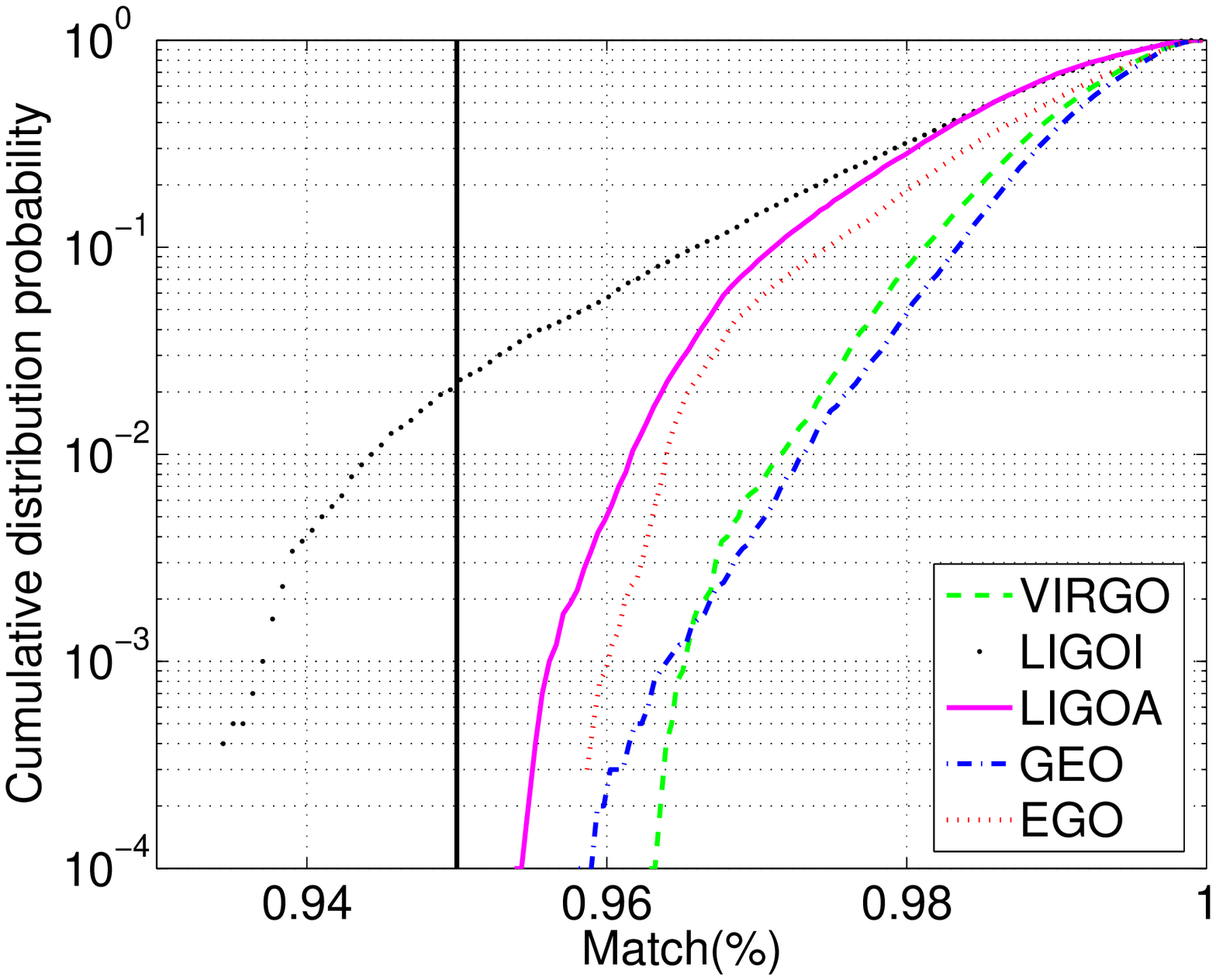}}
\caption{\label{fig:TaylorT1} Hexagonal template bank efficiencies using
TaylorT1 model. From left to right, results of the BNS, BHNS, and BBH
injections.}
\end{figure*}

\begin{figure*}[tbh]
\centering
{\includegraphics[trim =0 55mm 0 0, clip, width=0.3\textwidth]
{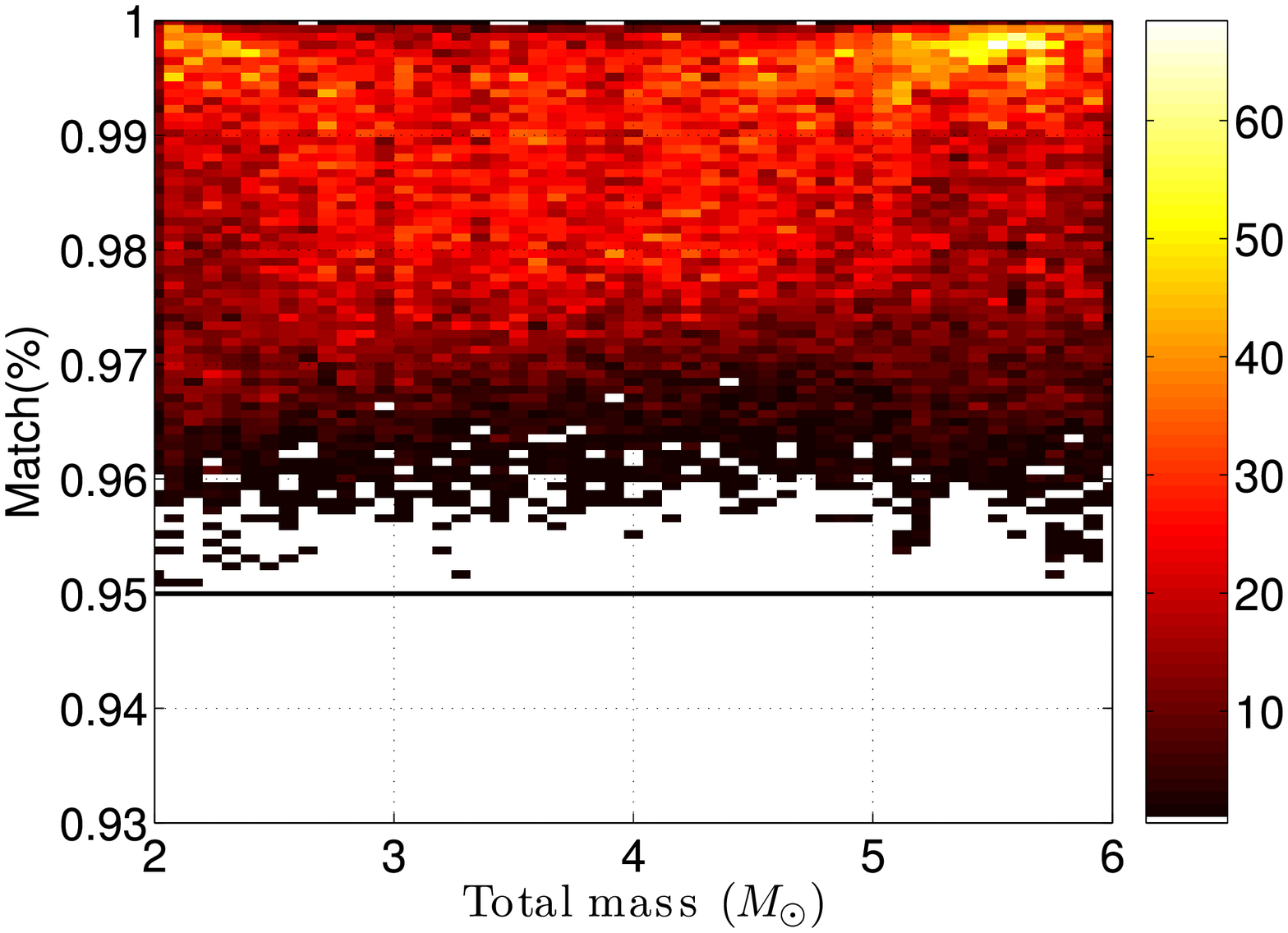}}
{\includegraphics[trim =0 55mm 0 0, clip, width=0.3\textwidth]
{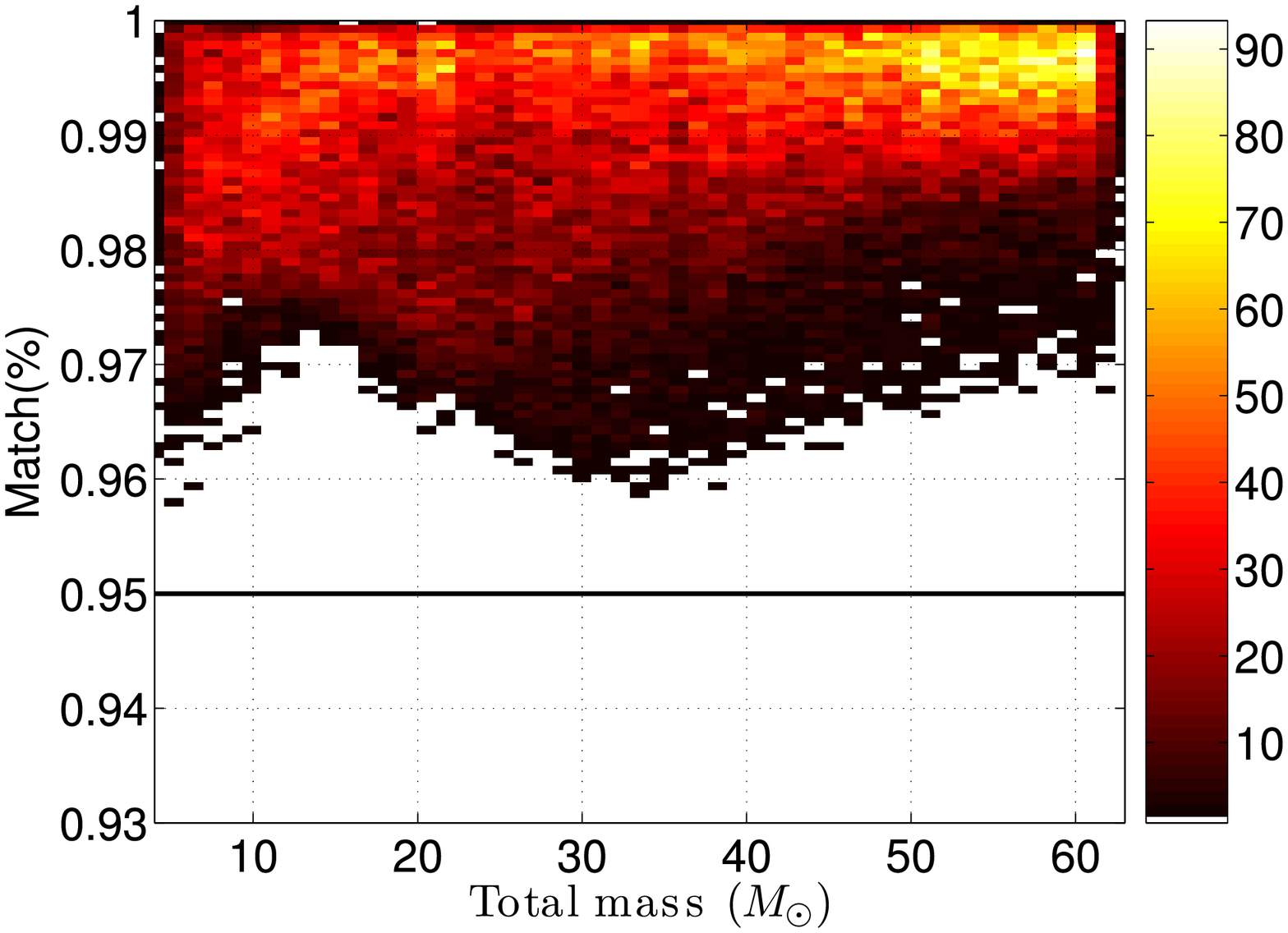}}
{\includegraphics[trim =0 55mm 0 0, clip, width=0.3\textwidth]
{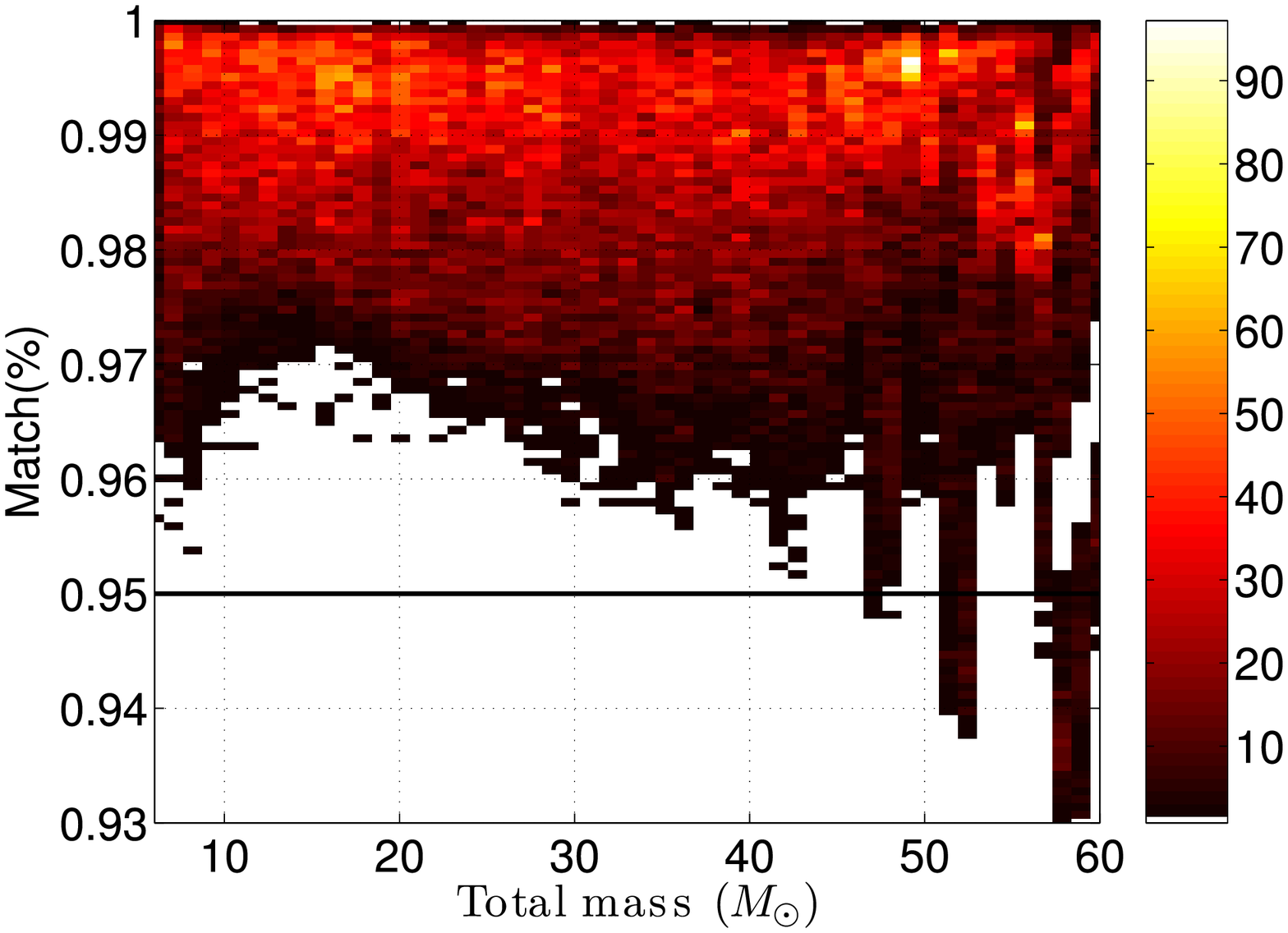}}\\
{\includegraphics[width=0.3\textwidth]
{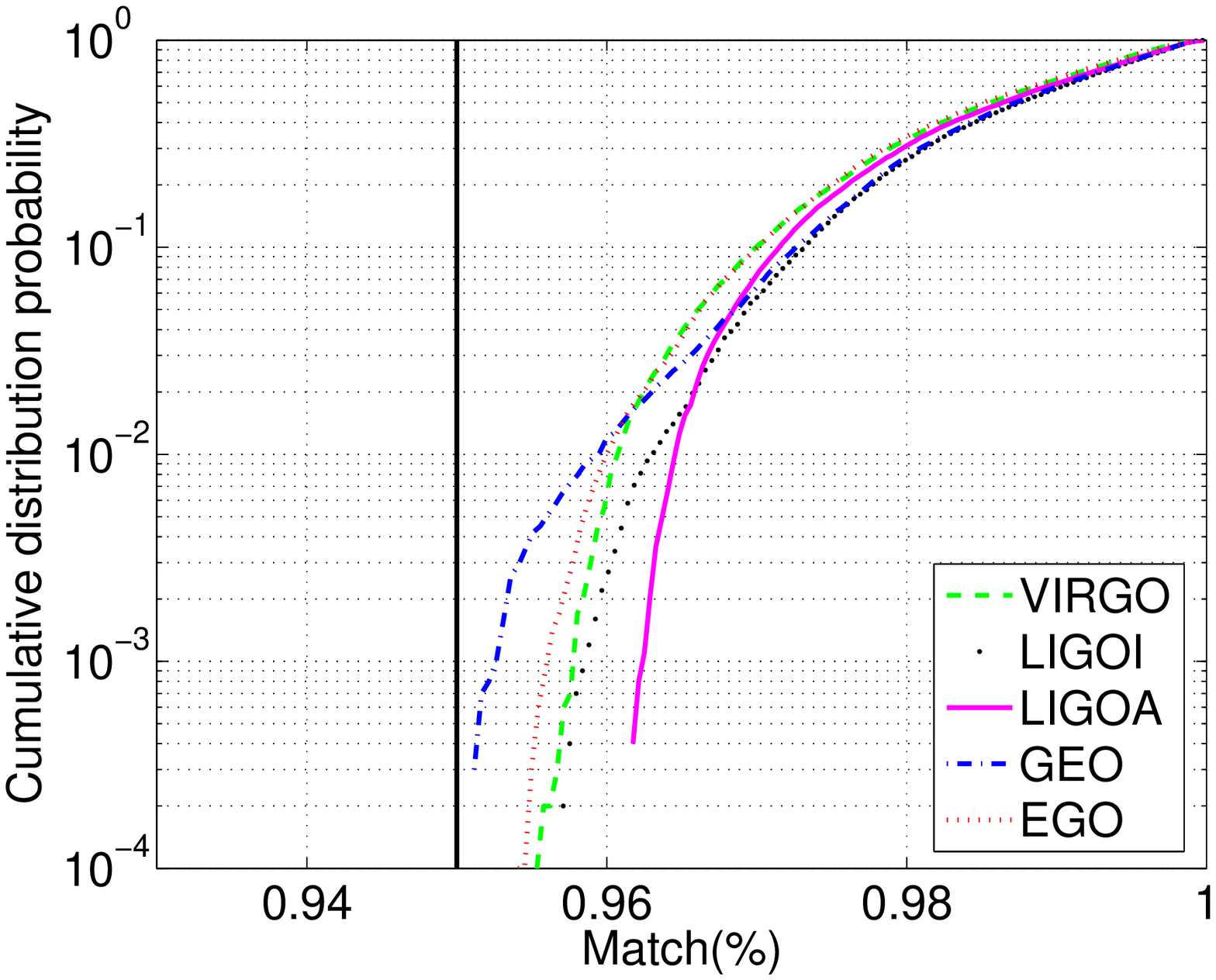}}
{\includegraphics[width=0.3\textwidth]
{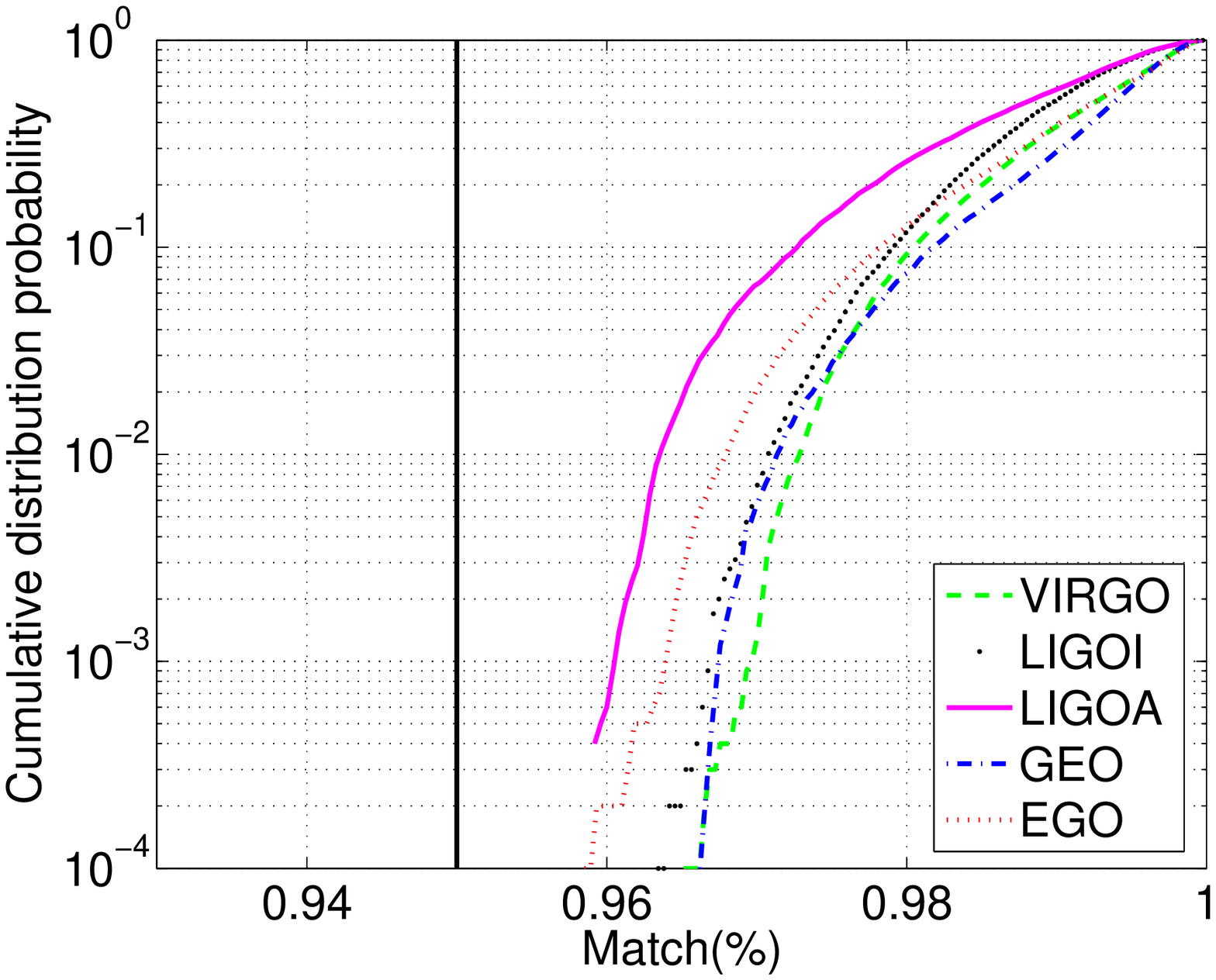}}
{\includegraphics[width=0.3\textwidth]
{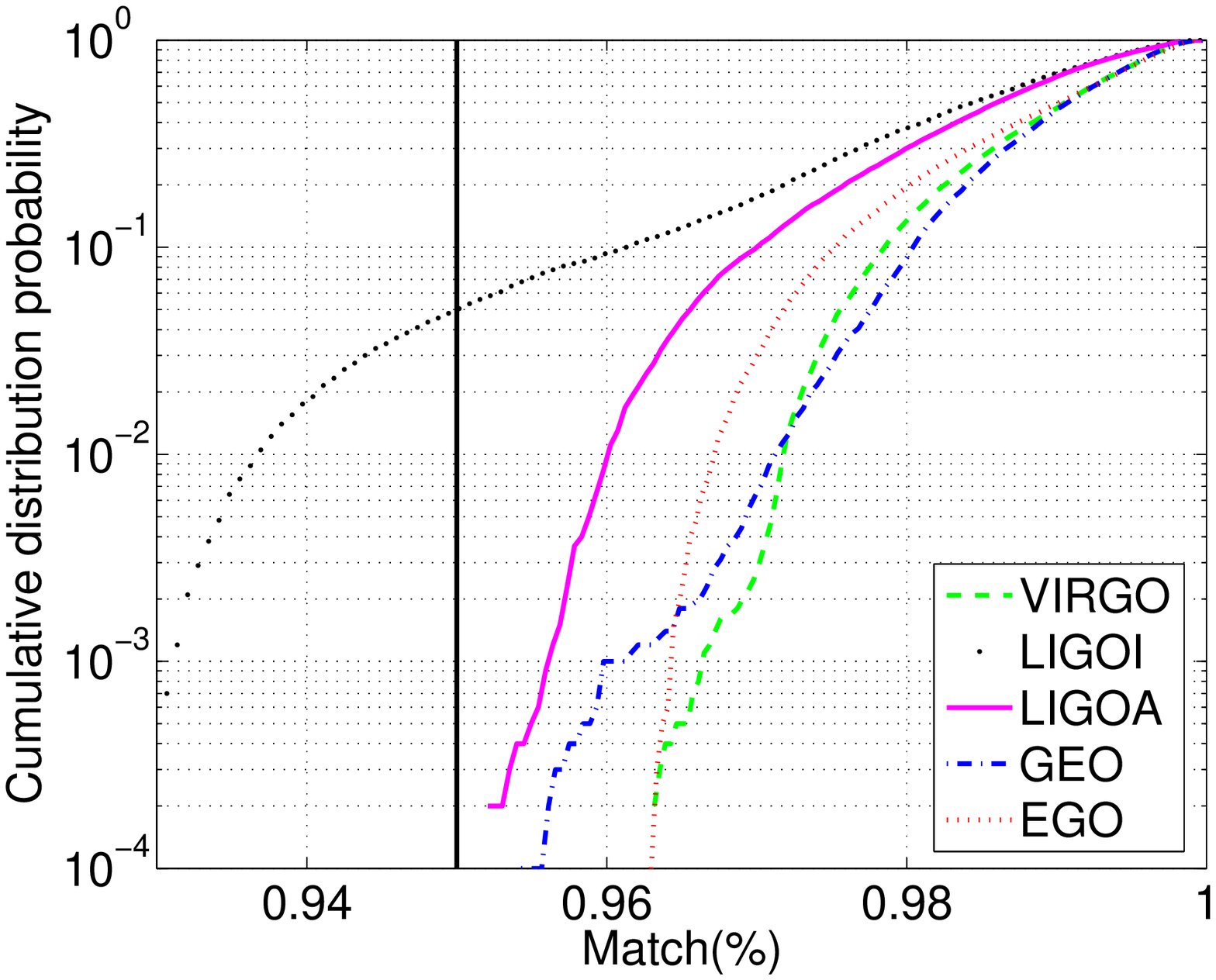}}
\caption{\label{fig:TaylorT3} Hexagonal template bank efficiencies using
TaylorT3 model. From left to right, results of the BNS, BHNS, and BBH
injections.}
\end{figure*}

\begin{figure*}[tbh]
\centering
{\includegraphics[trim =0 55mm 0 0, clip, width=0.3\textwidth]
{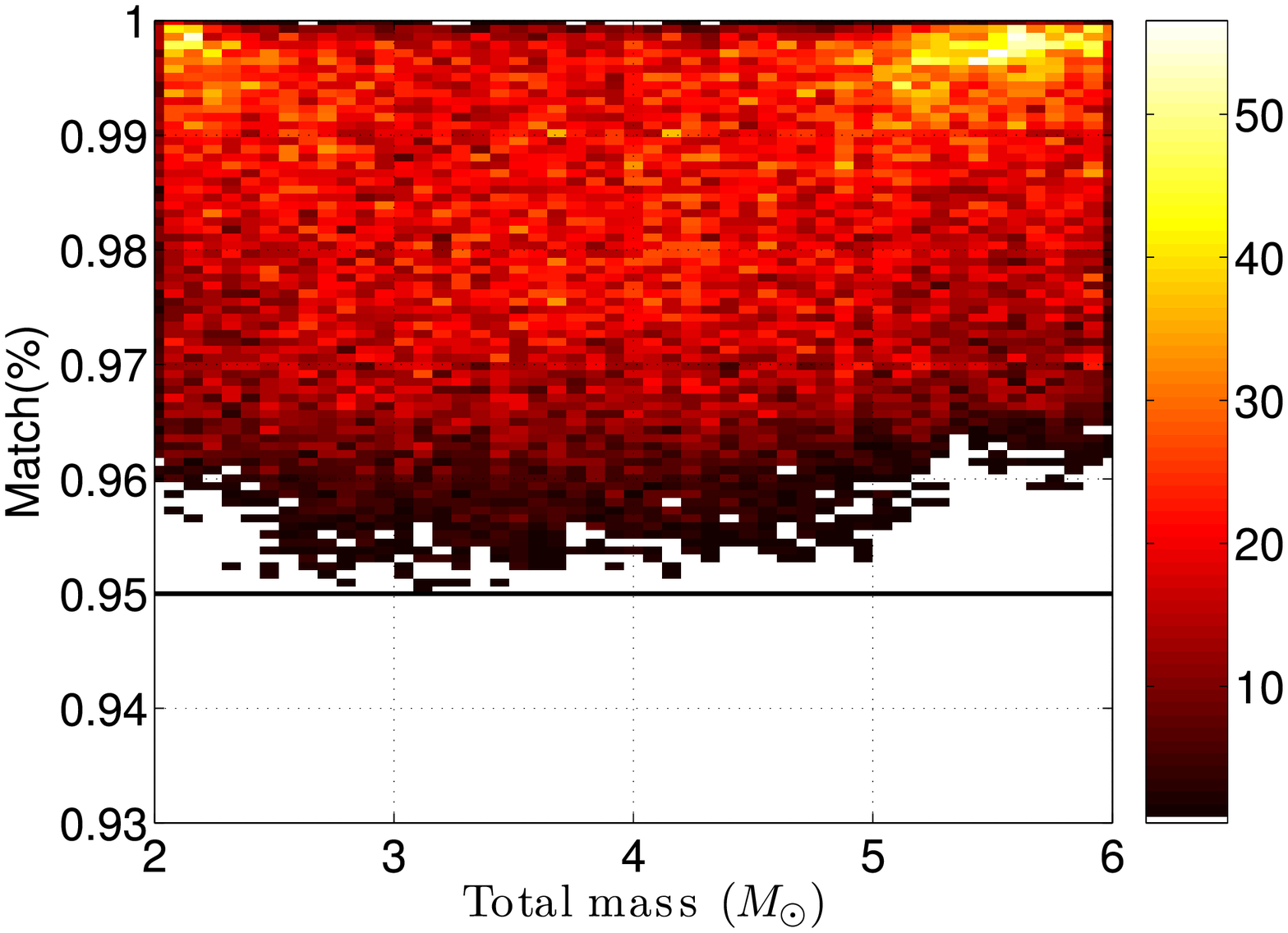}}
{\includegraphics[trim =0 55mm 0 0, clip, width=0.3\textwidth]
{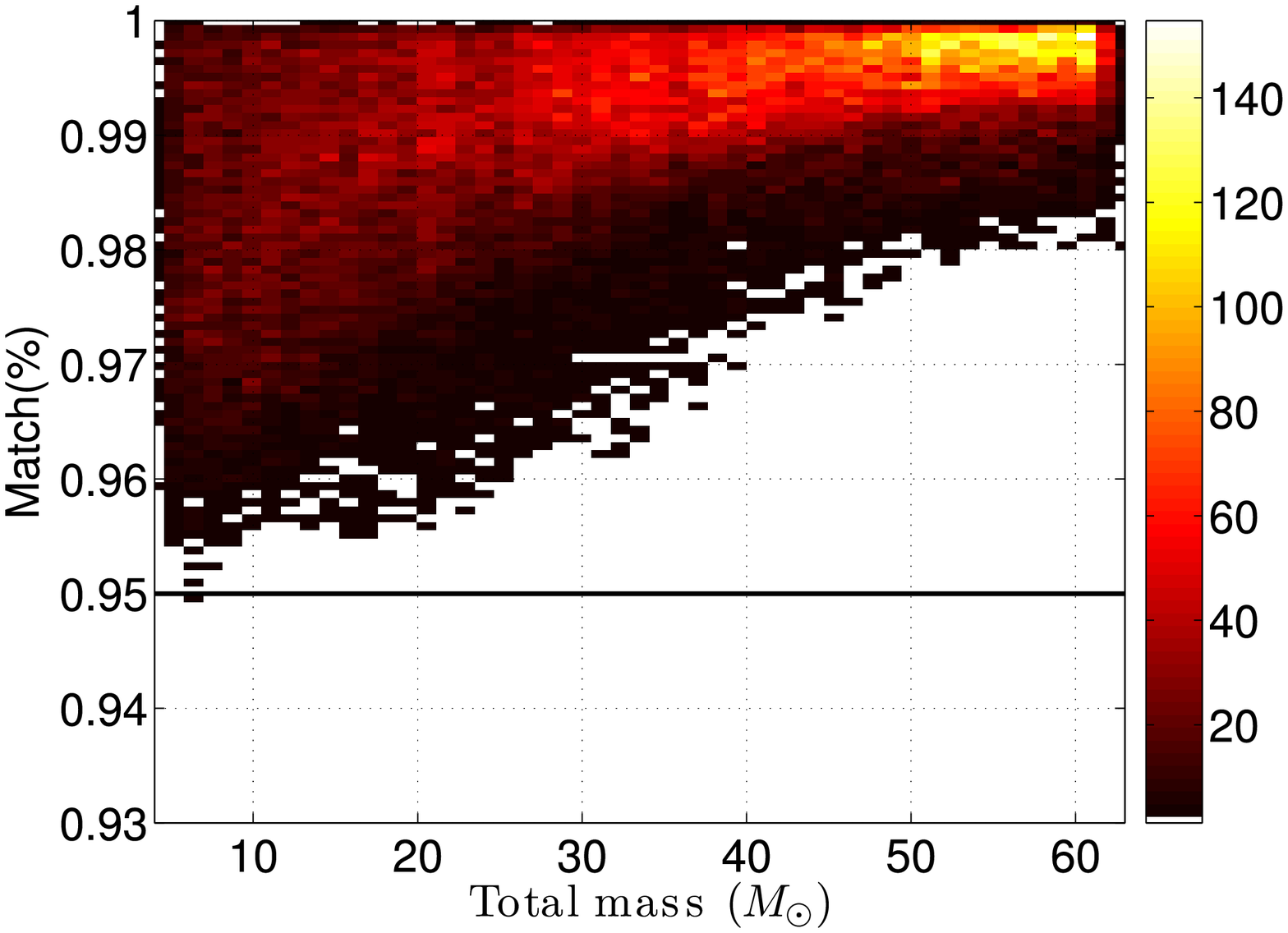}}
{\includegraphics[trim =0 55mm 0 0, clip, width=0.3\textwidth]
{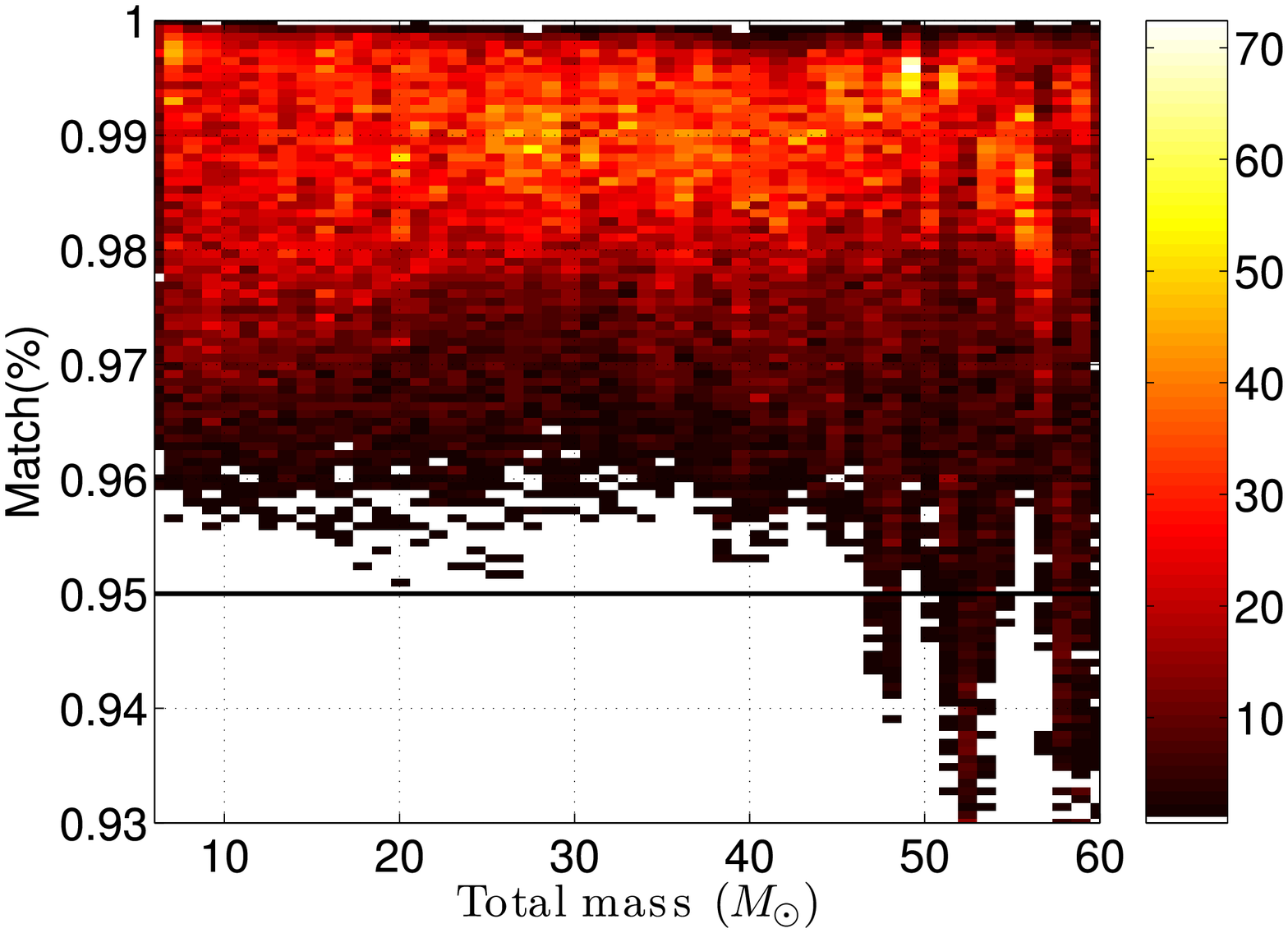}}\\
{\includegraphics[width=0.3\textwidth]
{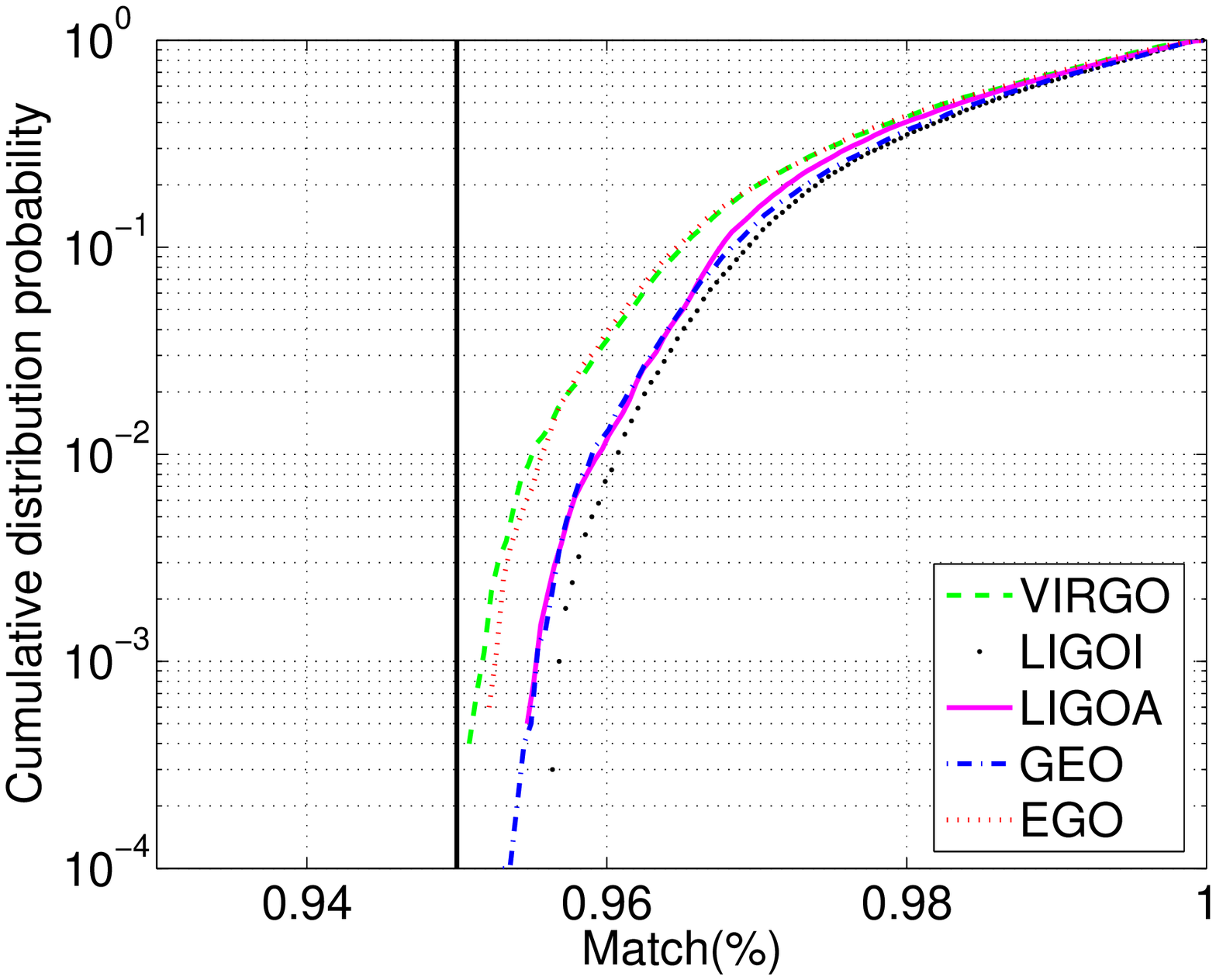}}
{\includegraphics[width=0.3\textwidth]
{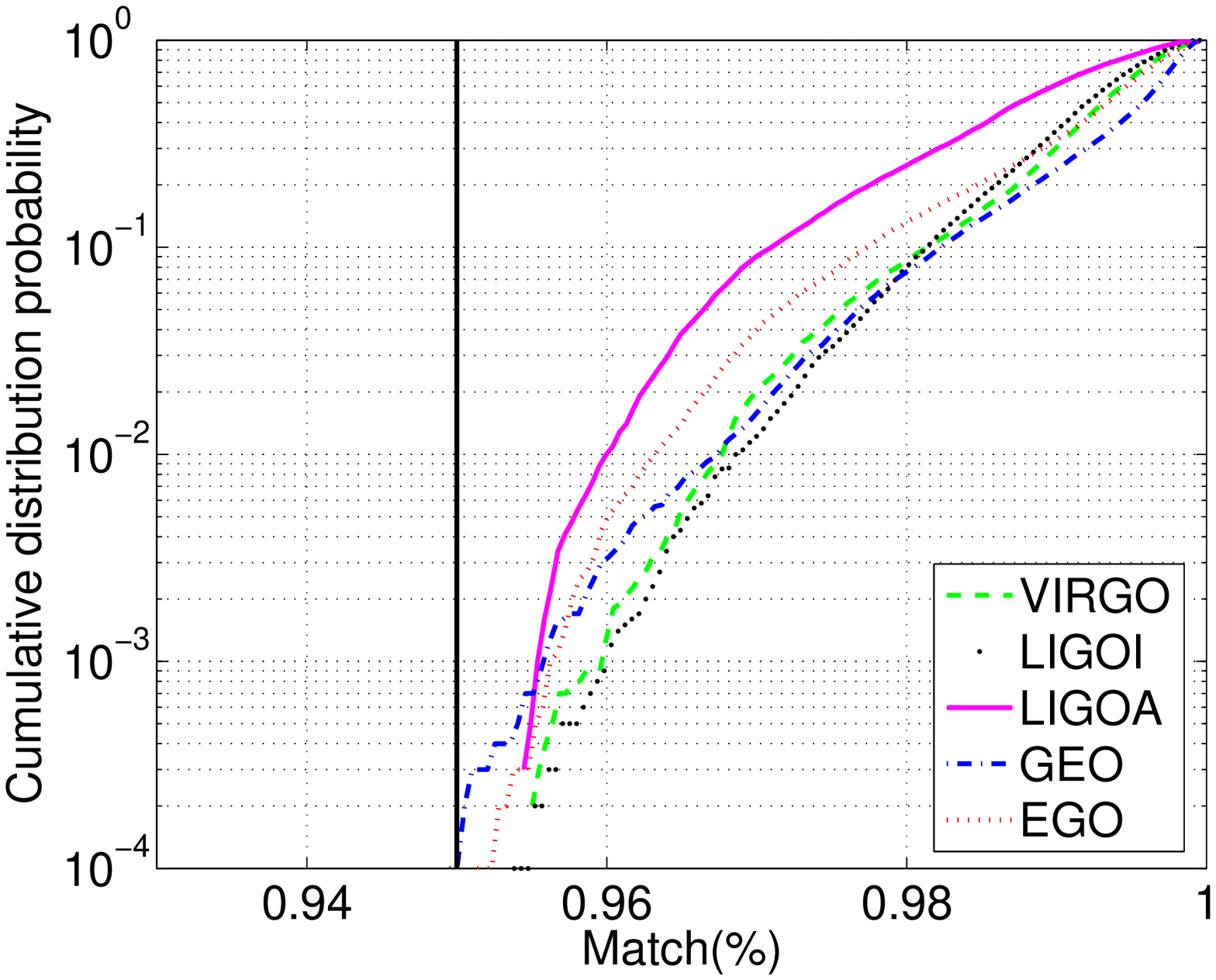}}
{\includegraphics[width=0.3\textwidth]
{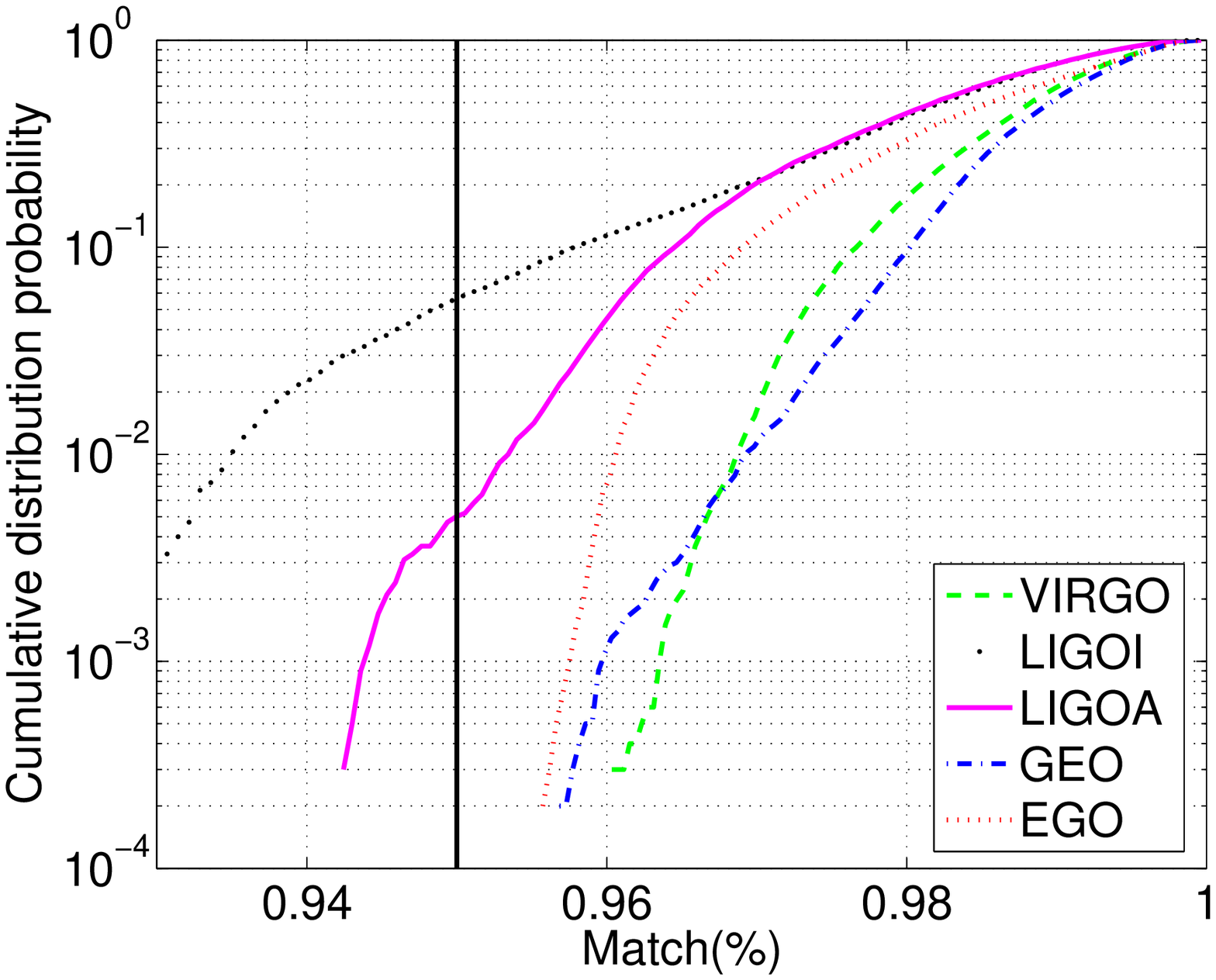}}
\caption{\label{fig:PadeT1} Hexagonal template bank efficiencies using PadeT1
model. From left ot right, results of the BNS, BHNS, and BBH
injections.}
\end{figure*}

The TaylorT1, TaylorT3 and PadeT1 models give very similar
results that are summarized in the Fig.~\ref{fig:TaylorT1},
\ref{fig:TaylorT3} and \ref{fig:PadeT1}. The safeness is greater than the
minimal match for the BNS and BHNS searches, for all  three waveforms.
More
precisely, $\mathcal{S}_\mathcal{R}\approx 95\%$ for BNS case, and it is
slightly over-efficient for BHNS case for total mass
above $20~M_\odot$, especially in the case of PadeT1 model. 
In the {\BBH} case, the bank is efficient between $[6\textrm{--}45]~M_\odot$.
Then, matches drop to 93\% for the same reason as in the case of {\SPA}
discussion. 
Therefore, we conclude that the proposed template bank is also efficient for
TaylorT1, TaylorT3 and PadeT1 models. 

\subsubsection{EOB}
\begin{figure*}[tbh]
\centering
{\includegraphics[trim =0 55mm 0 0, clip, width=0.3\textwidth]
{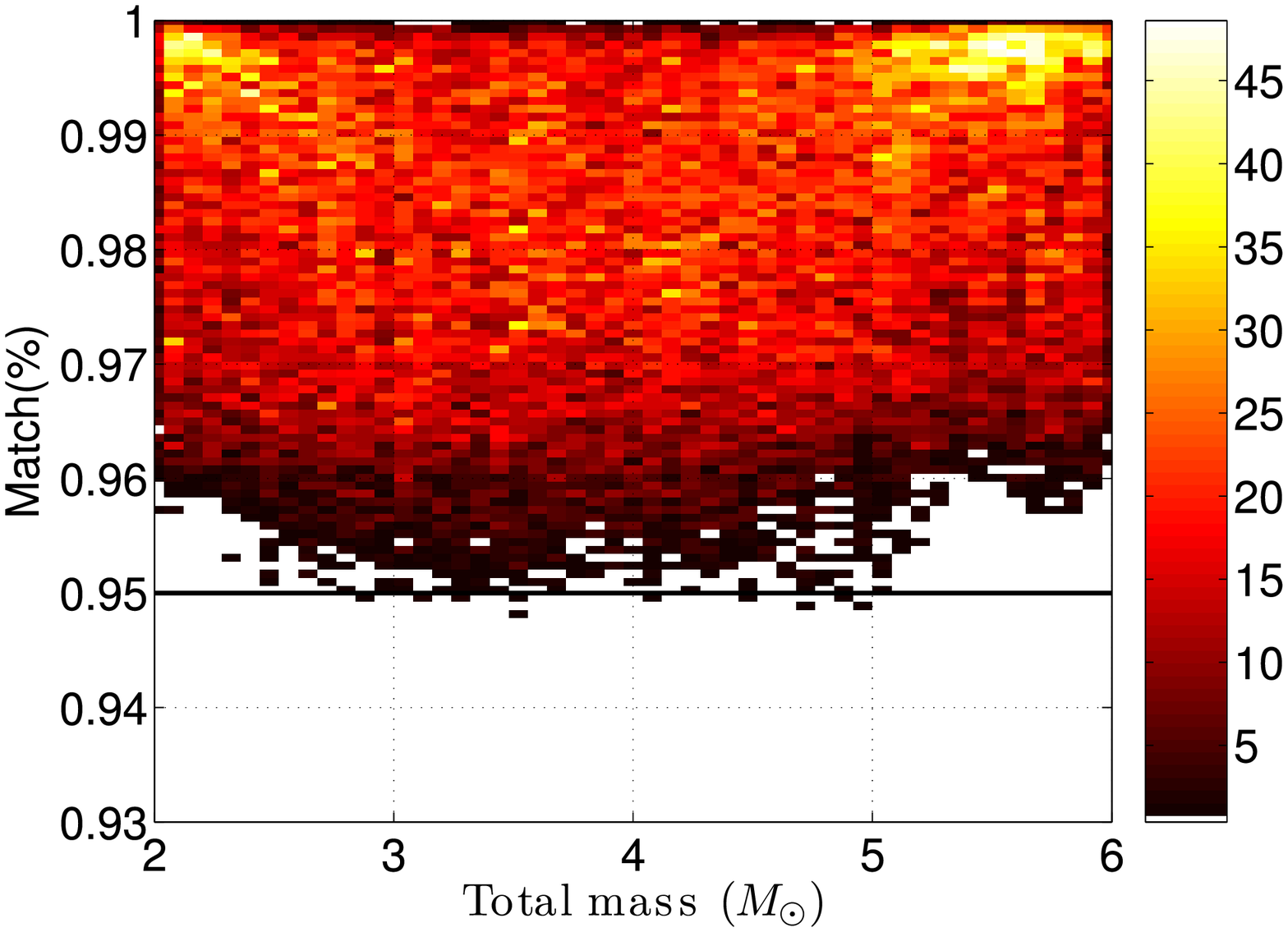}}
{\includegraphics[trim =0 55mm 0 0, clip, width=0.3\textwidth]
{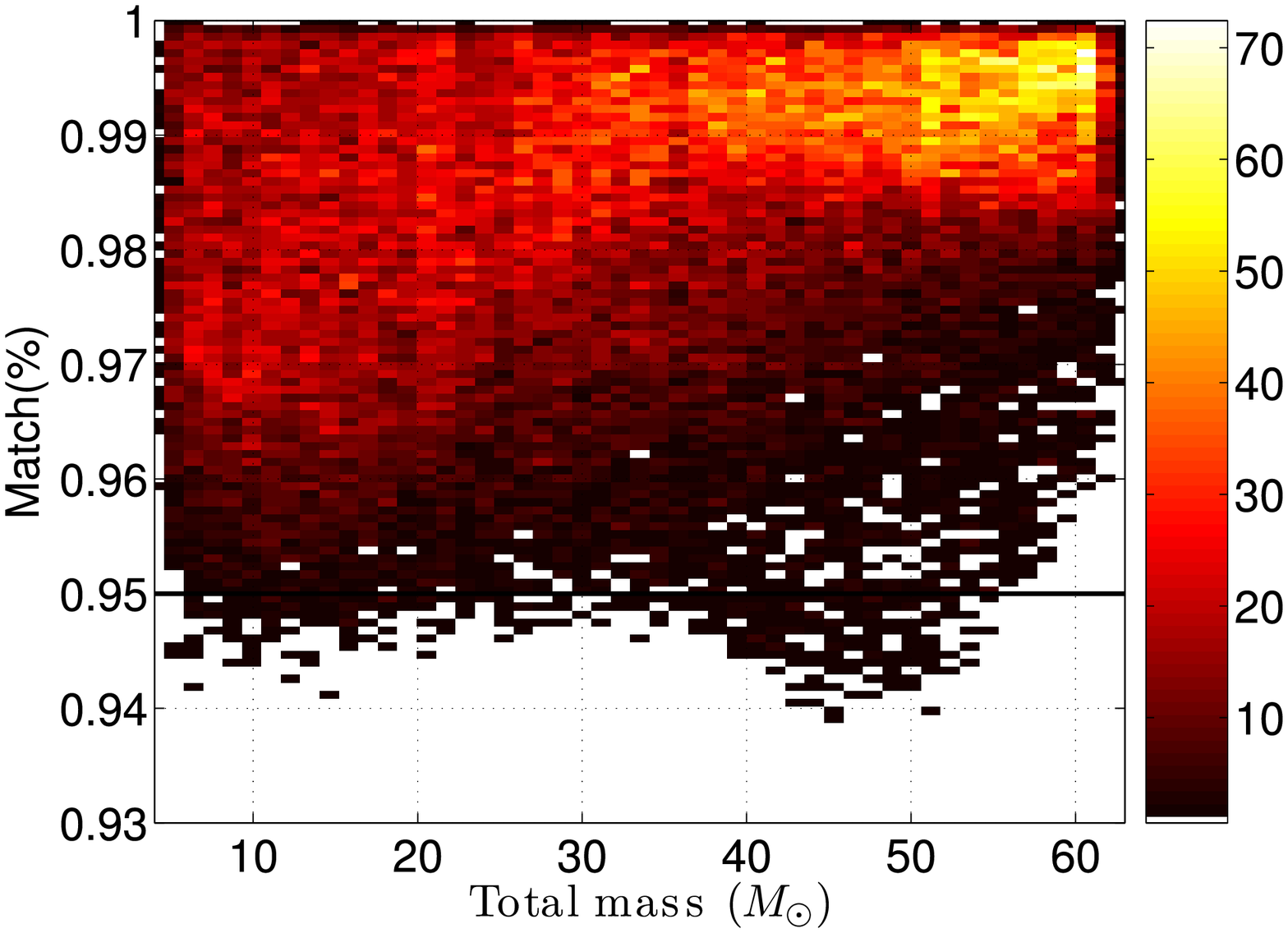}}
{\includegraphics[trim =0 55mm 0 0, clip, width=0.3\textwidth]
{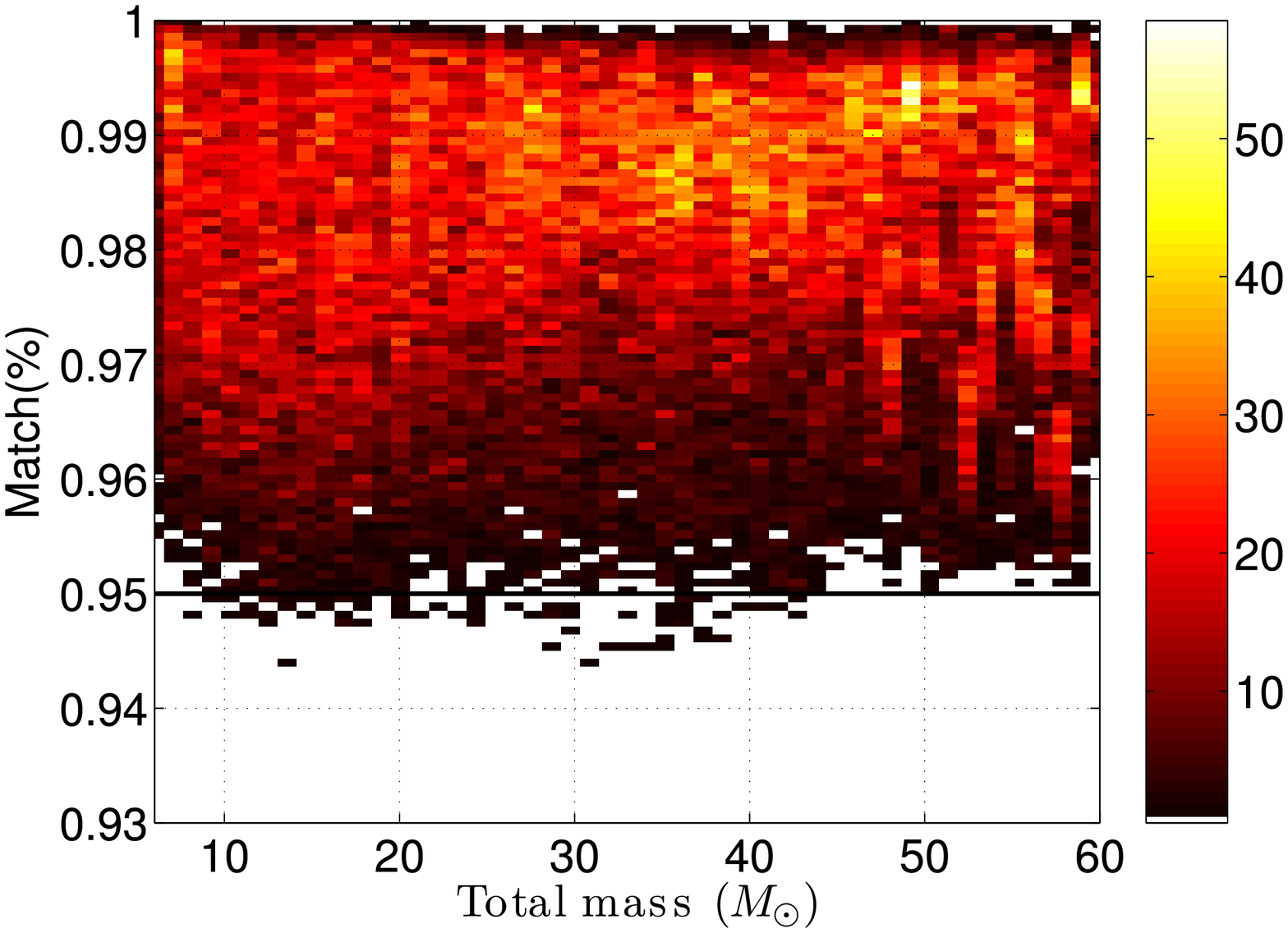}}\\
{\includegraphics[width=0.3\textwidth]
{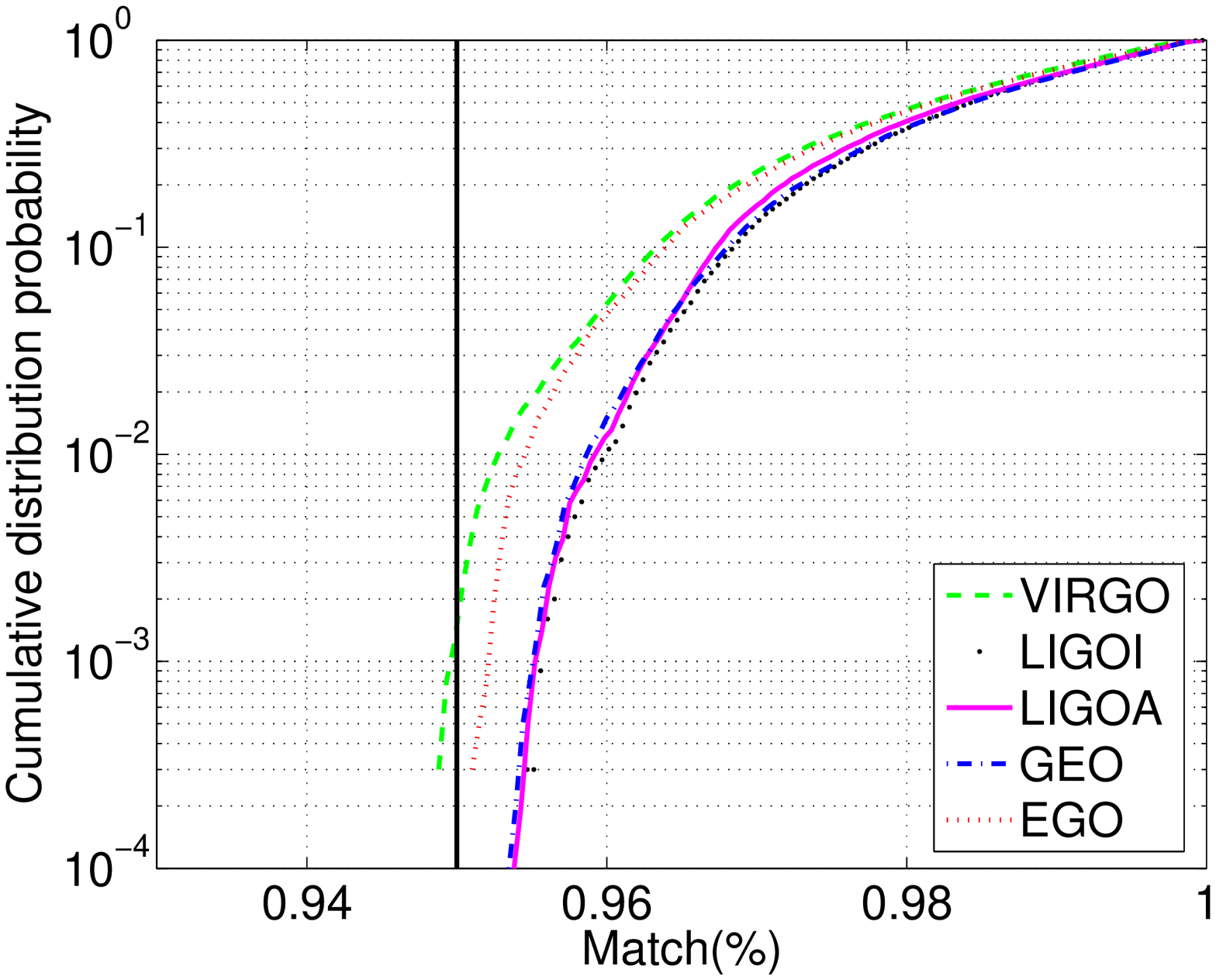}}
{\includegraphics[width=0.3\textwidth]
{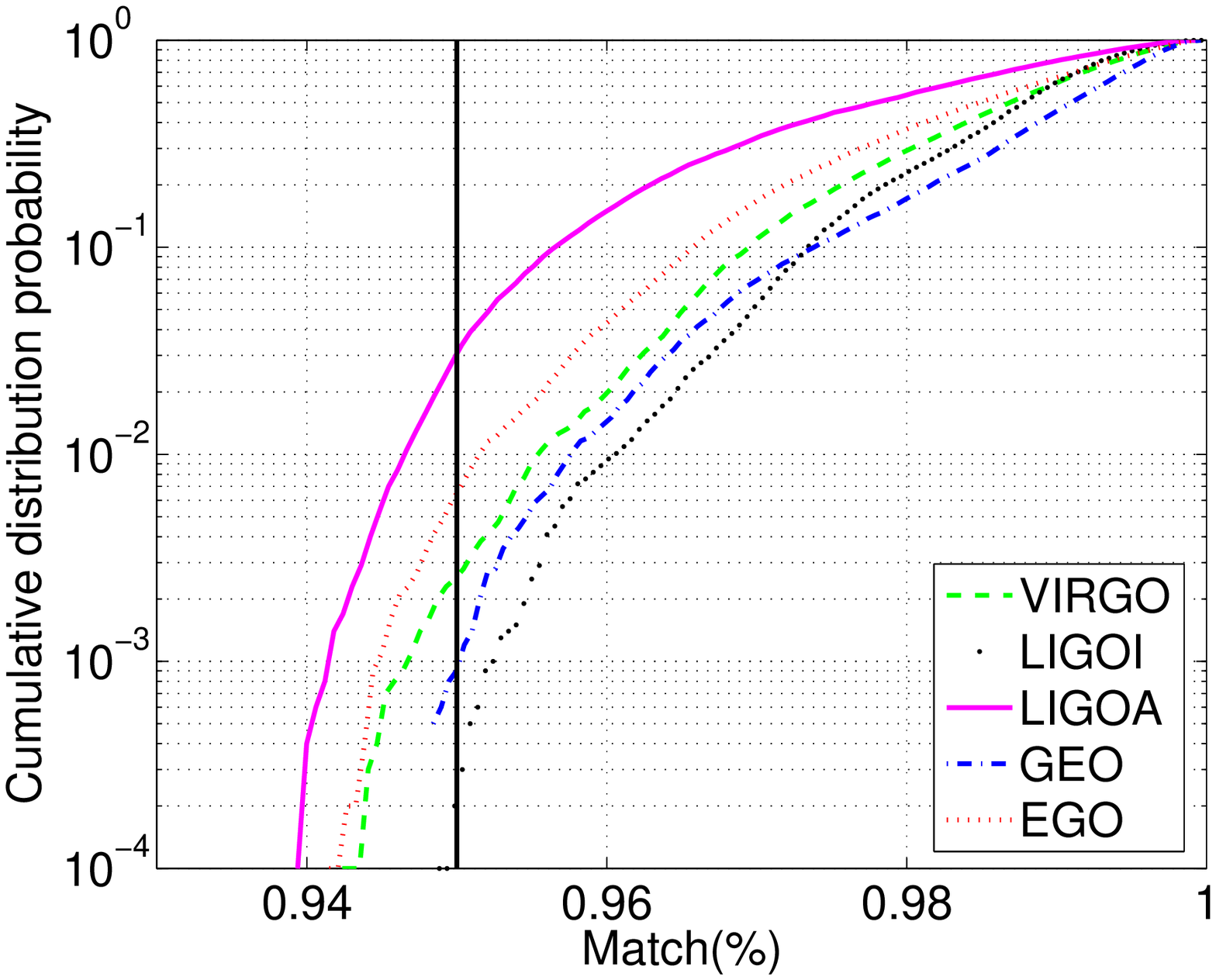}}
{\includegraphics[width=0.3\textwidth]
{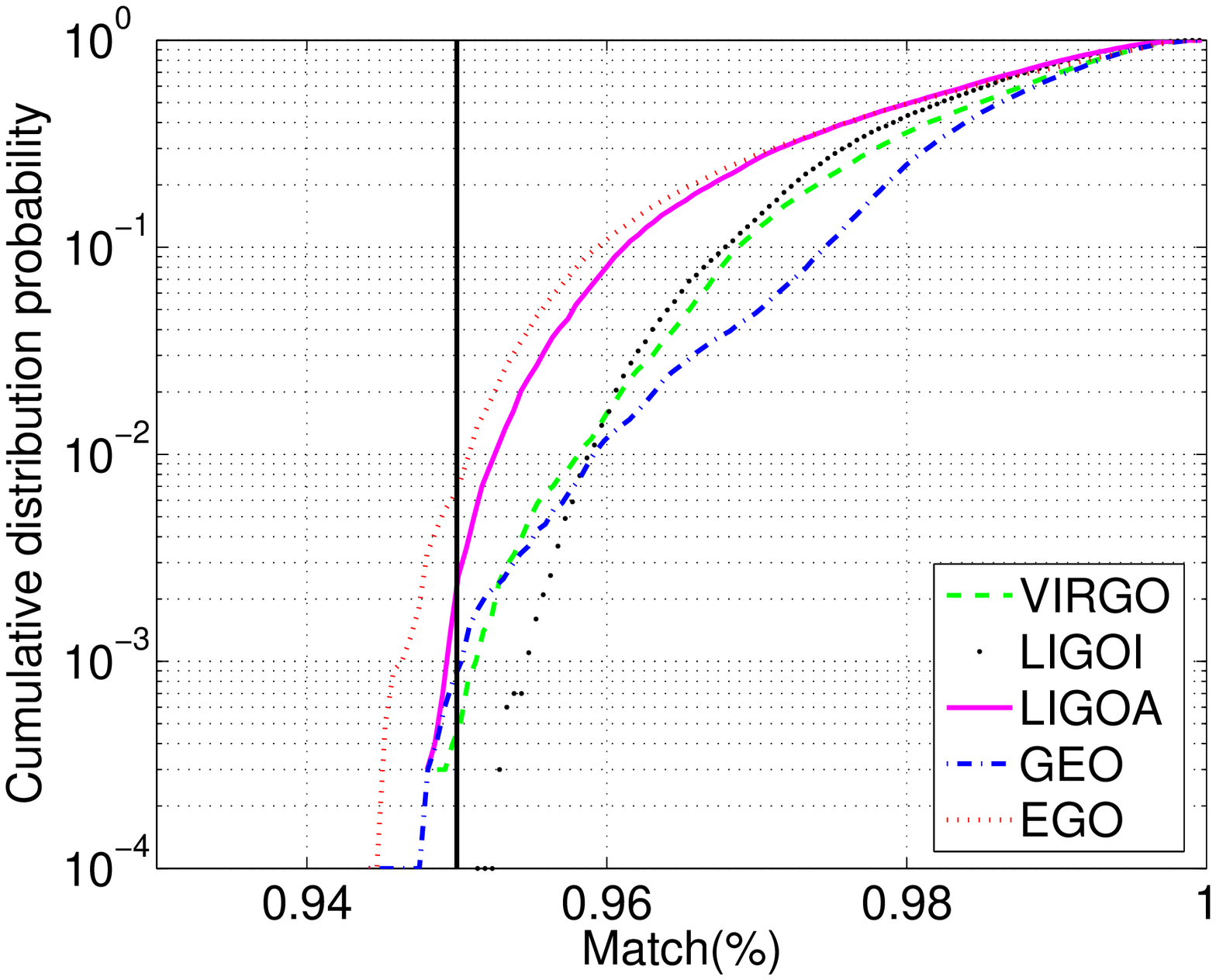}}
\caption{\label{fig:EOB} Hexagonal template bank efficiencies using EOB
model. From left ot right, results of the BNS, BHNS, and BBH
injections.}
\end{figure*}

We also investigate the efficiency of the hexagonal template bank using EOB templates and
signals. The EOB model is intrinsically different
from the previous models. The results  are summarized in Fig.~\ref{fig:EOB}. The
safeness is slightly under the requested minimal match ($\mathcal{S}_\mathcal{R}
= 94.5\% \approx 95\%$). The template bank is efficient for {\BNS},
{\BHNS} and {\BBH} cases. There is no over-efficiency noticed in any of the mass
range considered. We can also notice that the cumulative $\mathcal{E}_M$ drops quickly
and therefore we think that the proposed bank can be used with EOB model as well.
	
\section{Discussion and Conclusions}
In this paper, we described a hexagonal template bank placement for the search
of non-spinning inspiralling compact binaries in ground-based interferometers such as
LIGO. The placement is based on a metric computed on the signal manifold of
a stationary phase approximation model. The proposed hexagonal template bank size is
about 40\% smaller than the square template placement that was previously used to analyze
LIGO science runs (i.e., \cite{LIGOS3S4}). Yet, the matches between signal
and templates  are above the required minimal match. Therefore, the template bank
described in this paper is not over-efficient: it behaves as required. The main
consequence is a reduction of 40\% of the computational cost required to search
for inspiralling compact binaries with respect to previous searches.

The proposed template bank is not unique. Several parameters can be tuned such as the
sampling frequency, the final frequency used in the computation of the moments, the
placement of the template along one eigen-vector or the other, each of which can be
investigated in more detail. 

The bank was tested with the aid of many simulations that use design sensitivity
curves for advanced and current detectors, and various inspiralling compact
binaries with total mass between $[0.6\textrm{--}63] M_\odot$. We used a model
based on stationary phase approximation and showed that the template bank is efficient
for most of the parameter space considered. The higher end of the mass range was
slightly under efficient in the BBH case but this is partly related to the shortness of
the signal and templates considered. 

The proposed template bank can be used for various template families, not only the
stationary phase approximation family. In particular, we tested
the TaylorT1, TaylorT3, PadeT1, and EOB models at 2PN order, that
have been used for simulated injections in the various LIGO science runs. It
is interesting to see that the proposed template bank is efficient for most of the
models considered in this paper. It is also worth noticing that in some
cases the template bank is still over-efficient even though the bank size is
already reduced by 40\% (e.g., high mass  BHNS injections). 

The models that have been investigated in this paper are all based
on 2PN order, therefore template families based on higher PN-order should be
investigated. In the future, we also plan to consider the case of amplitude
corrected waveforms. All simulations presented in this paper use the same model for
both the template and signal generation. It would be interested to see how the
template bank performs when templates are based on one model (say, Pad\'e) and the
signals are from another (say, EOB).

This hexagonal template bank is currently used within the LIGO project to search for
non-spinning inspiralling compact binaries in the fifth science run.

\begin{acknowledgments}
This research was supported partly by Particle Physics and Astronomy
Research Council, UK, grant PP/B500731. The author thanks
Stas Babak for suggested the test of the bank with various template
families, and B.S.~Sathyaprakash and Gareth Jones for useful comments,
discussions, and corrections to this work. This paper has LIGO Document
Number LIGO-P070073-00-Z.
\end{acknowledgments}

\begin{appendix}
\section{Detector's Power Spectral Densities}\label{annex:PSDs}
The simulations that we performed use different PSD curves that are used to
compute the inner products (Eq.~\ref{eq:innerproduct}). The different
expressions provided uses the quantity $x=f/f_0$,
where $f$ is the frequency and $f_0$ is a constant. We summarize the
different design sensitivity curves that have been used in our simulations together
with the lower cut-off frequency $f_L$:
\begin{itemize}
\item The EGO PSD~\cite{VDB} is given by 
\begin{equation}
\begin{split}
S_h(f) =& S_0 \left\{  x^{p_1} + a_1x^{p_2} +\frac{}{}\right.\\
 & \left.a_2\frac{1+b_1x+b_2x^2+b_3x^3+b_4x^4+b_5x^5+b_6x^6}
{1+c_1x+c_2x^2+c_3x^3+c_4x^4}\right\}
\end{split}
\end{equation}
where $S_0=1.61 \times 10^{-51}$ and $f_0=200$~Hz. The other parameters are :\\
$p_1=-4.05$, $p_2=-0.69$, $a_1=185.62$, $a_2=232.52$,
$b_1=31.184$, $b_2=-64.72$, $b_3=52.24$, $b_4=-42.16$, $b_5=10.17$, $b_6=11.53$,
$c_1=13.58$, $c_2=-36.46$, $c_3=18.56$, and $c_4=27.43$.

The lower cut-off frequency is $f_L=20$~Hz.

\item The GEO PSD is given by~\cite{PSDs} 
\begin{equation}
\begin{split}
   S_h(f) &= S_0 \left\{ 10^{-46}x^{-30} +
            34 x^{-1} +  \frac{}{} \right. \\
& \left. 20\frac{\left [1 - x^2 + 0.5x^4 \right ] }{ 1 + 0.5x^2}  \right\}
\end{split} 
\end{equation}
where $S_0=10^{-46}$ and  $f_0 = 150$~Hz.
The lower cut-off frequency is $f_L=40$~Hz.

\item The LIGO-I PSD~\cite{PSDs} is given by
\begin{equation}
\begin{split}
   S_h(f) &= S_0 \left\{
	    (4.49x)^{-56} + \right.\\
           & \left. 0.16  x^{-4.52} + 0.52 +
            0.32 x^2
	\right\},
\end{split}
\end{equation}
where $S_0=9 \times 10^{-46}  $ and  $f_0 = 150$~Hz.
The lower cut-off frequency is $f_L=40$~Hz.

\item The advanced LIGO PSD is based on data provided in~\cite{PSDs} and given
by 
\begin{equation}
\begin{split}
    S_h(f) &= S_0\left\{  x^{-4.14} -
5x^{-2} + \right. \\
& \left. 111 \left(\frac{1 -     
x^2 + 0.5 x^4}{1 +
0.5x^2} \right)\right\},
\end{split}
\end{equation}
where  $S_0 = 10^{-49}$  and $f_0 = 215$~Hz.
The lower cut-off frequency is $f_L=20$~Hz.

\item Finally, the VIRGO PSD is based on data provided by J-Y.~Vinet and is
approximated by
\begin{equation}
   S_h(f) =
   S_0 \left\{ (7.87x)^{-4.8} + \frac{6}{17}x^{-1}
   + \left(1 + x^2 \right) \right\},
\end{equation}
where $S_0=10.2 \times 10^{-46}$
          with $f_0 = 500$~Hz.
The lower cut-off frequency is $f_L=20$~Hz.

\end{itemize}

\section{Parameter Space Tools}\label{annex:tools}
\subsection{Basic Relations}
Here is a summary of the relationship between individual masses $m_1,\;m_2$, and
the two chirptime parameters $\tau_0$ and $\tau_3$, that are given by  
\begin{equation}\label{eq:t0t3}
\tau_0 = \frac{5}{256 \pi f_L \eta} \left(\pi  M f_L\right)^{-5/3}, \ \
\tau_3 = \frac{1}{8  f_L \eta}  \left( \pi  M  f_L\right)^{-2/3},
\end{equation}
where $f_L$ is the lower cut-off frequency of the  template/signal,
$M=m_1+m_2$, and $\eta = m_1m_2/M^2$. The inversion is straightforward; $M$ and
$\eta$ are 
given by
\begin{equation}\label{eq:Meta}
M = \frac{5}{32 \pi^2 f_L} \frac{\tau_3}{\tau_0}, \ \
\eta = \frac{1}{8 f_L \tau_3}
\left (\frac{32\pi \tau_0}{5 \tau_3} \right )^{2/3}.
\end{equation}
It is convenient to introduce the constants $A_0$ and $A_3$ given by 
\begin{equation}
A_0 = \frac{5}{256 \left( \pi f_L  \right) ^{8/3}}, \ \
A_3 = \frac{\pi}{8\left(\pi f_L \right)^{5/3}} \;, 
\end{equation}
so that Eq.~\ref{eq:t0t3} becomes
\begin{equation}\label{eq:tau03Constants}
\tau_0 = \frac{A_0}{\eta} M^{-5/3}, \ \
\tau_3 = \frac{A_3}{\eta} M^{-2/3} \;.
\end{equation}
Finally, the chirp mass, $\mathcal{M}$, is given by
\begin{equation}
\mathcal{M} = \eta^{3/5} M 
\end{equation}
that allow $\tau_0$ to be expressed as a function of chirp mass only:
\begin{equation}
\tau_0 = A_0 \mathcal{M}^{-5/3}.
\end{equation}

\subsection{Parameter Space Boundaries relations}
The parameter space is defined by three boundaries (see Fig.~\ref{fig:space}).
On each of these boundaries, we want to express $\tau_3$ as a function of
$\tau_0$.  Using \ref{eq:tau03Constants}, we can eliminate $M$ and express
$\tau_3$ as a function of $\tau_0$ and $\eta$:
\begin{equation}\label{eq:t3t0eta}
\tau_3 =    \frac{  A_3}{\eta}  \left( \frac{\eta \tau_0}{A_0} \right)^{2/5}. 
\end{equation}
We can also eliminate $\eta$, and express $\tau_3$ as a function of $\tau_0$ and $M$:
\begin{equation}\label{eq:t3t0M}
\tau_3 = \frac{A_3}{A_0} \tau_0 M.
\end{equation}

The lower boundary corresponds to $m_1=m_2$, or $\eta=1/4$. Using Eq.~\ref{eq:t3t0eta}, we can express $\tau_3$ as a function of $\tau_0$ only
\begin{equation}
\tau_3 =     \left[ 4 A_3  \left( \frac{\tau_0}{4 A_0}
\right)^{2/5} \right]_{\eta=1/4}.
\end{equation}

The second boundary is defined by $m_1=m_{\rm min}$ and $m_2$ in $[m_{\rm min}-m_{\rm max}]$. 
The third boundary is defined by $m_1=m_{\rm max}$ and $m_2$ in $[m_{\rm min}-m_{\rm max}]$. 
On those two boundaries, we can assume that $m_1$ is set to one of the extremity of the mass range, denoted $m_{\rm e}$. Then $m_2= M-m_{\rm e}$, and $\eta=(M m_{\rm e})/(M-m_{\rm e})^2$. 
Starting from
\begin{equation}
\tau_0 =    \frac{  A_0}{\eta}  \left( M \right)^{-5/3}, 
\end{equation}
we replace $\eta$ by its expression as a function of $M$ and $m_{\rm e}$, 
and obtain after some algebra a cubic equation of the form
\begin{equation}
x^3 - px + q=0
\end{equation}
where $x = M^{1/3}$, $p = -A_0/(\tau_0/m_{\rm e})$ and $q= - m_{\rm
e}=0$, where $ m_{\rm e}$ is either set to $m_{\rm Min}$ or $m_{\rm Max}$
depending on which side of the parameter space we are. The solution for $x$
is standard and is given by 
\begin{eqnarray}
x &=&  \left(-\frac{q}{2}-\frac{1}{2}\sqrt{\frac{27 q^2 + 4
p^3}{27}}\right)^{\frac{1}{3}}+\\\nonumber
   && \left(-\frac{q}{2}+\frac{1}{2}\sqrt{\frac{27 q^2 + 4
p^3}{27}}\right)^{\frac{1}{3}}
\end{eqnarray}

We replace, $M=x^3$ in Eq.~\ref{eq:t3t0M} to obtain the value of $\tau_3$ on the
boundaries when $\tau_0$ is provided. 

\section{Flow Chart of the Hexagonal
Placement Algorithm}\label{annex:algorithm}
\begin{figure}[h]
{\includegraphics[width=0.48\textwidth,height=0.4\textheight]
{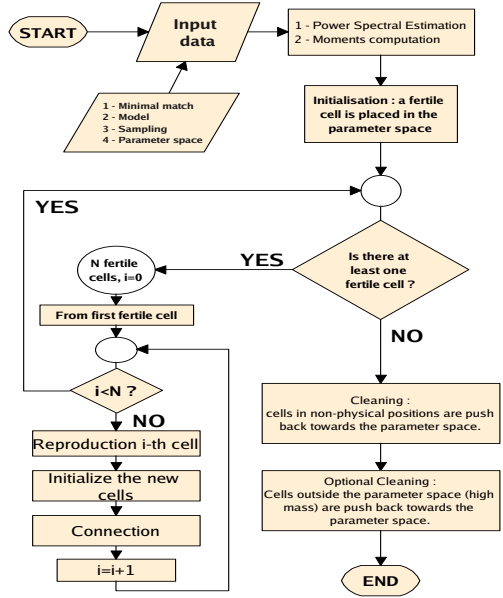}}
\caption{Flow chart of the hexagonal placement algorithm. See the text for
detailled description of the initialization, reproduction, and connection
process.}
\end{figure}

\end{appendix}

\label{theend}

\end{document}